\documentclass[10pt,aps,pra,showkeys,twocolumn,superscriptaddress]{revtex4-2}
\usepackage{amsmath, amssymb}
\usepackage{mathrsfs}
\usepackage{array}
\usepackage{booktabs}
\usepackage{tabu}
\usepackage{dcolumn}
\usepackage{amsmath}
\usepackage{amsfonts}
\usepackage{float}
\usepackage{amssymb}
\usepackage{graphicx,color}
\usepackage[colorlinks={true}]{hyperref}
\hypersetup{colorlinks=true,linkcolor=red,citecolor=blue,urlcolor=blue}
\usepackage{graphicx}
\usepackage{graphicx}% Include figure files
\usepackage{dcolumn}% Align table columns on decimal point
\usepackage{bm}% bold math
\usepackage{pstricks}
\usepackage{braket}
\usepackage{orcidlink}

\DeclareMathOperator{\csch}{csch}

\def\be{\begin{equation}}
  \def\ee{\end{equation}}
\def\bea{\begin{eqnarray}}
\def\eea{\end{eqnarray}}
\def\f{\frac}
\def\n{\nonumber}
\def\l{\label}
\def\p{\phi}
\def\o{\over}
\def\R{\rho}
\def\pa{\partial}
\def\om{\omega}
\def\na{\nabla}
\def\P{\Phi}
\begin{document} 
%%%%%%%%%%%%%%%%%%%%
\title{Magnetic Dipolar Quantum Battery with Spin-Orbit Coupling} 

\author{Asad Ali\orcidlink{0000-0001-9243-417X}} 
\email{asal68826@hbku.edu.qa}
\affiliation{Qatar Center for Quantum Computing, College of Science and Engineering, Hamad Bin Khalifa University, Doha, Qatar}

\author{Samira Elghaayda\orcidlink{0000-0002-6655-0465}}
\email{samira.elghaayda-etu@etu.univh2c.ma}
\affiliation{Laboratory of Mechanics and High Energy Physics, Department of Physics, Faculty of Sciences of Aïn Chock, Hassan II University, Casablanca 20100, Morocco} 

\author{Saif Al-Kuwari\orcidlink{0000-0002-4402-7710}}
\email{smalkuwari@hbku.edu.qa}
\author{M.I. Hussain\orcidlink{0000-0000-0000-0000}}
\author{M.T. Rahim\orcidlink{0000-0003-1529-928X}}
\author{Hashir~Kuniyil\orcidlink{0000-0003-0338-1278}} 
\affiliation{Qatar Center for Quantum Computing, College of Science and Engineering, Hamad Bin Khalifa University, Doha, Qatar}

\author{Tim Byrnes}
\affiliation{New York University Shanghai, NYU-ECNU Institute of Physics at NYU Shanghai, Shanghai Frontiers Science Center of Artificial Intelligence and Deep Learning, Shanghai 200126, China}
\affiliation{State Key Laboratory of Precision Spectroscopy, School of Physical and Material Sciences, East China Normal University, Shanghai 200062, China}
\affiliation{Center for Quantum and Topological Systems (CQTS), NYUAD Research Institute, New York University Abu Dhabi, UAE}
\affiliation{Department of Physics, New York University, New York, NY 10003, USA}

\author{James Q. Quach}
\affiliation{The University of Adelaide, SA 5005, Australia}

\author{Mostafa Mansour\orcidlink{0000-0003-0821-0582}} 
\email{mostafa.mansour.fsac@gmail.com}
\affiliation{Laboratory of Mechanics and High Energy Physics, Department of Physics, Faculty of Sciences of Aïn Chock, Hassan II University, Casablanca 20100, Morocco}

\author{Saeed Haddadi\orcidlink{0000-0002-1596-0763}} 
\email{haddadi@ipm.ir}
\affiliation{School of Particles and Accelerators, Institute for Research in Fundamental Sciences (IPM), P.O. Box 19395-5531, Tehran, Iran}

\date{\today}% It is always \today, today,
\def\be{\begin{equation}}
  \def\ee{\end{equation}}
\def\bea{\begin{eqnarray}}
\def\eea{\end{eqnarray}}
\def\f{\frac}
\def\n{\nonumber}
\def\l{\label}
\def\p{\phi}
\def\o{\over}
\def\R{\rho}
\def\pa{\partial}
\def\om{\omega}
\def\na{\nabla}
\def\P{$\Phi$}
\begin{abstract}
We investigate a magnetic dipolar system influenced by the $z$-component of Zeeman splitting, Dzyaloshinsky–Moriya (DM) interaction, and Kaplan–Shekhtman–Entin-Wohlman–Aharony (KSEA) exchange interaction, with emphasis on the role of quantum resources in both closed and open settings. By analyzing the Gibbs thermal state and solving the Lindblad master equation, we study the behavior of quantum coherence, discord, and entanglement under thermal equilibrium and dephasing noise. After exploring these resources, we apply the model to a closed quantum battery (QB). Our results show that while Zeeman splitting degrades quantum resources in noisy and thermal regimes, it enhances QB performance by improving ergotropy, anti-ergotropy, storage capacity, and coherence during cyclic charging. The axial parameter further amplifies performance, leading to coherence saturation and persistent ergotropy growth, in line with the notion of incoherent ergotropy. KSEA interaction and the rhombic term consistently preserve coherence and entanglement under noise, thereby strengthening QB functionality. DM interaction mitigates thermal degradation of resources in the Gibbs state and improves performance, though its effect is limited under Pauli-$X$ dephasing. We reveal diverse behaviors, including increased ergotropy without coherence and the coexistence of coherence with zero extractable work. Finally, we propose Nuclear Magnetic Resonance (NMR) as a feasible platform for experimental implementation.
\end{abstract}

\keywords{{magnetic dipolar systems, quantum coherence, spin-orbit interaction, quantum battery, ergotropy, anti-ergotropy, capacity of quantum battery.}}
\maketitle

\section{INTRODUCTION}\label{sec:1} Energy is the fundamental requirement for the occurrence and progression of all physical processes in nature. Therefore, the pursuit of innovative techniques for efficient energy flow and storage is always a worthwhile endeavor. In this context, the search for a quantum advantage in battery technology has inspired the development of quantum batteries (QBs), which have the potential to revolutionize the energy and power industries \cite{quach2020organic,kamin2020entanglement,kamin2020non,arjmandi2022enhancing,arjmandi2023localization,yang2023battery,ferraro2018high,campaioli2018quantum,binder2015quantacell,ali2024ergotropy,Ali_2024,alicki2013entanglement,quach2022superabsorption,PRXQuantum.5.030319,PhysRevLett.132.210402,Rodríguez_2024,Gyhm2024,Wang01,Wang02,Wang03,mojaveri2024extracting,Haseli2024RINP,HaddadiQB2024,PhysRevA.107.032218,PhysRevA.109.032201,PhysRevA.111.042216,PhysRevA.110.052601,PhysRevLett.133.243602,PhysRevLett.133.197001,GRAZI2025116383}. Compared to traditional batteries \cite{dell2001understanding,luo2015overview}, QBs harness the principles of quantum mechanics for energy storage in quantum systems where quantum superposition allows the simultaneous use of multiple quantum states, providing higher energy densities, fast charging speed, and enhanced lifespans \cite{ferraro2018high,campaioli2018quantum,binder2015quantacell,ali2024ergotropy}. Understanding energy transfer at the fundamental level may also provide new insights into quantum thermodynamics and applications in quantum computing. 

Experimental realizations of QBs are limited, however, recent studies demonstrated a QB with superconducting qutrits \cite{Hu_2022,AsadAliAQT2025}. They showed that such QBs are optimized for stable charging and feature a self-discharge mechanism akin to supercapacitors, suggesting efficient energy storage in superconducting circuits. Similarly, other experimental evidence demonstrated quantum advantage in QB charging using Nuclear Magnetic Resonance (NMR) star-topology spin systems \cite{joshi2022experimental}, topological QBs \cite{lu2024topologicalquantumbatteries}, and the charging behavior in organic QBs \cite{quach2022superabsorption}. Recently, QB charging with single photons in a linear optics setup has been investigated, and demonstrations using single photons further highlight their practical potential \cite{huang2023demonstration}. 

\begin{figure}[!]
    \centering
\includegraphics[width=.95\columnwidth]{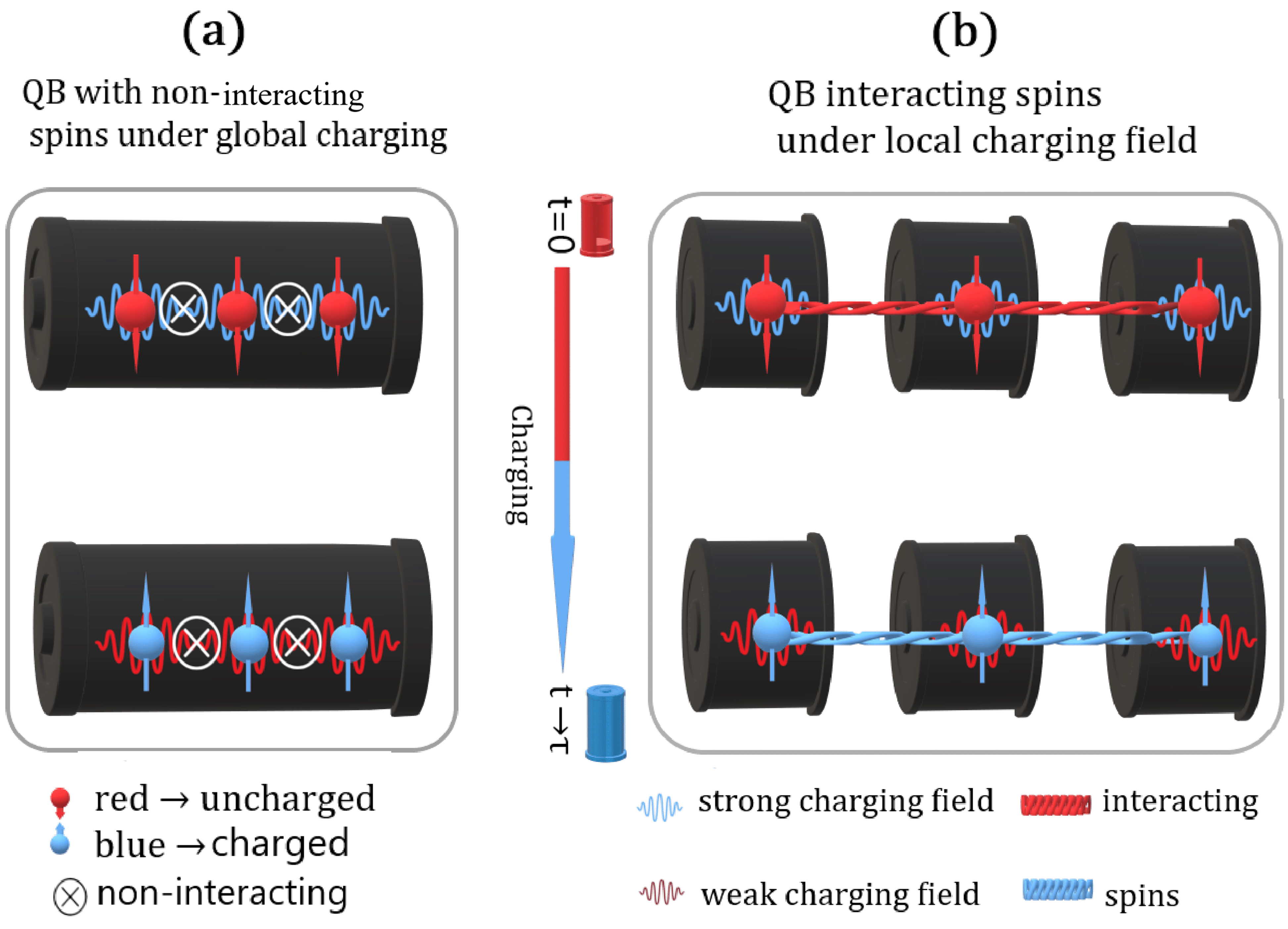}
    \caption{(a) Non-interacting quantum cell-based QB with a global charging from a common cavity field. (b) Interacting quantum cell-based QB with a local charging field stored in each cavity, including cell interactions.}
    \label{f1}
    \end{figure}

QBs are realized through isolated or interacting quantum cell designs \cite{le2018spin}, as illustrated in Fig. \ref{f1}.{ For isolated quantum cells, each cell operates independently. However, for interacting quantum cells, spin-chain models can be used to simulate many-body systems.} Although QBs offer transformative potential for energy storage, overcoming challenges related to decoherence and dissipation is the key to their development and advancements \cite{barra2019dissipative}. Addressing these issues through control strategies and reservoir engineering is essential for robust quantum energy storage \cite{carrega2020dissipative}.

Magnetic dipolar systems have garnered significant interest due to their quantum correlations and practical applications \cite{hu2018effect,bovca2006magnetic,reis2013fundamentals,ren2017ground,batista2024magnetic}. These systems, including Josephson qubits, polar molecular crystals, cold atoms, and cold polar molecules, benefit from controllable dipole-dipole interactions \cite{kruckenhauser2020quantum,peter2012anomalous, martinis2005decoherence, zhou2015quantum, rabl2007molecular, buchler2007three}. The Dzyaloshinsky-Moriya (DM) interaction \cite{dzyaloshinsky1958thermodynamic, moriya1960anisotropic} and the Kaplan–Shekhtman–Entin-Wohlman–Aharony (KSEA) interaction \cite{kaplan1983single, shekhtman1992moriya, shekhtman1993bond} play crucial roles in the evolution of quantum correlations. The KSEA interaction has been observed in materials like Yb$_4$As$_3$ and La$_2$CuO$_4$ \cite{shiba2000effective, ex2, ex3}, while the DM interaction's unique features, such as chiral N\'eel  domain barriers and skyrmions, suggest potential applications in spin models \cite{ex5, ex6}.

\subsection{ Contribution}
\label{subsec1A}
This paper explores magnetic dipolar spin systems as magnetic dipolar QBs, leveraging their quantum coherence and correlations for improved energy injection, extraction, and capacity. We have modeled two magnetic dipoles with symmetric and antisymmetric exchange interactions (KSEA and DM) under the uniform Zeeman splitting field in the $z$-direction.

The motivations for this study are twofold: (i) to thoroughly analyze  and correlations under thermodynamic equilibrium at finite temperature and Lindbladian dynamics at absolute zero, ensuring a solid understanding of quantum resources in the Hamiltonian of working medium before modeling a QB out of it—much like assessing available ``fuel" before starting a journey; (ii) to investigate the Gibbs state as the QB’s initial state, charged by an external transverse magnetic field aligned with the Zeeman splitting field, while tracking quantum resources through the quantum coherence, akin to monitoring ``fuel” levels throughout the trip.
This study also covers whether key model parameters affecting quantum resources similarly impact QB performance indicators. Given that QB performance hinges on quantum resources like superposition and entanglement for energy storage and delivery, we analyze these resources under equilibrium, dephasing, and during QB charging with the same model Hamiltonian as the working medium. This approach ensures a coherent integration of QB modeling with quantum resource evaluation. Finally, we have suggested that the NMR platform holds the potential for effectively simulating these magnetic dipolar QBs \cite{doronin2007dipolar,doronin2007multiple,fel2024relaxation,levitt2008spin}.

{
\subsection{Organization}
\label{subsec1B}The structure of this paper is organized as follows. In Sec.~\ref{sec2}, we investigate the magnetic dipolar spin model, providing a detailed analysis of the interactions, the derivation of the thermal equilibrium state, and the solution of the Lindblad master equation under Pauli-$X$ noise. Building upon the theoretical framework established in Sec.~\ref{sec2}, Sec.~\ref{sec4} presents a comprehensive examination of different quantum resources constituting a hierarchical structure, considering both Lindbladian dephasing processes and thermal equilibrium conditions, followed by a detailed analysis of quantum resources in magnetic dipolar systems. In Sec.~\ref{sec5}, we analyze magnetic dipolar systems as closed QB and provide a thorough investigation of their properties. Finally, Sec.~\ref{sec6} concludes with a summary of our key findings and outlines promising directions for future research.}
{
\section{Magnetic Dipolar–Spin Model with Spin-Orbit Interaction}\label{sec2}
QBs exploit coherent many-body dynamics to outperform classical energy storage in charging speed and energy density. We suggest that magnetic dipolar systems offer a natural platform for QBs due to their long-range interactions and ability to generate entangled states crucial for quantum-enhanced performance.

In materials with broken inversion symmetry or significant spin-orbit coupling, additional anisotropic interactions emerge—most notably the antisymmetric DM and symmetric KSEA interactions. These spin-orbit driven terms, when combined with dipolar coupling, enrich the dynamical landscape and enable tunable Hamiltonians favorable for QB operation in NMR platform \cite{micadei2019reversing,kuprov2023spin,chizhik2014magnetic,anikeevajournal,reis2013fundamentals}.

The classical dipolar interaction between two magnetic dipoles $\vec{\mu}_1$ and $\vec{\mu}_2$ separated by vector $\vec{r}$ is
\begin{equation}
H_{\text{cl}} = \frac{\mu_0}{4\pi r^3}\left[\vec{\mu}_1 \cdot \vec{\mu}_2 - 3(\vec{\mu}_1 \cdot \hat{r})(\vec{\mu}_2 \cdot \hat{r})\right],
\end{equation}
with $\hat{r} = \vec{r}/r$ and $\mu_0$ is magnetic permeability of free space. In quantum systems, magnetic moments map to spin operators: $\hat{\vec{\mu}}_i = -g\mu_B\hat{\vec{S}}_i = -\tfrac{g\mu_B\hbar}{2} \hat{\vec{\sigma}}_i$, where $\hat{\vec{\sigma}}_i$ are Pauli vectors and the constant $\mu_B$ represents Bohr magneton. The resulting quantum dipolar Hamiltonian becomes \cite{joseph2025decoupling,reis2013fundamentals}
\begin{equation}
\hat{H}_{\text{dip}} = \frac{\mu_0(g\mu_B)^2}{16\pi r^3} \left[\hat{\vec{\sigma}}_1 \cdot \hat{\vec{\sigma}}_2 - 3(\hat{\vec{\sigma}}_1 \cdot \hat{r})(\hat{\vec{\sigma}}_2 \cdot \hat{r})\right].
\end{equation}

This can be compactly written using a symmetric traceless tensor 
\begin{equation}
\hat{P} = \mathrm{diag}\{\Delta - 3\epsilon,\, \Delta + 3\epsilon,\,-2\Delta\},
\end{equation}
yielding the form \cite{kuprov2023spin,chizhik2014magnetic,anikeevajournal,reis2013fundamentals}

\begin{equation}
\hat{H}_{\text{dip}} = \hat{\vec{\sigma}}_1^\intercal \cdot \hat{P} \cdot \hat{\vec{\sigma}}_2,
\end{equation}
where $\Delta$ controls axial and $\epsilon$ controls rhombic anisotropy respectively.

In systems with strong spin-orbit coupling or broken inversion symmetry, the magnetic dipolar interactions can be supplemented by anisotropic terms. The DM interaction, allowed in such symmetry conditions, takes the form \cite{micadei2019reversing,kuprov2023spin} 

\begin{equation}
\hat{H}_{\text{DM}} = \vec{d} \cdot (\hat{\vec{\sigma}}_1 \times \hat{\vec{\sigma}}_2),
\end{equation}
where $\vec{d} = (D_x, D_y, D_z)$ defines the DM vector. Expanding this yields
\begin{align}
\hat{H}_{\text{DM}} &= D_x(\hat{\sigma}_1^y \hat{\sigma}_2^z - \hat{\sigma}_1^z \hat{\sigma}_2^y) 
+ D_y(\hat{\sigma}_1^z \hat{\sigma}_2^x - \hat{\sigma}_1^x \hat{\sigma}_2^z) \nonumber\\
&\quad + D_z(\hat{\sigma}_1^x \hat{\sigma}_2^y - \hat{\sigma}_1^y \hat{\sigma}_2^x).
\end{align}

The symmetric KSEA interaction, also arising from spin-orbit coupling, preserves time-reversal symmetry and introduces spatial anisotropy:
\begin{equation}
\hat{H}_{\text{KSEA}} = \hat{\vec{\sigma}}_1 \cdot \bm{G} \cdot \hat{\vec{\sigma}}_2,
\end{equation}
with
\begin{equation}
\bm{G} = \begin{pmatrix}
0 & G_z & G_y \\
G_z & 0 & G_x \\
G_y & G_x & 0
\end{pmatrix}.
\end{equation}
Expanded, this yields
\begin{align}
\hat{H}_{\text{KSEA}} &= G_x(\hat{\sigma}_1^y \hat{\sigma}_2^z + \hat{\sigma}_1^z \hat{\sigma}_2^y)
+ G_y(\hat{\sigma}_1^z \hat{\sigma}_2^x + \hat{\sigma}_1^x \hat{\sigma}_2^z) \nonumber\\
&\quad + G_z(\hat{\sigma}_1^x \hat{\sigma}_2^y + \hat{\sigma}_1^y \hat{\sigma}_2^x).
\end{align}

Both interactions can be synthetically engineered in NMR platforms via tailored pulse sequences using average Hamiltonian theory \cite{micadei2019reversing}.
The desired DM interaction in the $z$-direction $D_z=D$
\begin{equation}
\bar{H}_{\text{DM}}^{(z)} = D(\hat{\sigma}_1^x \hat{\sigma}_2^y - \hat{\sigma}_1^y \hat{\sigma}_2^x),
\end{equation} 
can be experimentally realized using a carefully designed pulse sequence. For instance, the sequence 
$(\pi/2)_x - \tau - (\pi/2)_y - \tau - (\pi/2)_{-x} - \tau - (\pi/2)_{-y} - \tau$
effectively generates this interaction. Here, $(\pi/2)_x$ and similar terms denote $\pi/2$ rotations around the specified axes, and $\tau$ represents free evolution periods under the natural scalar coupling. The sequence works by systematically transforming the scalar coupling $\hat{\sigma}_1^z \hat{\sigma}_2^z$ into the desired cross-product terms through a series of rotations. Specifically, the $(\pi/2)_x$ pulse rotates the spins to align along the $y$-axis, while subsequent $(\pi/2)_y$ and $(\pi/2)_{-x}$ pulses further manipulate the spin states to produce the antisymmetric $\hat{\sigma}_1^x \hat{\sigma}_2^y - \hat{\sigma}_1^y \hat{\sigma}_2^x$ form characteristic of the DM interaction. This method in NMR platform, as demonstrated in the experiment by \cite{micadei2019reversing}, leverages the interplay between free evolution and controlled RF pulses to engineer the effective Hamiltonian required. 
In the same spirit, one can generate the KSEA interaction in the $z$-direction $G_z=G$
\begin{equation}
\bar{H}_{\text{KSEA}}^{(z)} = G(\hat{\sigma}_1^x \hat{\sigma}_2^y + \hat{\sigma}_1^y \hat{\sigma}_2^x),
\end{equation}
with $D$ and $G$ tunable via pulse parameters.

Combining the $z$-components of DM and KSEA interactions with the dipolar term and external field, the full two-qubit Hamiltonian for the QB model is
\begin{align}\label{eq4}
\hat{\mathcal{H}}_{\text{QB}} &= D(\hat{\sigma}_1^x \hat{\sigma}_2^y - \hat{\sigma}_1^y \hat{\sigma}_2^x) 
+ G(\hat{\sigma}_1^x \hat{\sigma}_2^y + \hat{\sigma}_1^y \hat{\sigma}_2^x) \nonumber\\
&\quad - \tfrac{1}{3} \hat{\vec{\sigma}}_1^\intercal \cdot \hat{P} \cdot \hat{\vec{\sigma}}_2 
+ B(\hat{\sigma}_1^z \otimes \mathbb{I} + \mathbb{I} \otimes \hat{\sigma}_2^z),
\end{align}
where $B$ is the external magnetic field strength and $\mathbb{I}$ is the identity operator.

This Hamiltonian encapsulates the interplay between dipolar coupling, spin-orbit-induced anisotropic interactions, and Zeeman splitting. The tunable parameters $D$, $G$, and $B$ define operational regimes critical for optimizing charging protocols and coherence properties. NMR's high controllability and coherence times make it an ideal platform for implementing and benchmarking QB performance.
}

Now, the eigenvalues and corresponding normalized eigenstates of the Hamiltonian $\mathcal{\hat{H}_{\mathcal{QB}}}$ can be determined through a straightforward calculation, which would be given by
\begin{eqnarray}
\begin{aligned}[c]
\nu_{1}&=-2(\Delta + \kappa_{1})/3, \\
\nu_{2}&= 2(-\Delta + \kappa_{1})/3, \\
\nu_{3}&= 2(\Delta - 3\kappa_{2})/3, \\
\nu_{4}&= 2(\Delta + 3\kappa_{2})/3,
\end{aligned}
\qquad
\begin{aligned}[c]
|\psi_{1} \rangle &= q \left( -\eta |01\rangle + |10\rangle \right), \\
|\psi_{2}\rangle &= q \left( \eta |01\rangle + |10\rangle \right), \\
|\psi_{3}\rangle &= \chi_{2} \left( \delta_{1} |00\rangle + |11\rangle \right), \\
|\psi_{4}\rangle &= \chi_{1} \left( \delta_{2} |00\rangle + |11\rangle \right), \\
\end{aligned}
\label{eq2}
\end{eqnarray}
where $\kappa_{1} = \sqrt{9D^{2} + \Delta^{2}}$, $\kappa_{2} = \sqrt{B^{2} + G^{2} + \epsilon^{2}}$, $\eta = \frac{2iD - \Delta}{\kappa_{1}},$ 
$\delta_{1} = \frac{i(-B + \kappa_{2})}{G - i\epsilon},$ $\delta_{2} = -\frac{i(B + \kappa_{2})}{G - i\epsilon}$, 
$q = \frac{1}{\sqrt{1 + |\eta|^{2}}}$, $\chi_{1,2} = \frac{1}{\sqrt{1 + |\delta_{1,2}|^{2}}}$.
\begin{figure}[!t]
\begin{center}
\includegraphics[width=0.48\textwidth]{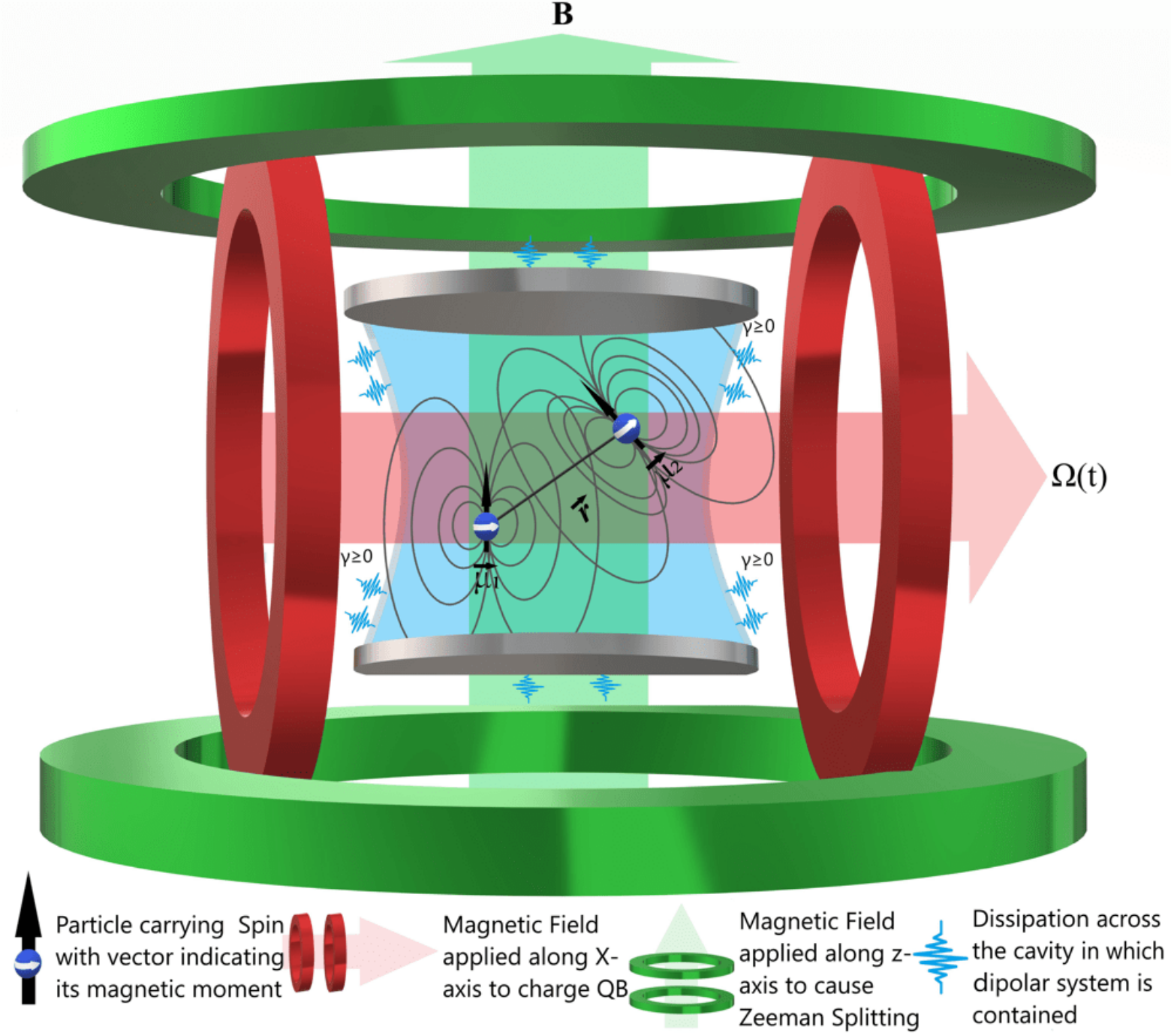}	
\end{center}
\caption{Two magnetic dipoles with moments $\vec{\mu}_1$ and $\vec{\mu}_2$ in 3D space, separated by distance $\vec{r}$. A Zeeman field $B$ is applied along the $z$-axis, and a time-dependent field $\Omega(t)$ is applied along the $x$-axis to charge the dipolar spins for QB.}
\label{f2}
\end{figure}
\subsection{At thermal equilibrium}
\label{subsec2A}
The Gibbs state at thermal equilibrium can be obtained by a general Hamiltonian $\mathcal{H}$ as follows:
\begin{equation}
\hat{\zeta }= \frac{1}{\mathcal{Z}} e^{-\beta \mathcal{H}} = \frac{1}{\mathcal{Z}}\sum_{s} \exp\left(-\beta\nu_{s}\right) \ket{\psi_{s}} \bra{\psi_{s}},
\label{eq3}
\end{equation}
where $\mathcal{Z} = \text{Tr}[\exp(-\beta \mathcal{H})]$ is the partition function, with $\nu_{s}$ denoting the eigenenergies corresponding to eigenstates $\ket{\psi_{s}}$. Moreover, $\beta=1/k_B T$ is the inverse temperature where $k_B$ denotes the Boltzmann constant, set to 1 for convenience, and $T$ is the absolute temperature.

{ The thermal density operator Eq. \eqref{eq3} on the computational basis of two qubits $\mathcal{B}={\{\ket{00}, \, \ket{01}, \, \ket{10}, \, \ket{11}\}
}$ can be determined using the spectral decomposition of $\mathcal{\hat{H}_{\mathcal{QB}}}$ presented in Eq. \eqref{eq2}. By considering $\mathcal{\hat{H}_{\mathcal{QB}}}$ and using Eqs. \eqref{eq2} and \eqref{eq3}, one can arrive at the following Gibbs thermal state in the standard computational basis $\mathcal{B}$, namely
\begin{equation}
\hat{\zeta} = \left(
\begin{array}{cccc}
 \zeta_{11} & 0 & 0 & \zeta_{14} \\
 0 & \zeta_{22} & \zeta_{23} & 0 \\
 0 & \zeta_{23}^* & \zeta_{22} & 0 \\
 \zeta_{14}^* & 0 & 0 & \zeta_{44} \\
\end{array}
\right).\label{EQ7}
\end{equation}
The nonzero elements of the above matrix, eigenvalues, and corresponding eigenstates are reported in Appendix \ref{App0}.}

\subsection{The role of dissipation}
\label{subsec2B}
The Lindblad master equation models the time evolution of the density matrix $\hat{\varrho}(t)$ in an open quantum system, accounting for coherent evolution by the Hamiltonian $\mathcal{\hat{H}_{\mathcal{QB}}}$ and dissipative effect \cite{bochkin2019exact,johansson2012qutip,ullah2023low}. It is given by (setting $\hbar =1$)

\begin{equation}
\small
\frac{d\hat{\varrho}(t)}{dt} = -i[\mathcal{\hat{H}_{\mathcal{QB}}}, \hat{\varrho}(t)] + \sum_{k} \left( \hat{C}_k \hat{\varrho}(t) \hat{C}_k^\dagger - \frac{1}{2} \left\{ \hat{C}_k^\dagger \hat{C}_k, \hat{\varrho}(t) \right\} \right),
\end{equation}
where $\hat{C}_k$ are collapse operators modeling dephasing. For our study, we have
\begin{equation}\label{eeq9}
\hat{C}_1 = \sqrt{\gamma} \, \hat{\sigma}_x \otimes \hat{\mathbb{I}}, \quad \hat{C}_2 = \sqrt{\gamma} \, \hat{\mathbb{I}} \otimes \hat{\sigma}_x,
\end{equation}
with $\gamma$ as the dephasing rate and $\hat{\sigma}_x$ as the Pauli-$X$ operator. The equation integrates coherent dynamics and dissipative effects to describe the time evolution of the system.

The use of Pauli-$X$ collapse operators in Eq.~\eqref{eeq9} reflects their direct alignment with dominant decoherence mechanisms in NMR systems, particularly transverse ($T_2$) relaxation. This form of noise models random phase shifts in the $x$–$y$ plane of the Bloch sphere, arising from magnetic field fluctuations due to nuclear spins and field inhomogeneities, which are the primary sources of decoherence in NMR experiments. These processes degrade quantum coherence, impacting QB charging and work extraction efficiency. Within the Lindblad framework, Pauli-$X$ noise captures the essential decoherence dynamics and supports the development of control protocols for optimizing QB performance under realistic conditions~\cite{fel2024relaxation,bochkin2020many}. Experimentally, it can be precisely engineered via tailored magnetic field gradients~\cite{krogmeier2024low,onizhuk2024decoherence}. Focusing on fast $T_2$ dynamics—rather than slower $T_1$ processes ensures a physically motivated and experimentally grounded model for studying QB optimization in decohering environments~\cite{fel2012solid}.

\section{Quantum resource dynamics}\label{sec4}

Before evaluating QB performance, we must understand how model parameters affect quantum resources measured by $l_1$-norm of quantum coherence (see Appendix \ref{App1}), quantum discord (Appendix \ref{App2}), and concurrence (Appendix \ref{App3}). This involves analyzing the dephasing and thermal equilibrium states to identify parameters that enhance quantum coherence and correlations, and their impact on anti-ergotropy, capacity and ergotropy. We aim to determine whether maximizing coherence and correlations optimizes the QB performance metrics, clarifying the relationship between these quantum resources and ergotropy extraction.

\subsection{Analysis of quantum resources}
Considering the dephasing dynamics of all quantum resources as shown in Fig.~\ref{figure3}, we find the following results:

\begin{figure*}[!t]
    \centering
    
    \begin{minipage}[b]{0.32\textwidth}
        \centering
        \includegraphics[width=\textwidth]{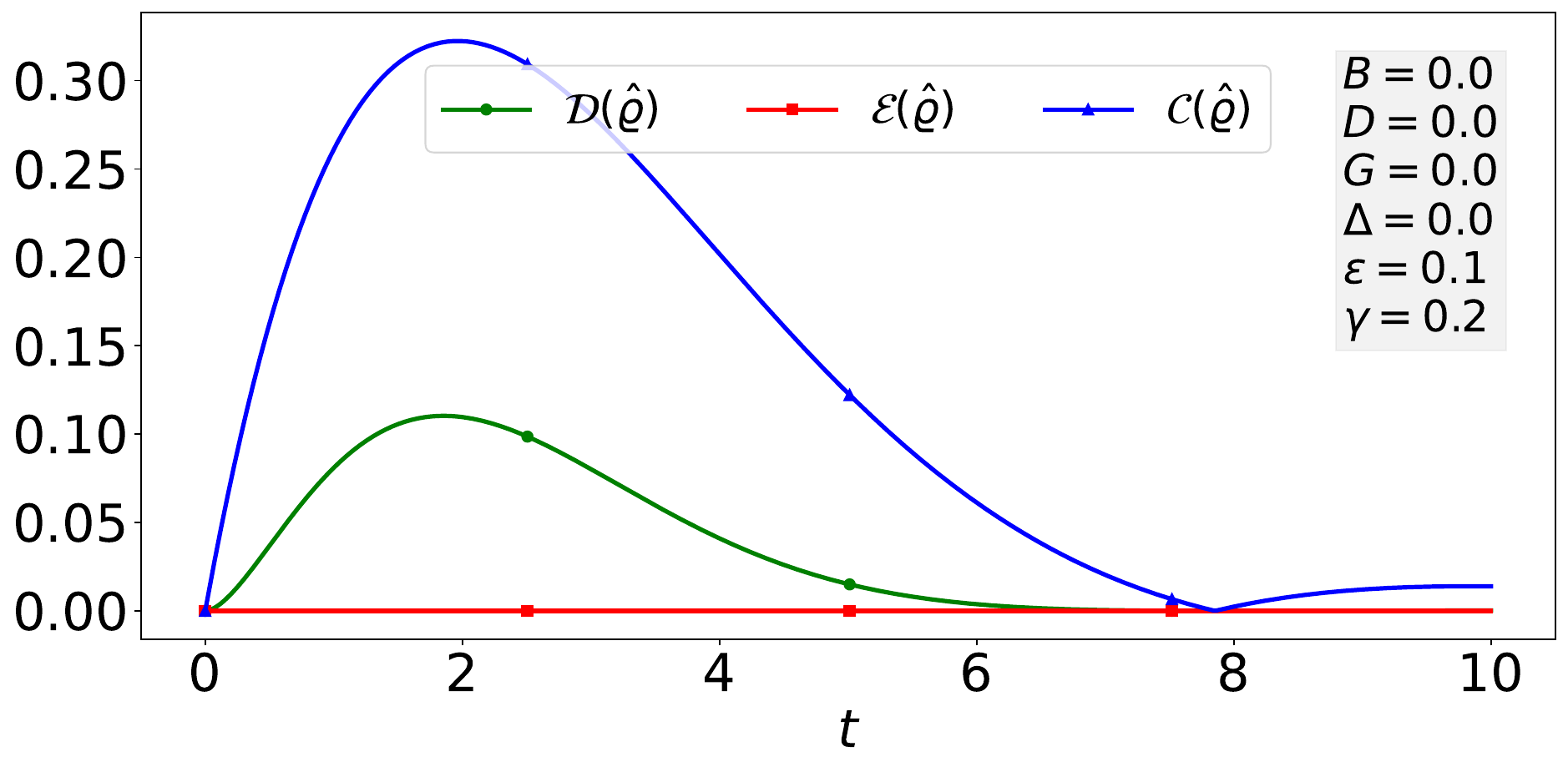}
        \put(-23,30){$(a)$}
    \end{minipage}%
    \begin{minipage}[b]{0.32\textwidth}
        \centering
        \includegraphics[width=\textwidth]{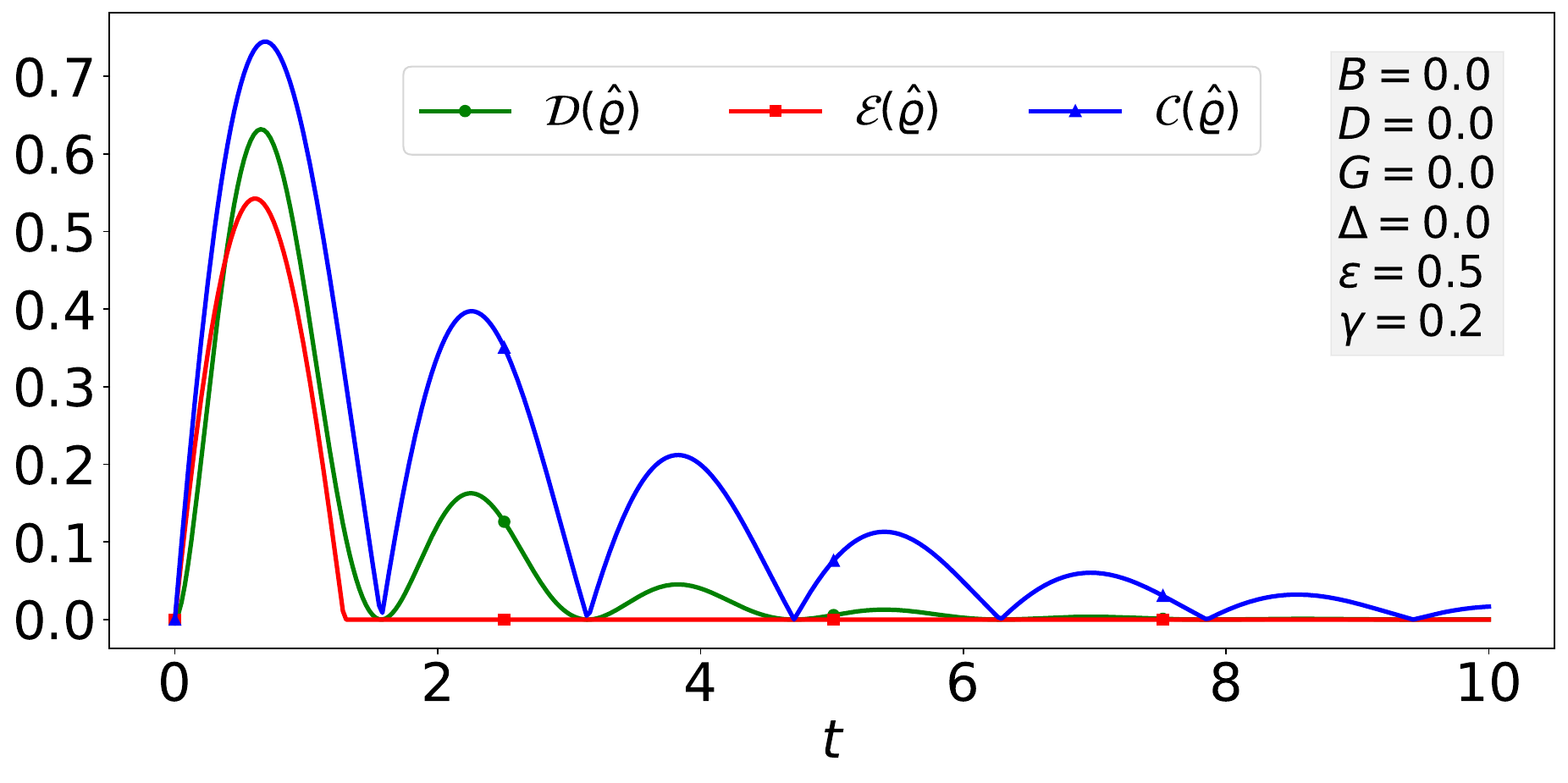}
        \put(-23,30){$(b)$}
    \end{minipage}%
    \begin{minipage}[b]{0.32\textwidth}
        \centering
        \includegraphics[width=\textwidth]{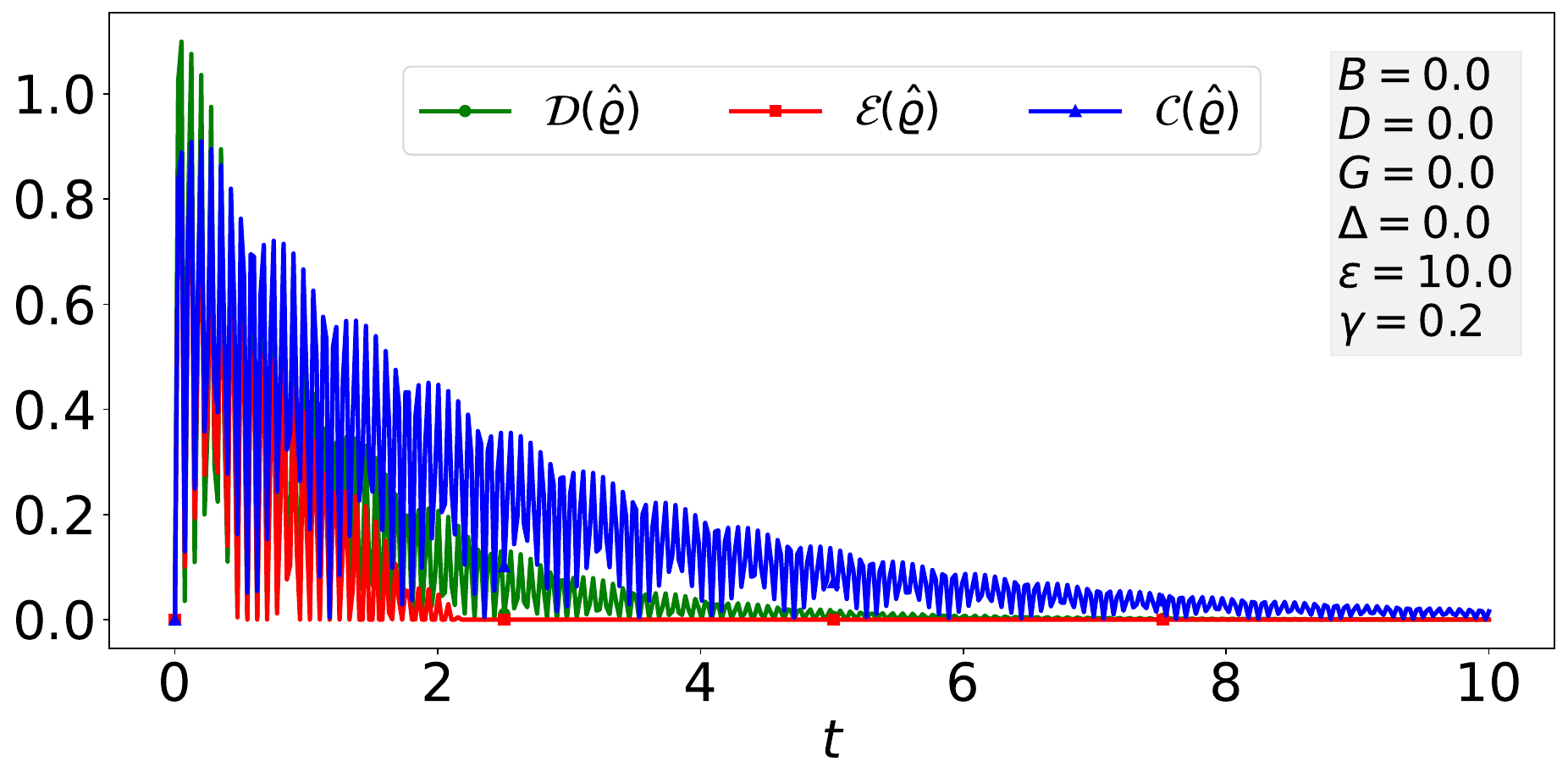}
        \put(-23,30){$(c)$}
    \end{minipage} \\

    \begin{minipage}[b]{0.32\textwidth}
        \centering
        \includegraphics[width=\textwidth]{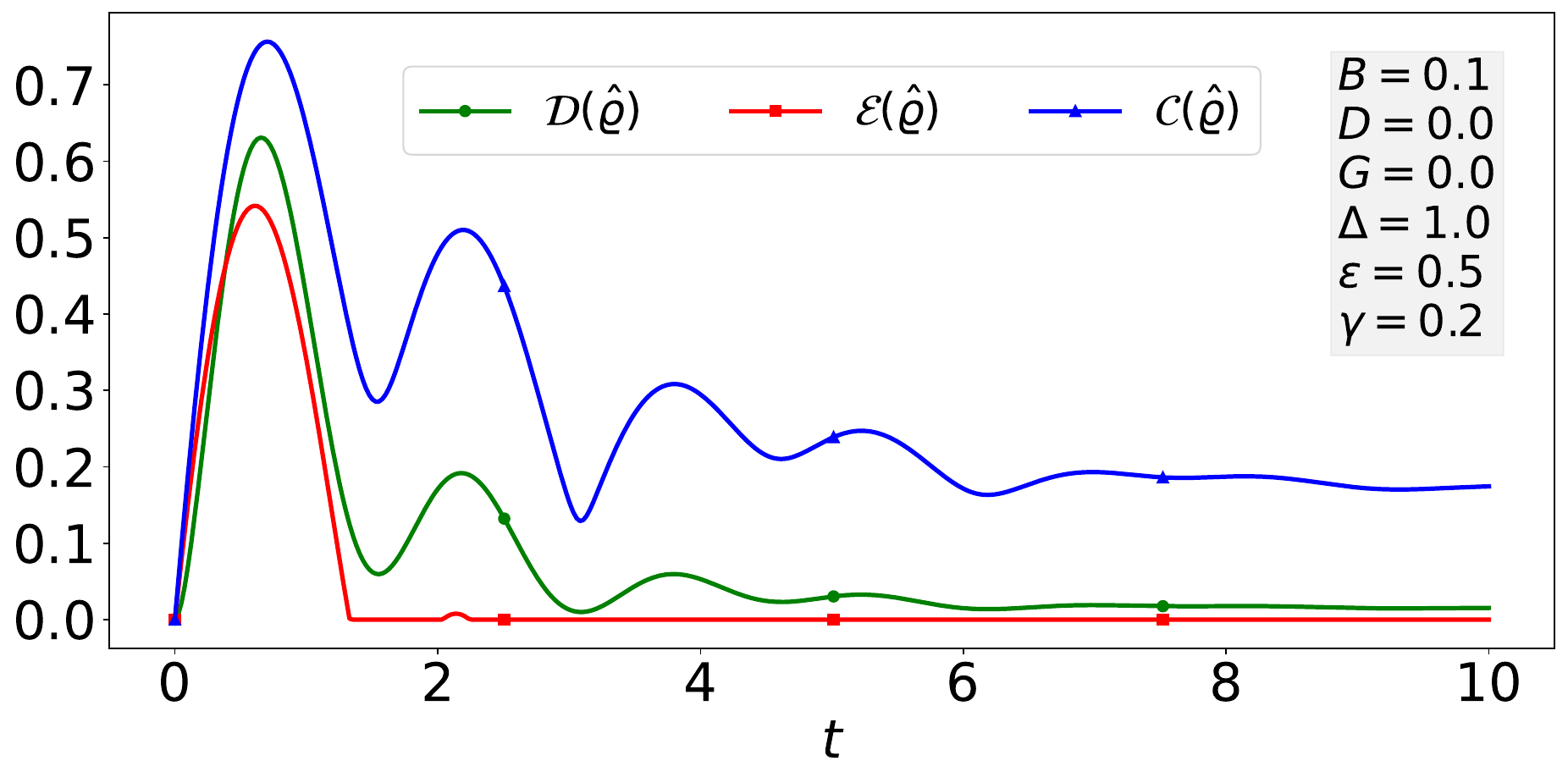}
        \put(-23,30){$(d)$}
    \end{minipage}%
    \begin{minipage}[b]{0.32\textwidth}
        \centering
        \includegraphics[width=\textwidth]{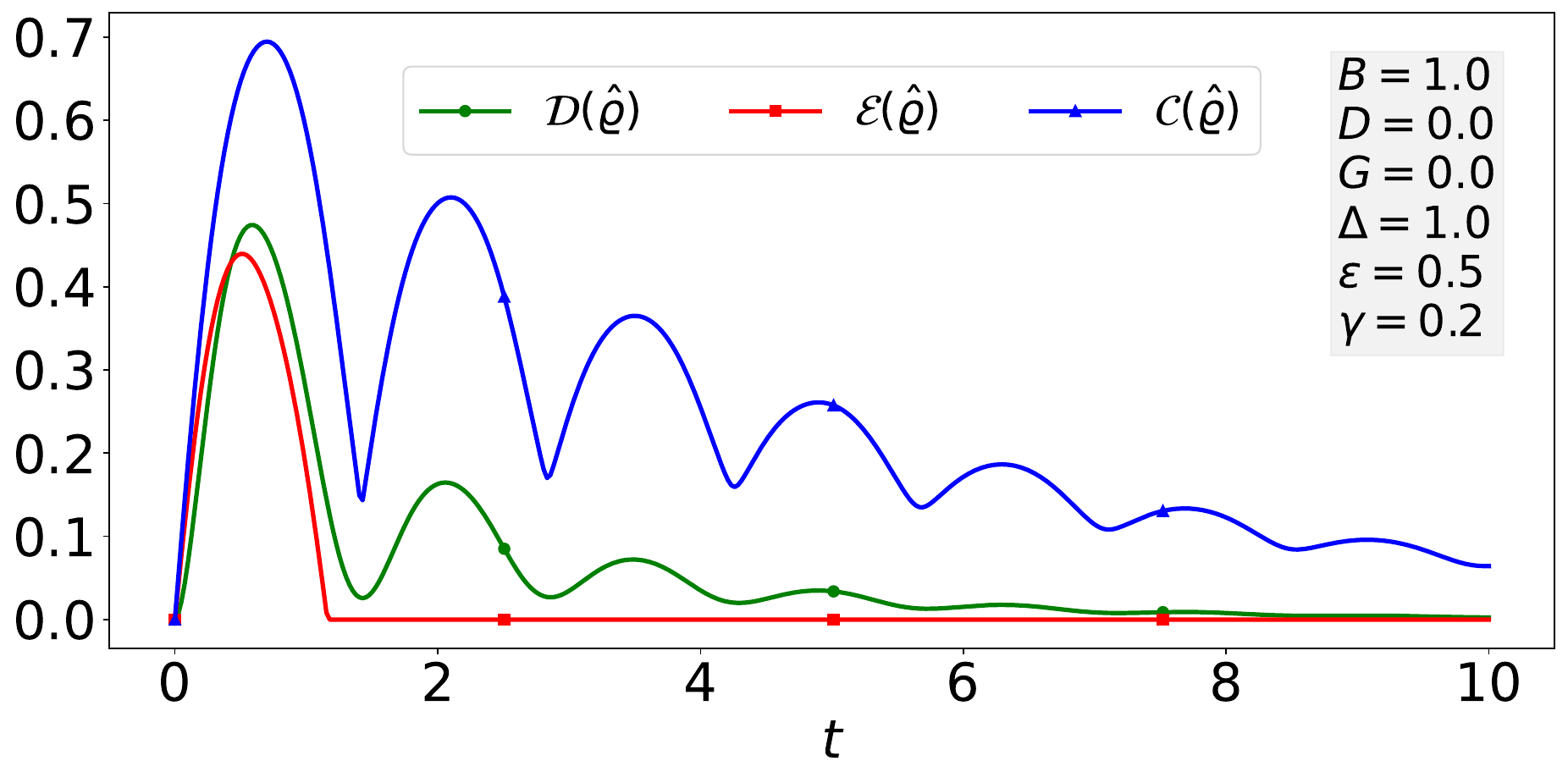}
        \put(-23,30){$(e)$}
    \end{minipage}%
    \begin{minipage}[b]{0.32\textwidth}
        \centering
        \includegraphics[width=\textwidth]{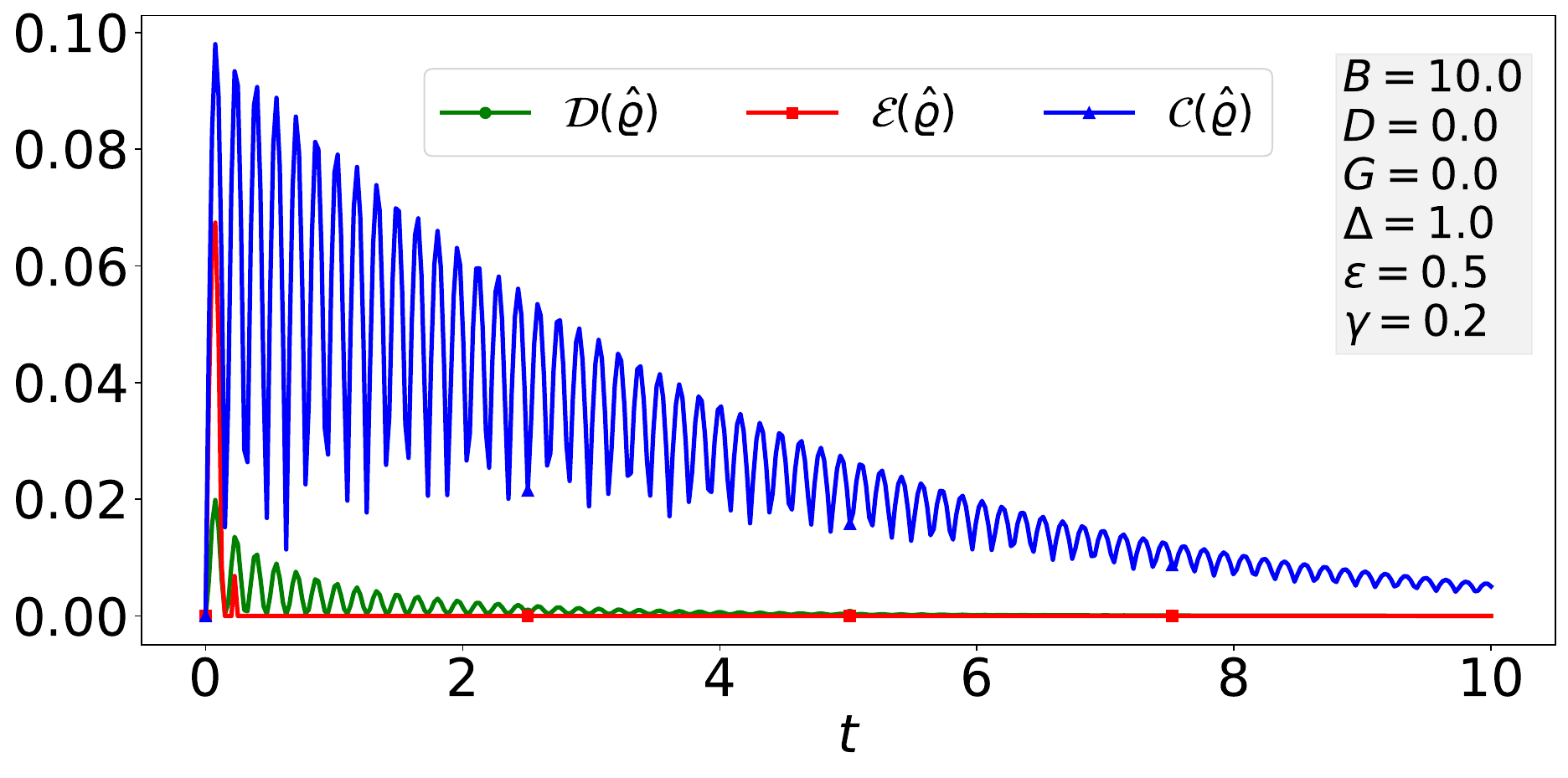}
        \put(-23,30){$(f)$}
    \end{minipage} \\

    \begin{minipage}[b]{0.32\textwidth}
        \centering
        \includegraphics[width=\textwidth]{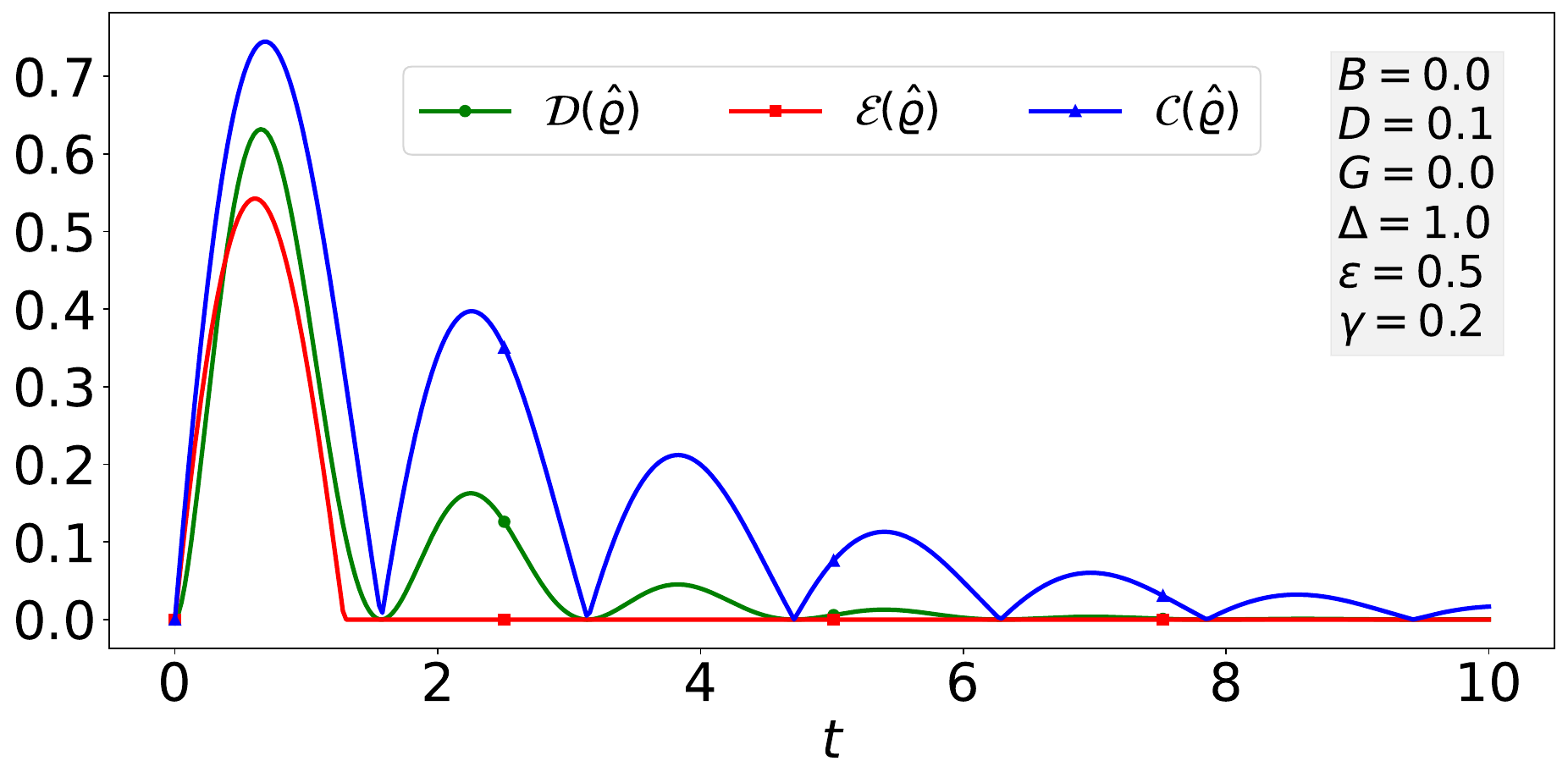}
        \put(-23,30){$(g)$}
    \end{minipage}%
    \begin{minipage}[b]{0.32\textwidth}
        \centering
        \includegraphics[width=\textwidth]{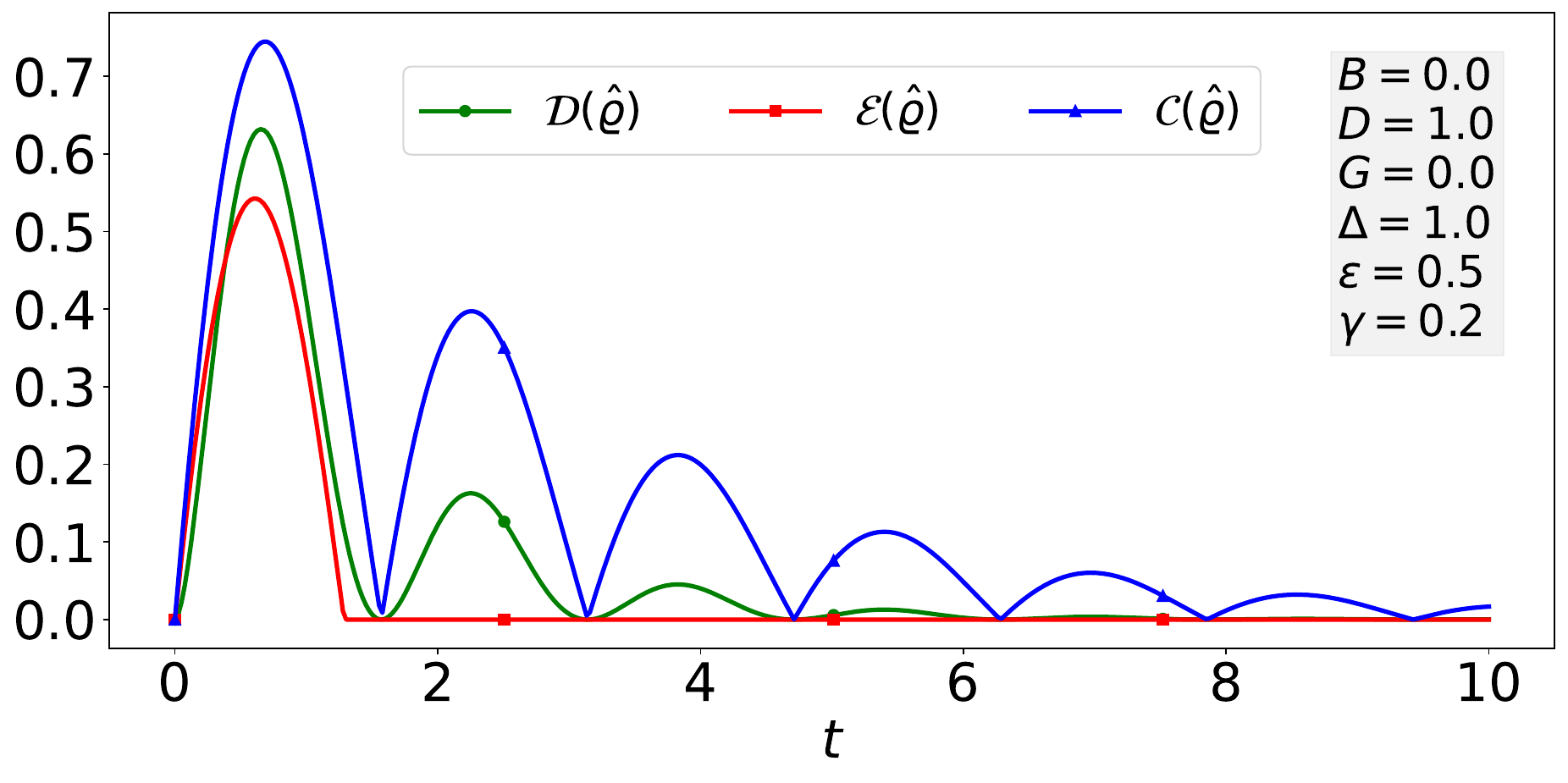}
        \put(-23,30){$(h)$}
    \end{minipage}%
    \begin{minipage}[b]{0.32\textwidth}
        \centering
        \includegraphics[width=\textwidth]{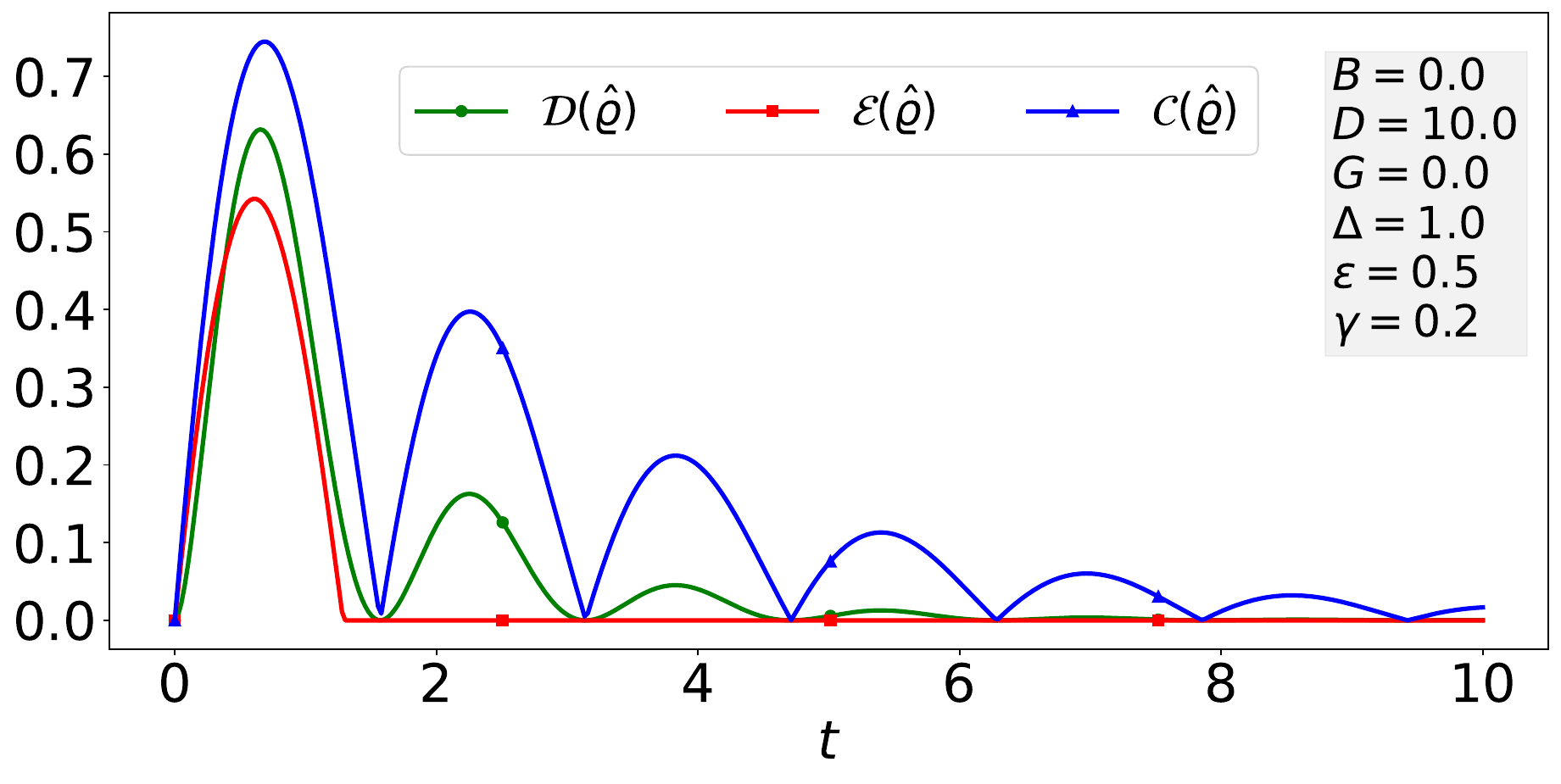}
        \put(-23,30){$(i)$}
    \end{minipage} \\

    \begin{minipage}[b]{0.32\textwidth}
        \centering
        \includegraphics[width=\textwidth]{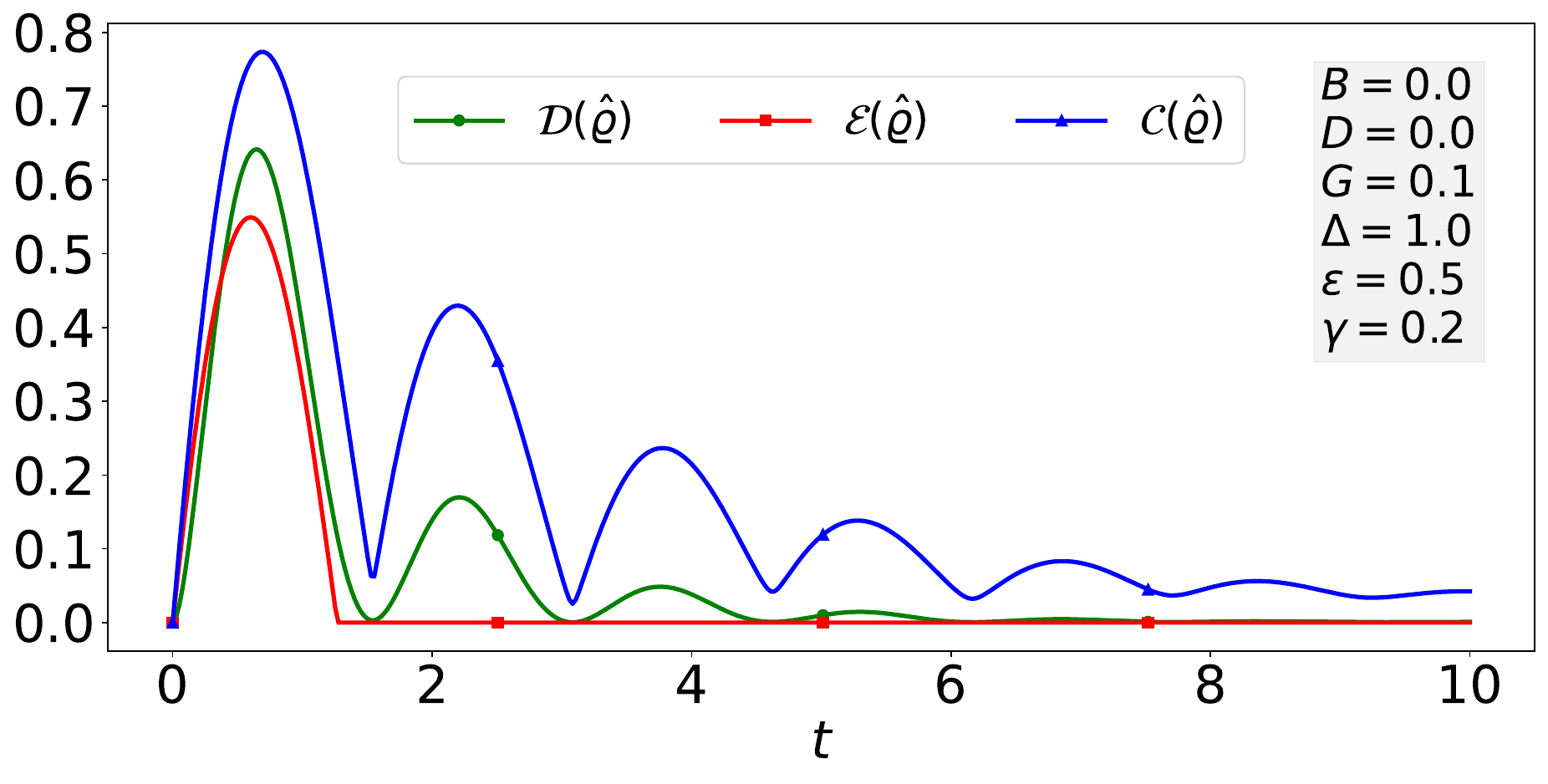}
        \put(-23,30){$(j)$}
    \end{minipage}%
    \begin{minipage}[b]{0.32\textwidth}
        \centering
        \includegraphics[width=\textwidth]{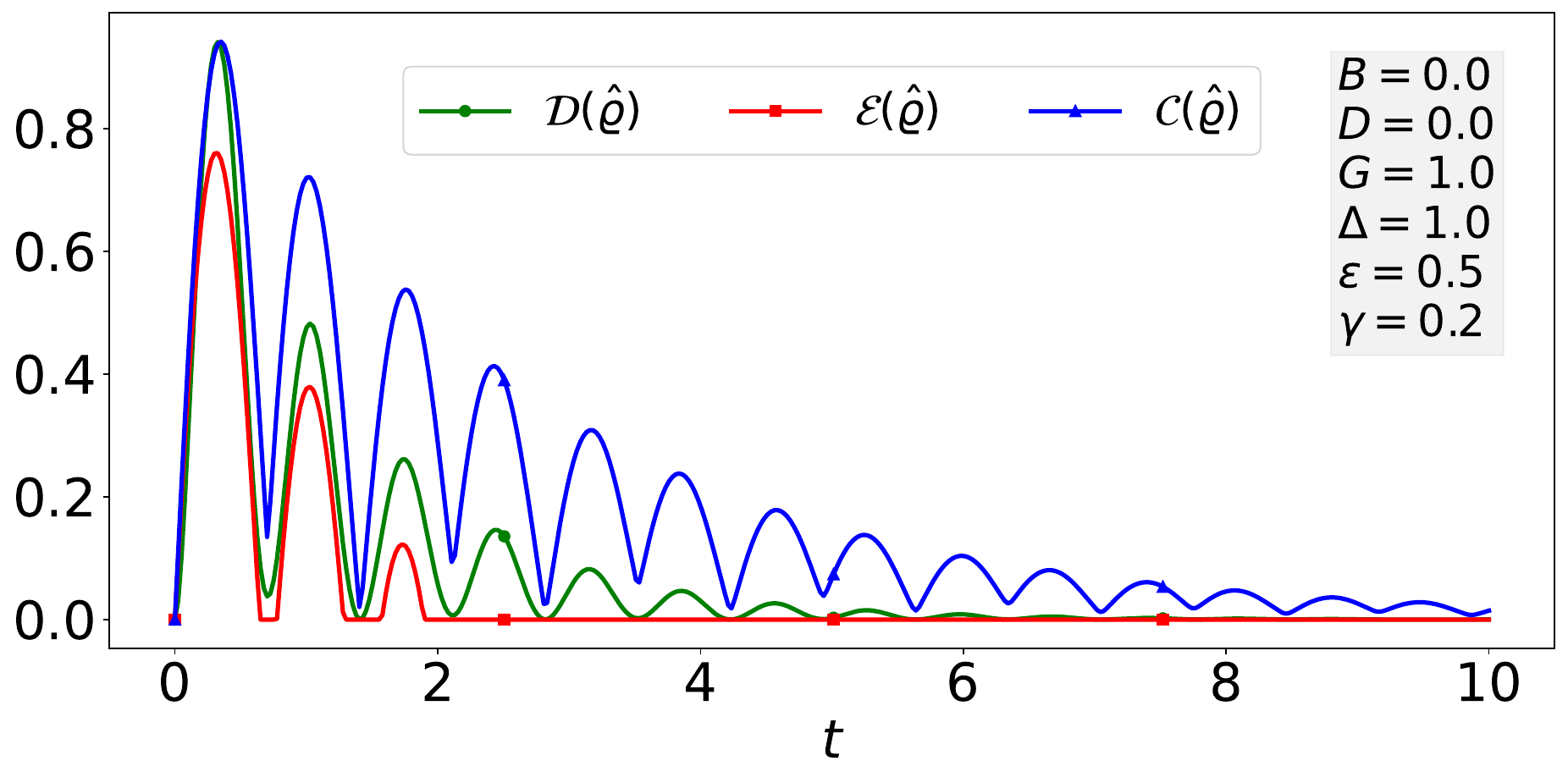}
        \put(-23,30){$(k)$}
    \end{minipage}%
    \begin{minipage}[b]{0.32\textwidth}
        \centering
        \includegraphics[width=\textwidth]{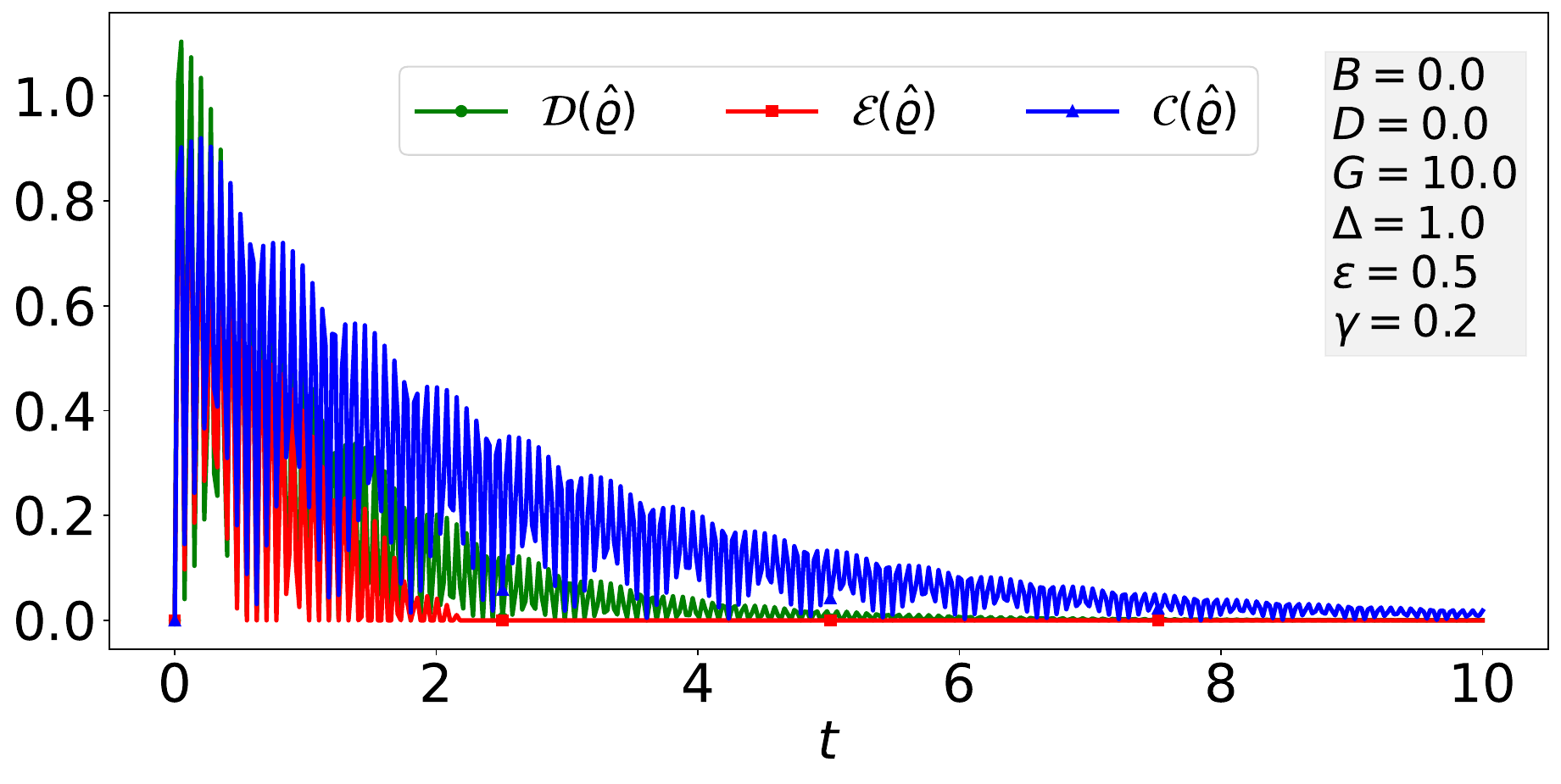}
        \put(-23,30){$(l)$}
    \end{minipage}  
        \begin{minipage}[b]{0.32\textwidth}
        \centering
        \includegraphics[width=\textwidth]{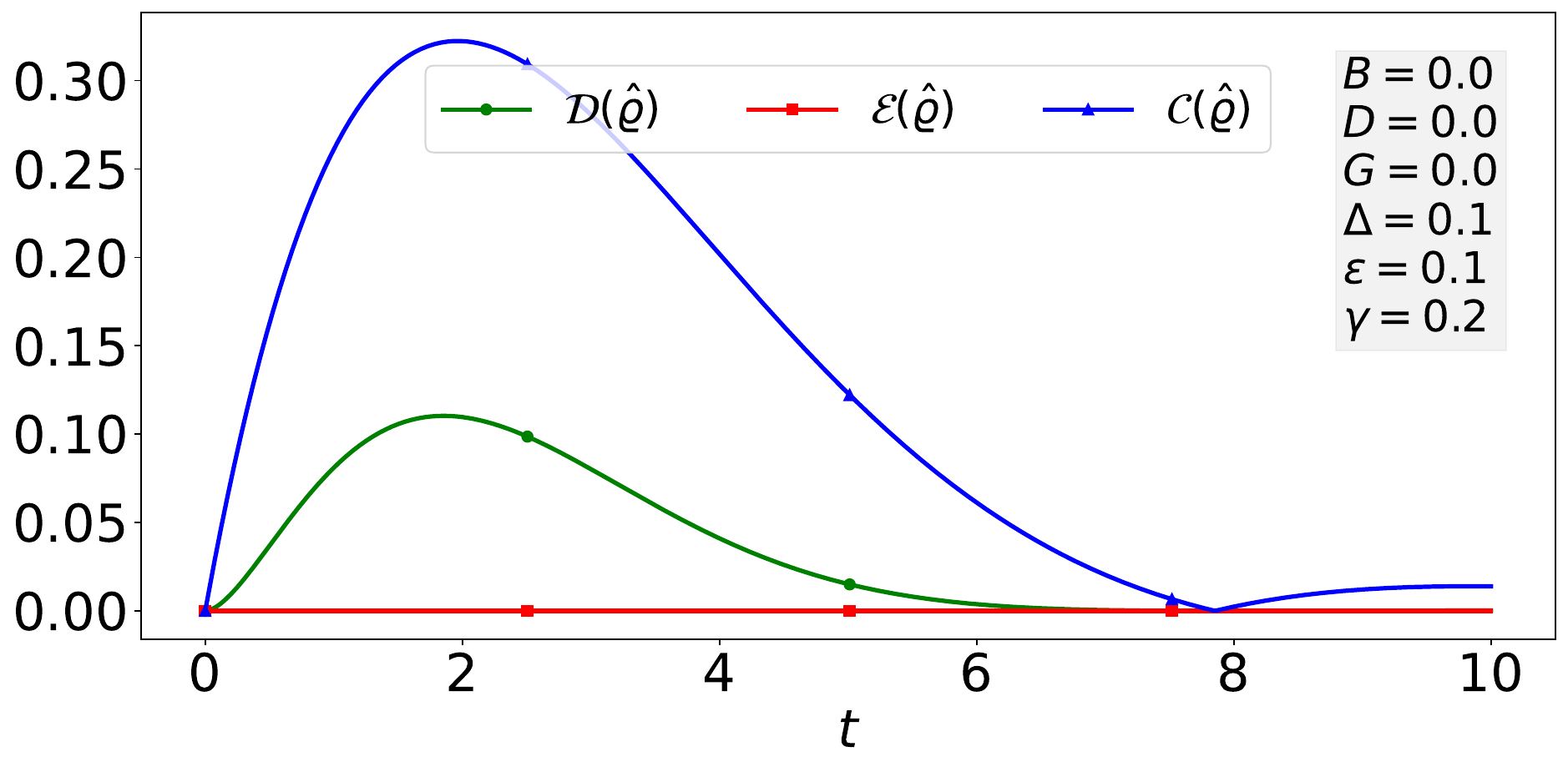}
        \put(-23,30){$(m)$}
    \end{minipage}%
    \begin{minipage}[b]{0.32\textwidth}
        \centering
        \includegraphics[width=\textwidth]{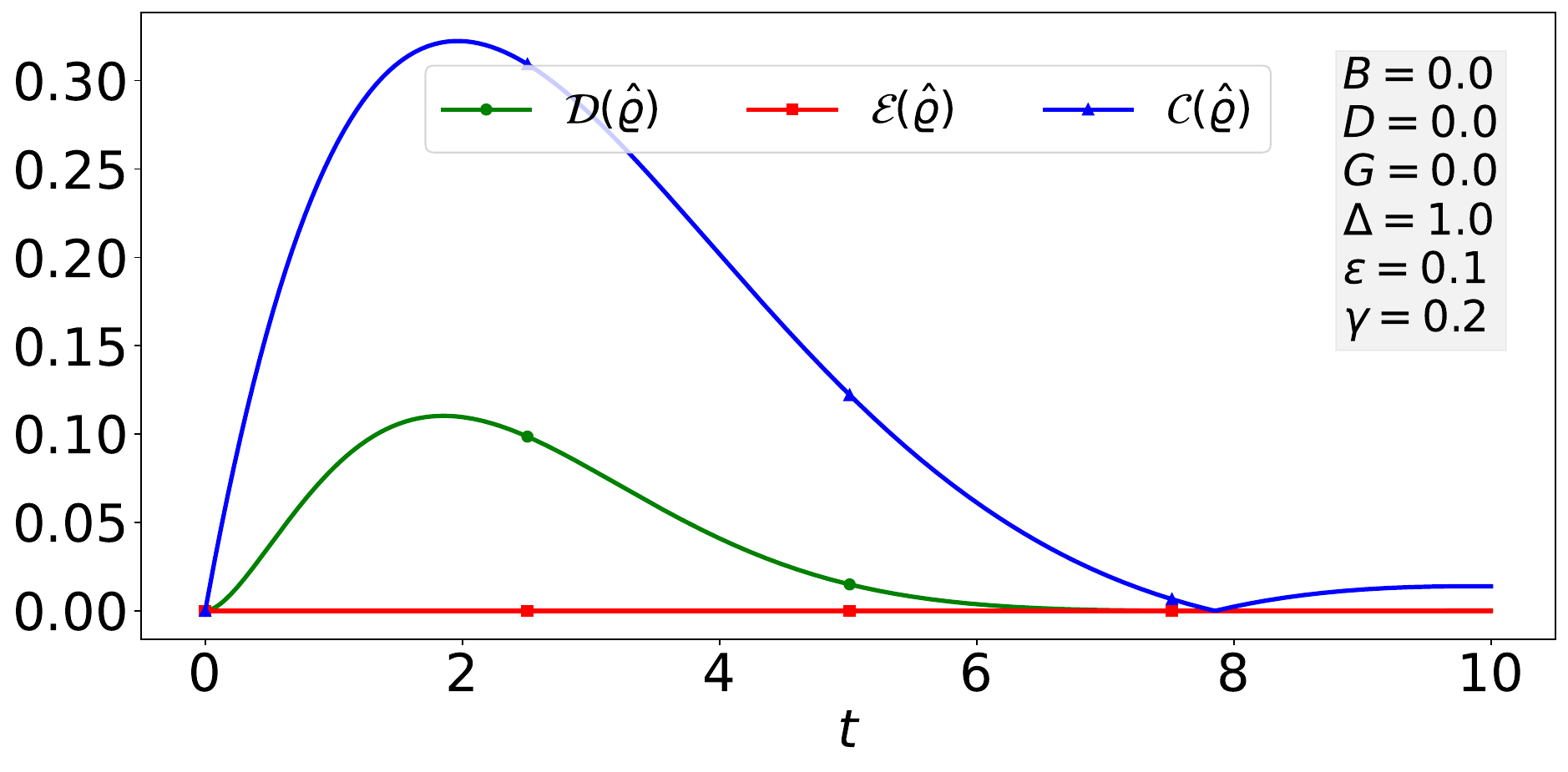}
        \put(-23,30){$(n)$}
    \end{minipage}%
    \begin{minipage}[b]{0.32\textwidth}
        \centering
        \includegraphics[width=\textwidth]{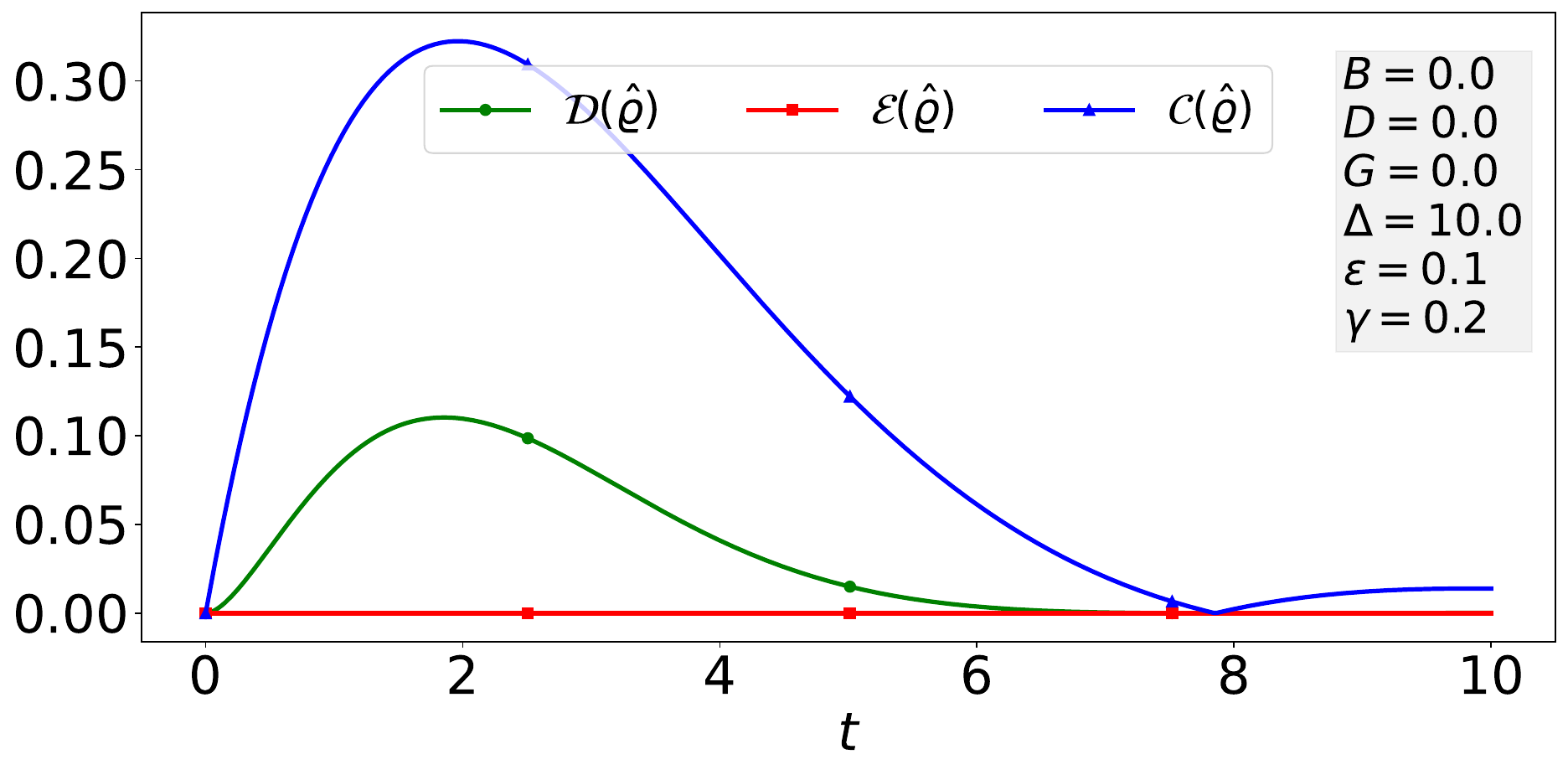}
        \put(-23,30){$(o)$}
    \end{minipage} 
\caption{Time evolution of quantum discord $\mathcal{D}(\hat{\varrho})$, concurrence $\mathcal{E}(\hat{\varrho})$,  and $l_1$-norm of  $\mathcal{C}(\hat{\varrho})$ for varying parameters. Panels ($a$-$c$) illustrate the dynamics for $\epsilon = 0.1$ ($a$), $\epsilon = 0.5$ ($b$), and $\epsilon = 10$ ($c$), with $D = G = B =\Delta = 0$. Panels ($d$-$f$) present the effects of varying $B$, with $ B=0.1$ ($d$), $ B=1.0$ ($e$), and $ B=10$ ($f$), under $D = G = 0$, $\Delta = 1$, and $\epsilon = 0.5$. Panels ($g$-$i$) examine the influence of $D$, showing results for $D = 0.1$ ($g$), $D = 1.0$ ($h$), and $D = 10$ ($i$), with $ B=0$, $G = 0$, $\Delta = 1$, and $\epsilon = 0.5$. Panels ($j$-$l$) show the influence of $G$, revealing results for $G = 0.1$ ($j$), $G = 1.0$ ($k$), and $G = 10$ ($l$), with $ B=0$, $D = 0$, $\Delta = 1$, and $\epsilon = 0.5$. Finally, panels ($m$-$o$) depict the mentioned behavior under varying $\Delta$, with $\Delta = 0.1$ ($m$), $\Delta = 1.0$ ($n$), and $\Delta = 10$ ($o$), keeping $ B= G = 0$ and $\epsilon = 0.1$. Fixed parameter is $\gamma = 0.2$.}
    \label{figure3}
\end{figure*}
\begin{enumerate}
    \item Among the measures, $l_1$-norm of quantum coherence $\mathcal{C}(\hat{\varrho})$ exhibits the highest robustness across all cases, followed by quantum discord $\mathcal{D}(\hat{\varrho})$ with moderate resilience, and concurrence $\mathcal{E}(\hat{\varrho})$, which shows the least robustness. Therefore, for brevity in the performance analysis of the QB, we will later focus solely on the $l_1$-norm of quantum coherence as the correlation metric.
    
    \item As depicted in Fig.~\ref{figure3}($a$-$c$), increasing the rhombic parameter $\epsilon$ enhances entanglement and elevates the peak values of all quantum resources. However, all measures decay over time due to dephasing dynamics. Notably, higher $\epsilon$ values also accelerate the rates of collapse and revival.
    
    \item As shown in Fig.~\ref{figure3}($d$-$f$), at small $B$ values, all measures exhibit high amplitude with low oscillation frequency. As $B$ increases, the amplitude decreases while the oscillation frequency rises. At large $B$ value, quantum coherence and correlation measures are significantly reduced, with $\mathcal{D}(\hat{\varrho})$ and $\mathcal{E}(\hat{\varrho})$ quickly approaching zero for very large $B$. Although $\mathcal{C}(\hat{\varrho})$ remains detectable for a longer period, it does so with diminished amplitude. Overall, increasing $B$ negatively impacts all quantum coherence and correlation measures.
    
    \item According to Fig.~\ref{figure3}($g$-$i$), increasing the { asymmetric spin-orbit interaction} along the $z$-direction, $D$, does not affect the temporal behavior of $\mathcal{C}(\hat{\varrho})$, $\mathcal{D}(\hat{\varrho})$, or $\mathcal{E}(\hat{\varrho})$. This indicates that under dephasing conditions, the DM interaction has no noticeable impact on quantum coherence and correlation metrics.
\begin{figure*}[!t]
    \centering
    \begin{minipage}[b]{0.32\textwidth}
        \centering
        \includegraphics[width=\textwidth]{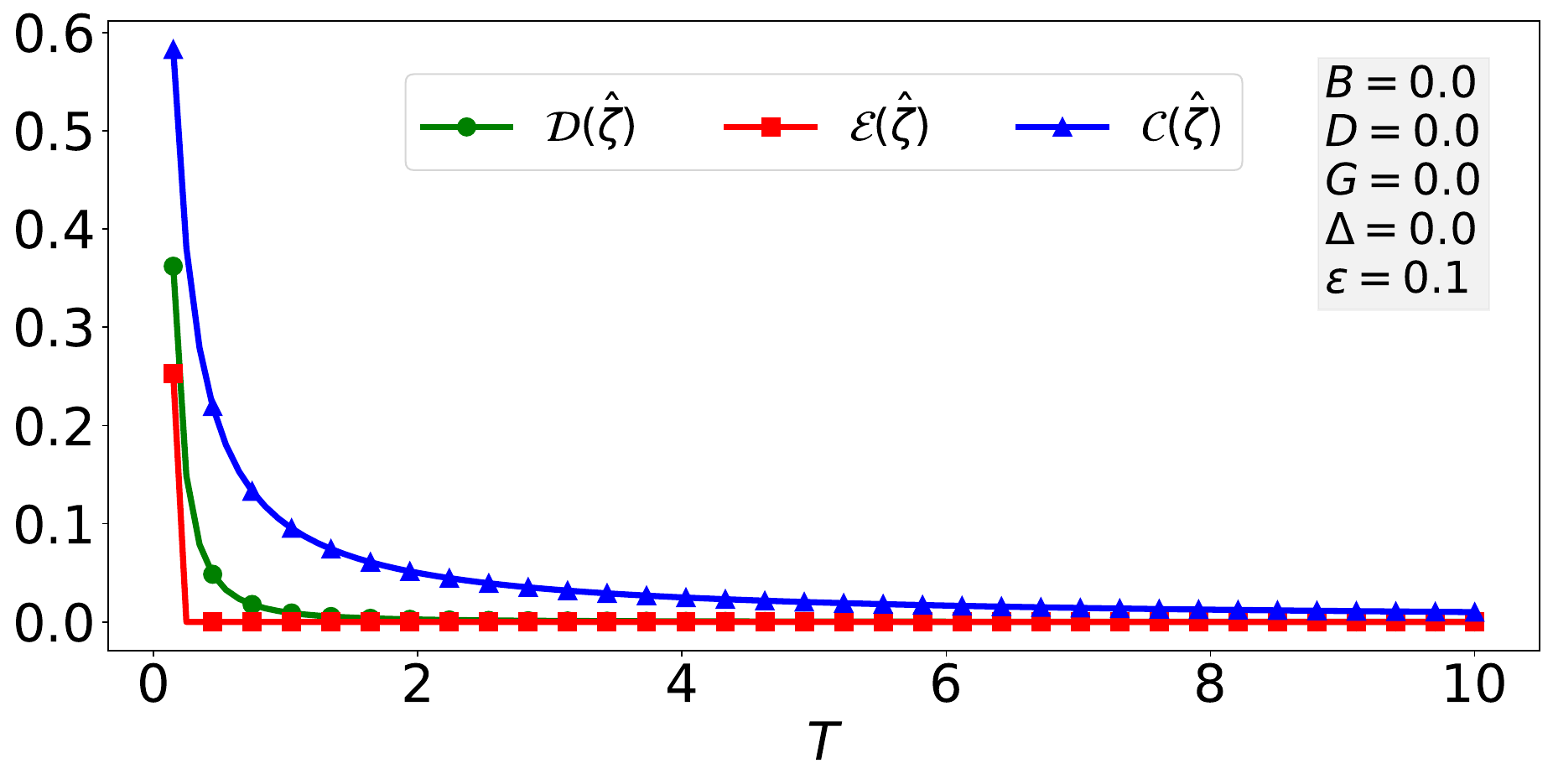}
        \put(-22,40){$(a)$}
    \end{minipage}%
    \begin{minipage}[b]{0.32\textwidth}
        \centering
        \includegraphics[width=\textwidth]{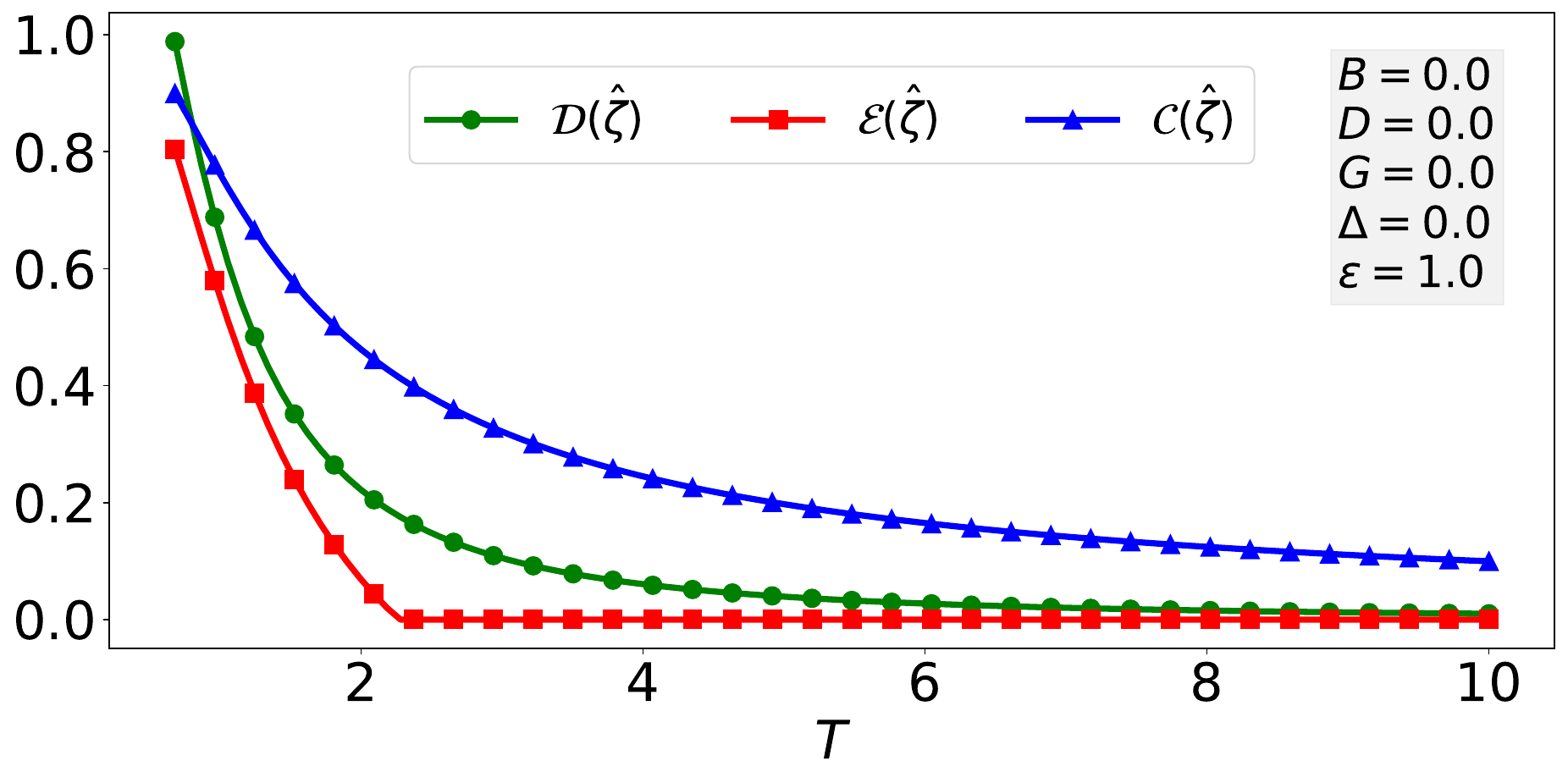}
        \put(-22,40){$(b)$}
    \end{minipage}%
    \begin{minipage}[b]{0.32\textwidth}
        \centering
        \includegraphics[width=\textwidth]{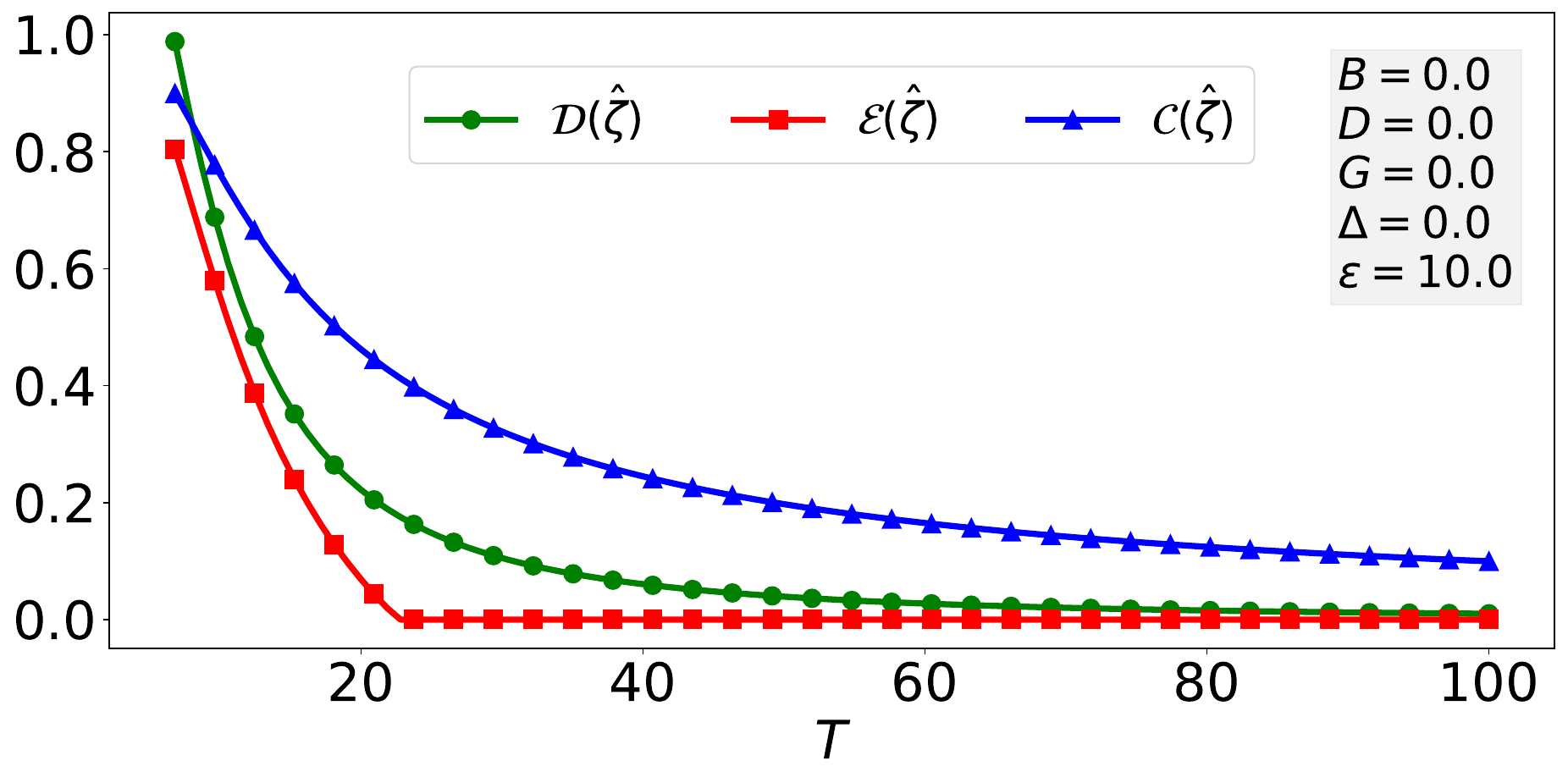}
        \put(-22,40){$(c)$}
    \end{minipage}
    
    \begin{minipage}[b]{0.32\textwidth}
        \centering
        \includegraphics[width=\textwidth]{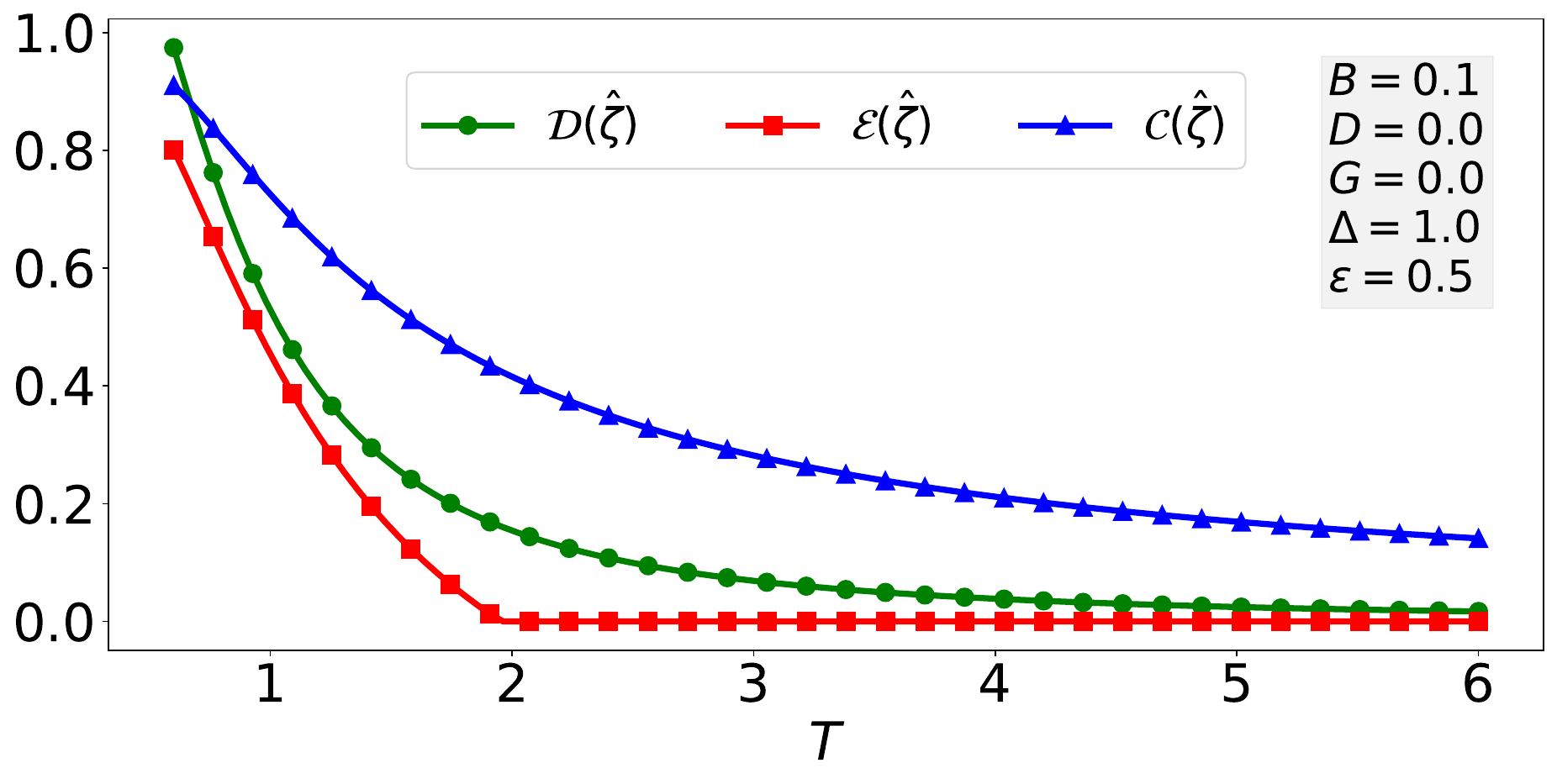}
        \put(-22,40){$(d)$}
    \end{minipage}%
    \begin{minipage}[b]{0.32\textwidth}
        \centering
        \includegraphics[width=\textwidth]{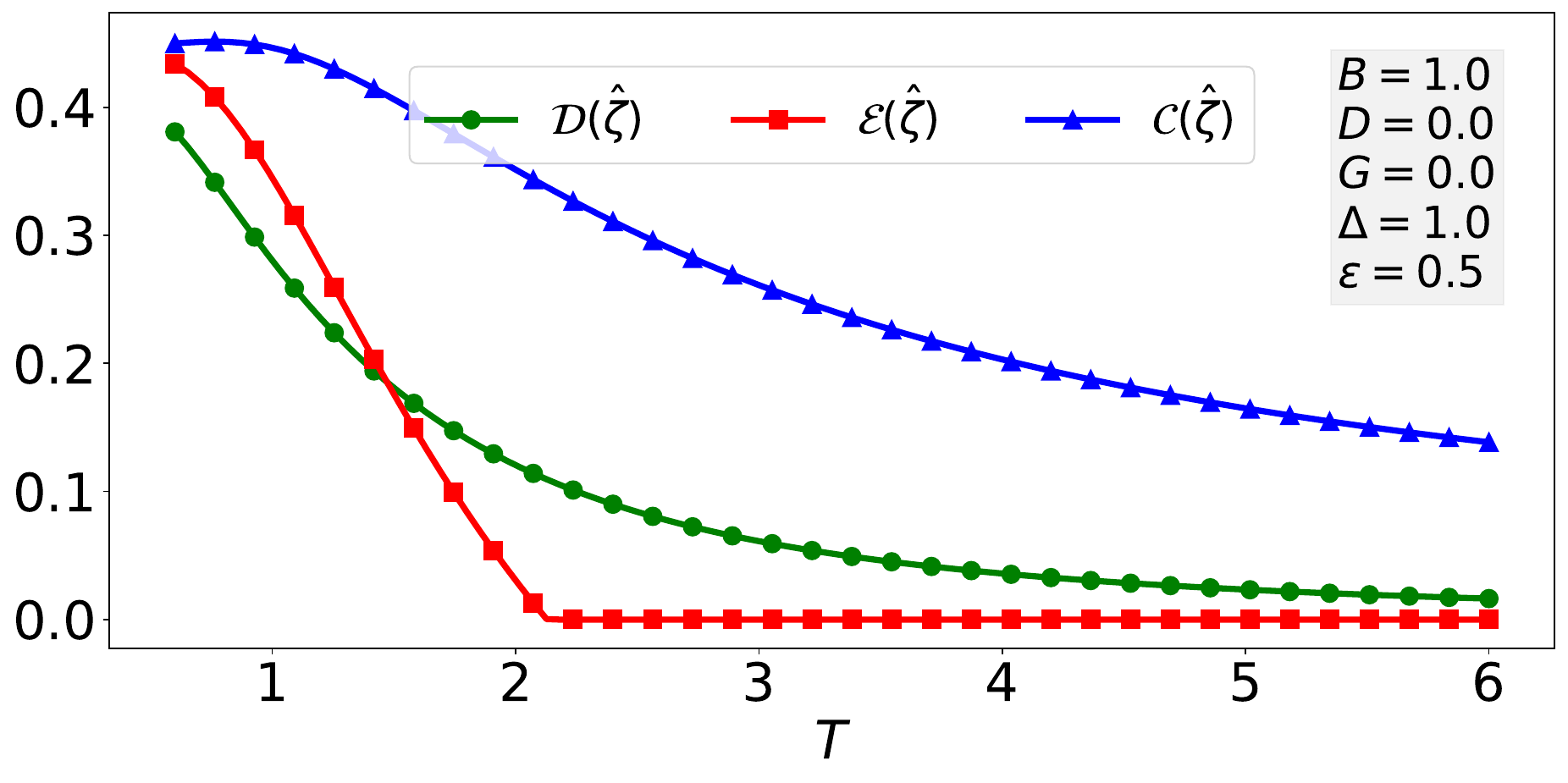}
        \put(-22,40){$(e)$}
    \end{minipage}%
    \begin{minipage}[b]{0.32\textwidth}
        \centering
        \includegraphics[width=\textwidth]{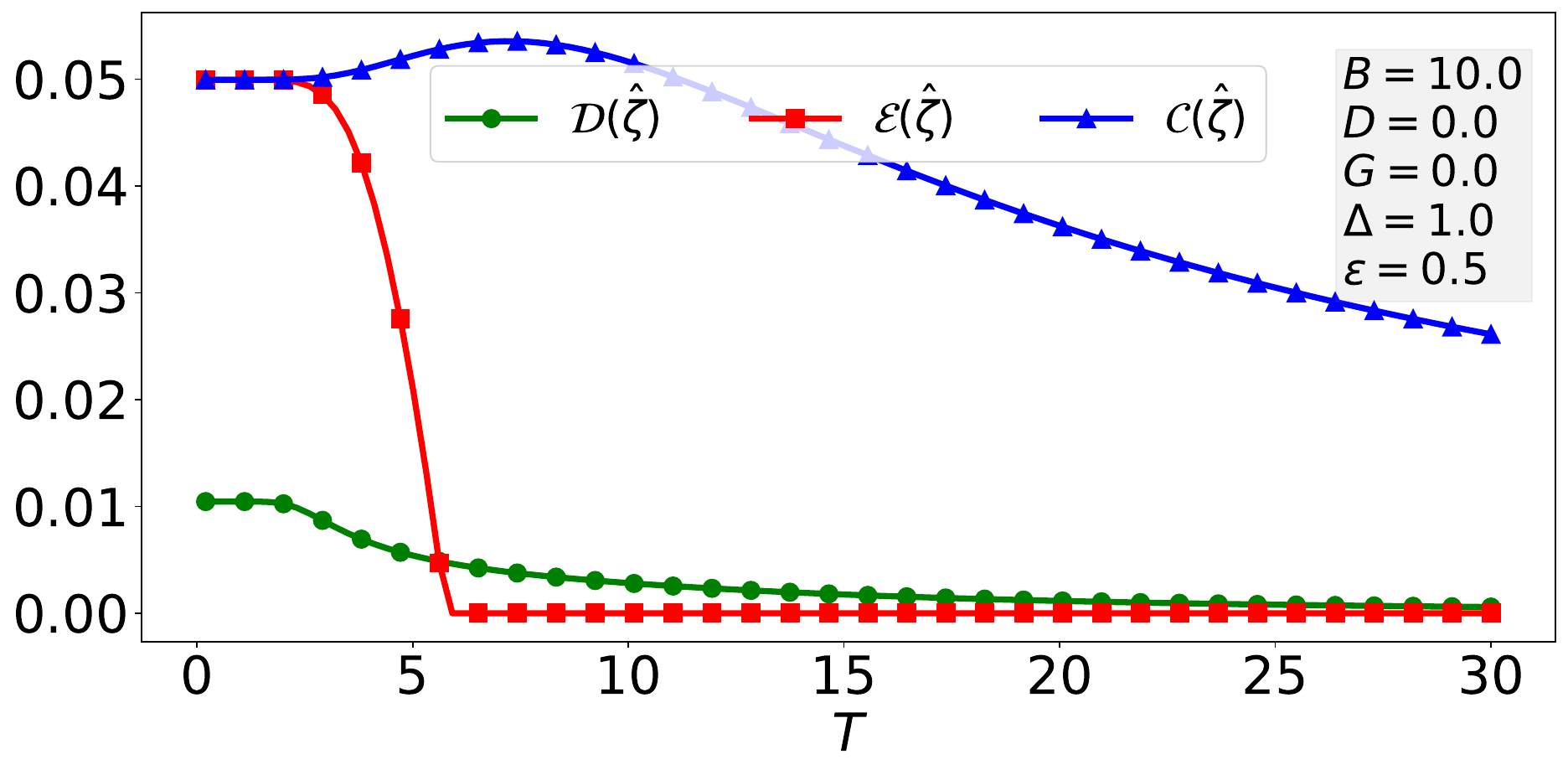}
        \put(-22,40){$(f)$}
    \end{minipage} \\

    \begin{minipage}[b]{0.32\textwidth}
        \centering
        \includegraphics[width=\textwidth]{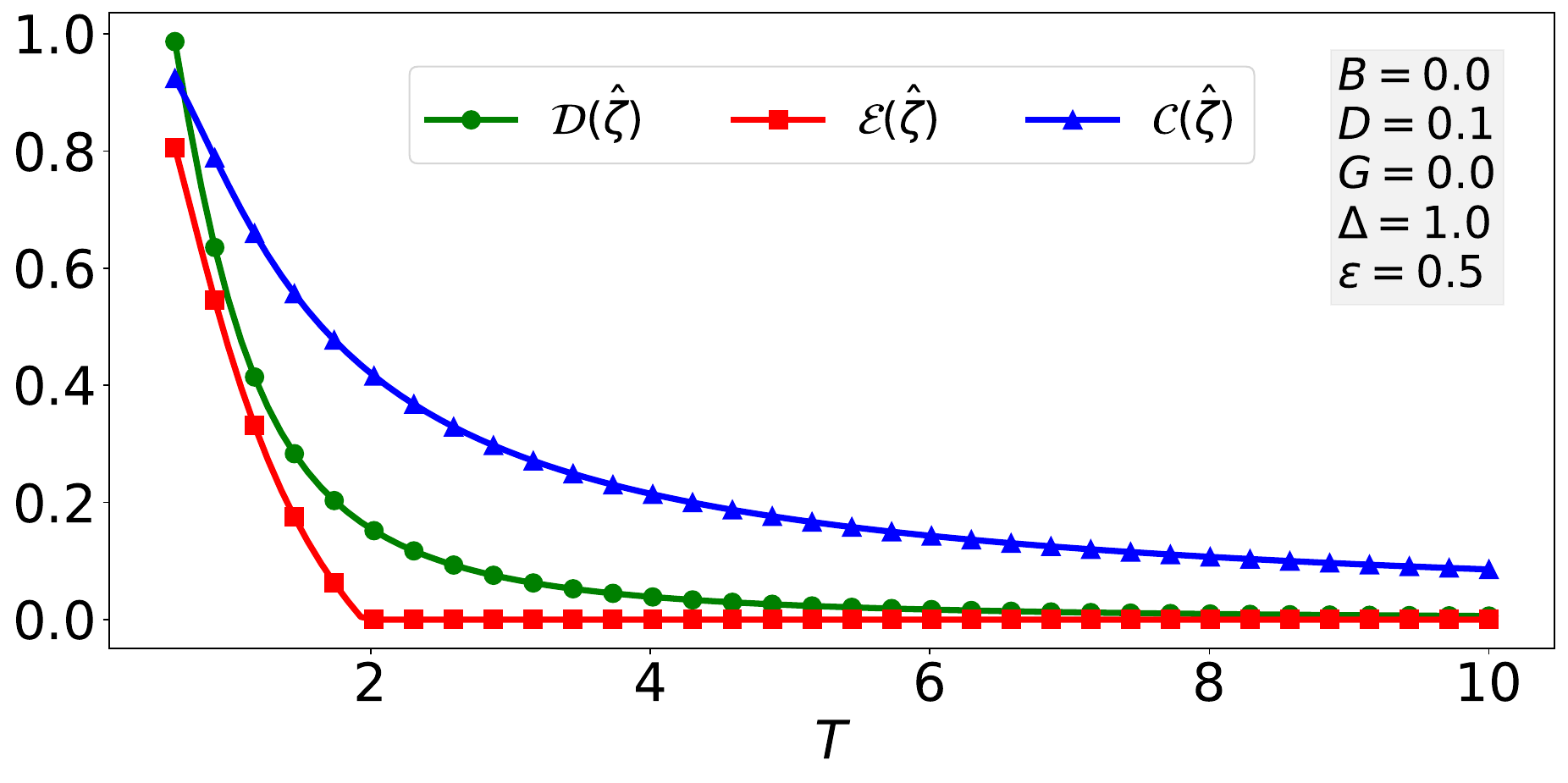}
        \put(-22,40){$(g)$}
    \end{minipage}%
    \begin{minipage}[b]{0.32\textwidth}
        \centering
        \includegraphics[width=\textwidth]{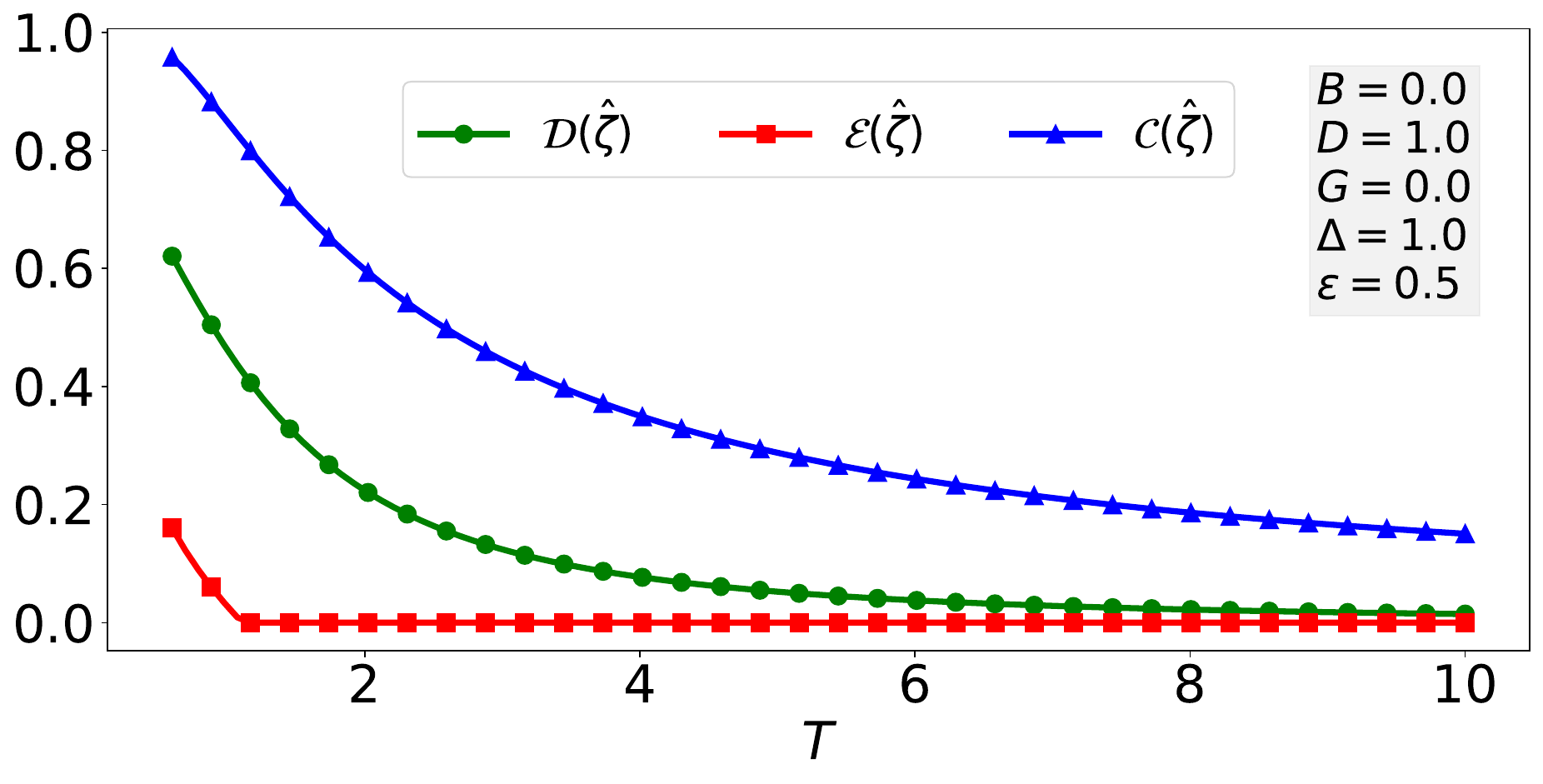}
        \put(-22,40){$(h)$}
    \end{minipage}%
    \begin{minipage}[b]{0.32\textwidth}
        \centering
        \includegraphics[width=\textwidth]{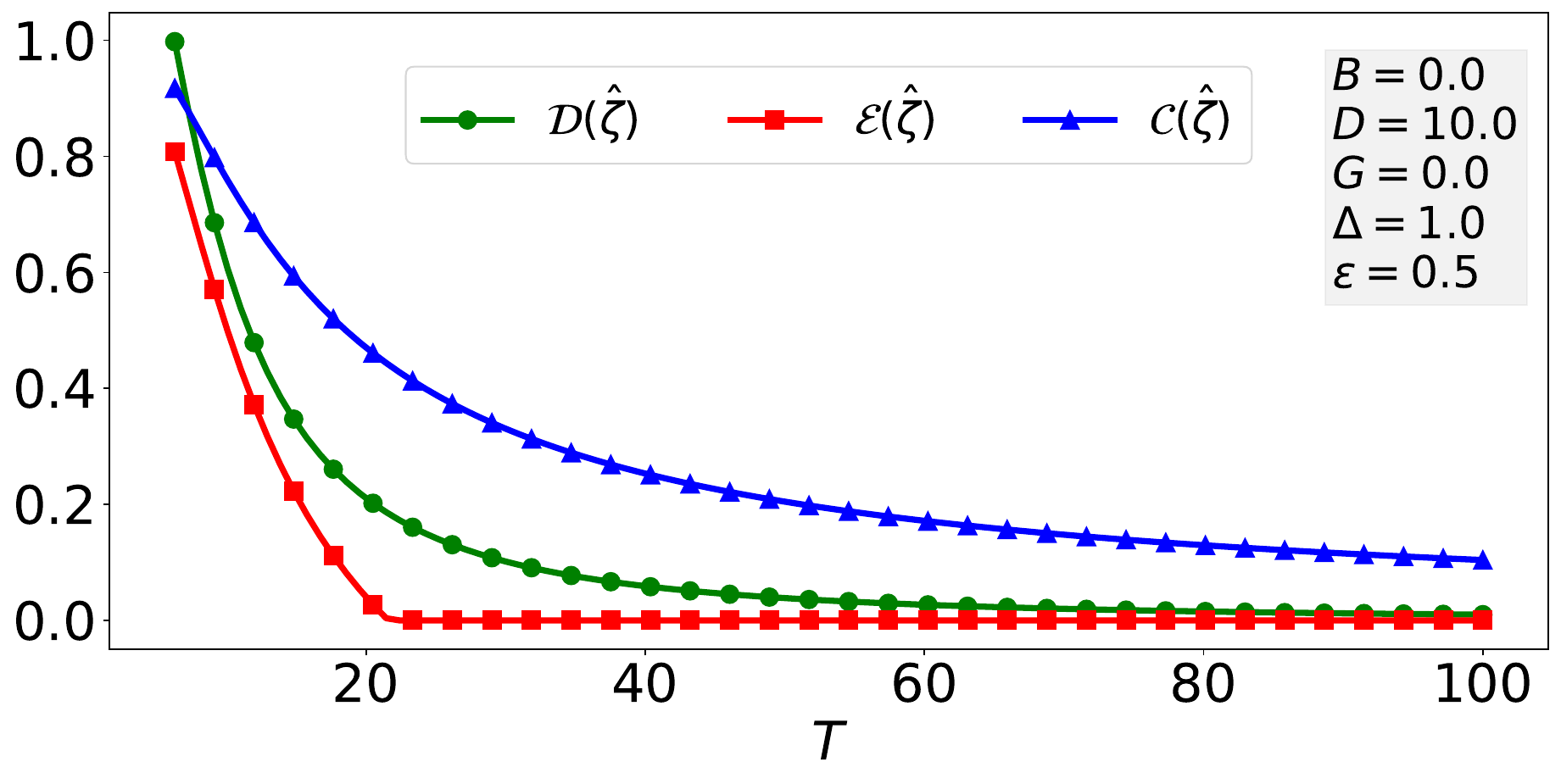}
        \put(-22,40){$(i)$}
    \end{minipage} \\

    \begin{minipage}[b]{0.32\textwidth}
        \centering
        \includegraphics[width=\textwidth]{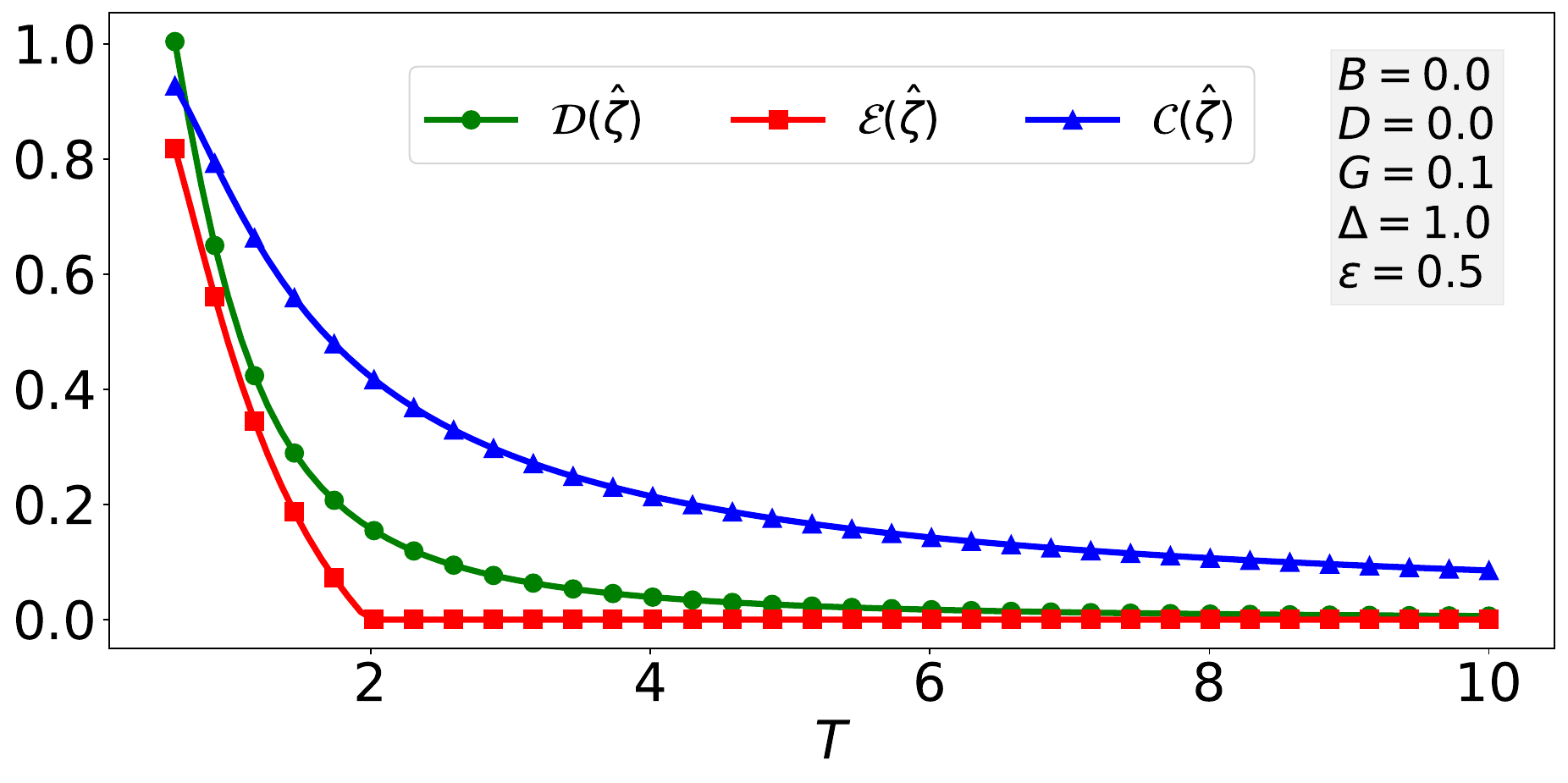}
        \put(-22,40){$(j)$}
    \end{minipage}%
    \begin{minipage}[b]{0.32\textwidth}
        \centering
        \includegraphics[width=\textwidth]{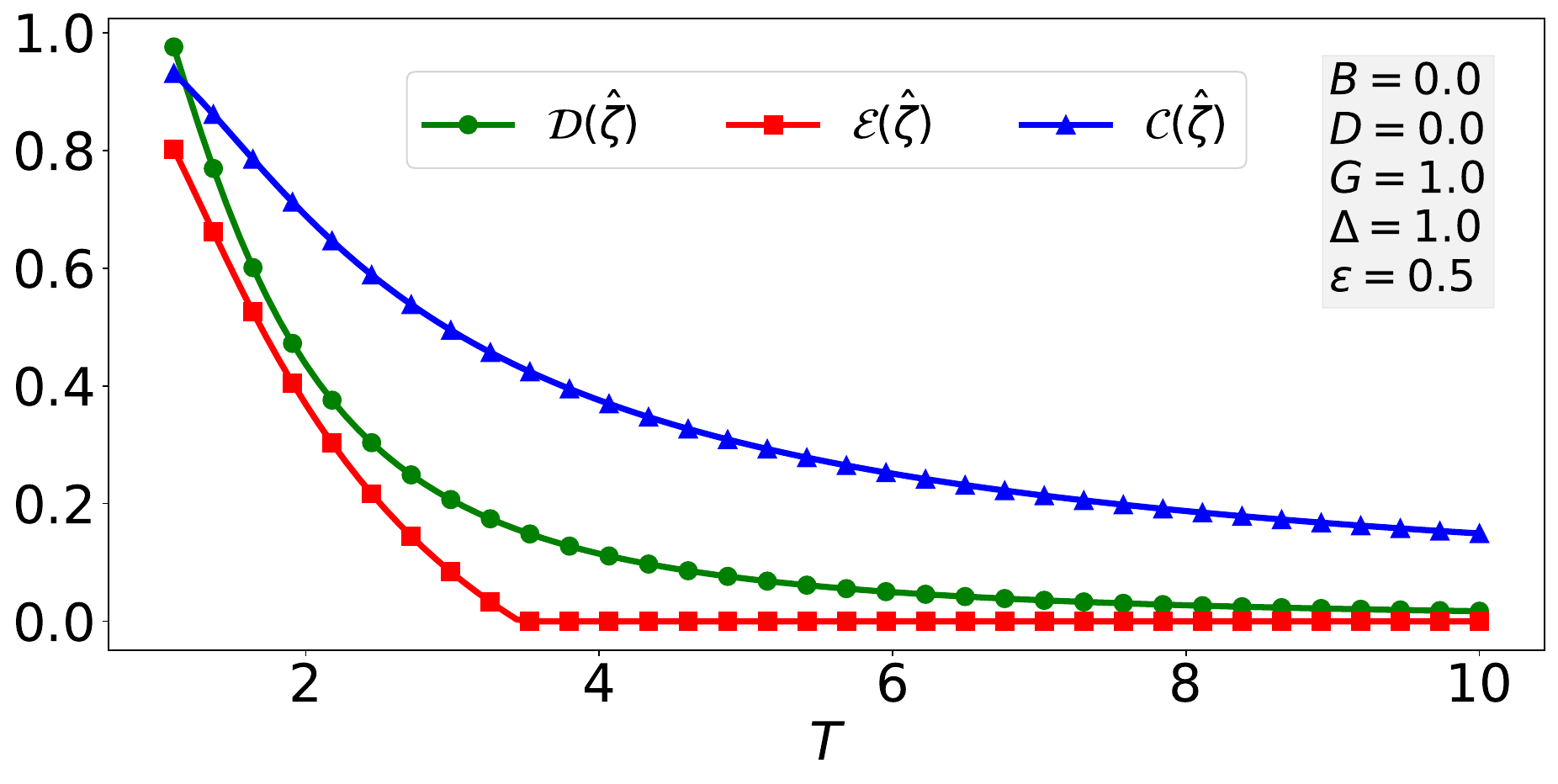}
        \put(-22,40){$(k)$}
    \end{minipage}%
    \begin{minipage}[b]{0.32\textwidth}
        \centering
        \includegraphics[width=\textwidth]{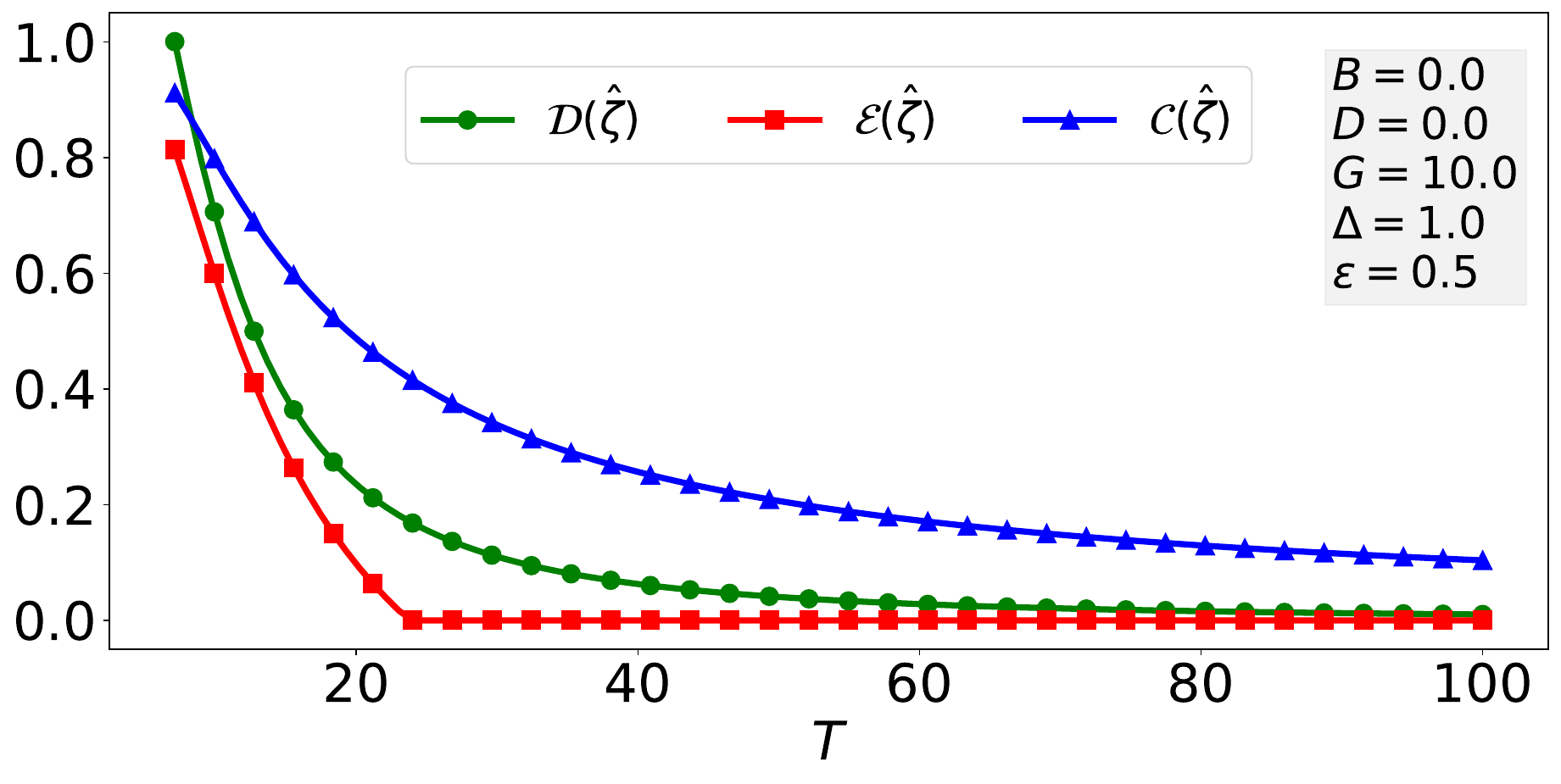}
        \put(-22,40){$(l)$}
    \end{minipage} \\

    \begin{minipage}[b]{0.32\textwidth}
        \centering
        \includegraphics[width=\textwidth]{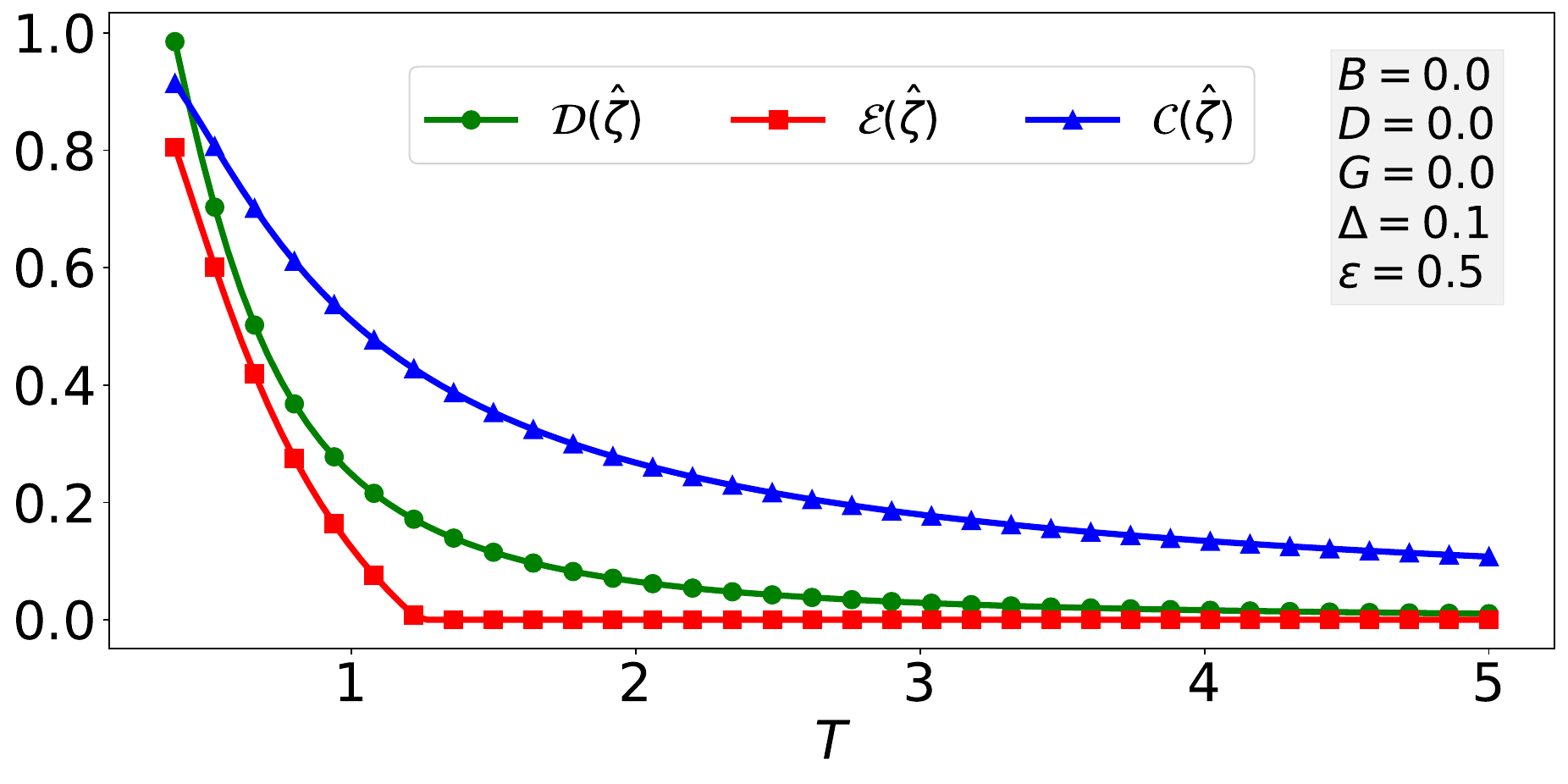}
        \put(-22,40){$(m)$}
    \end{minipage}%
    \begin{minipage}[b]{0.32\textwidth}
        \centering
        \includegraphics[width=\textwidth]{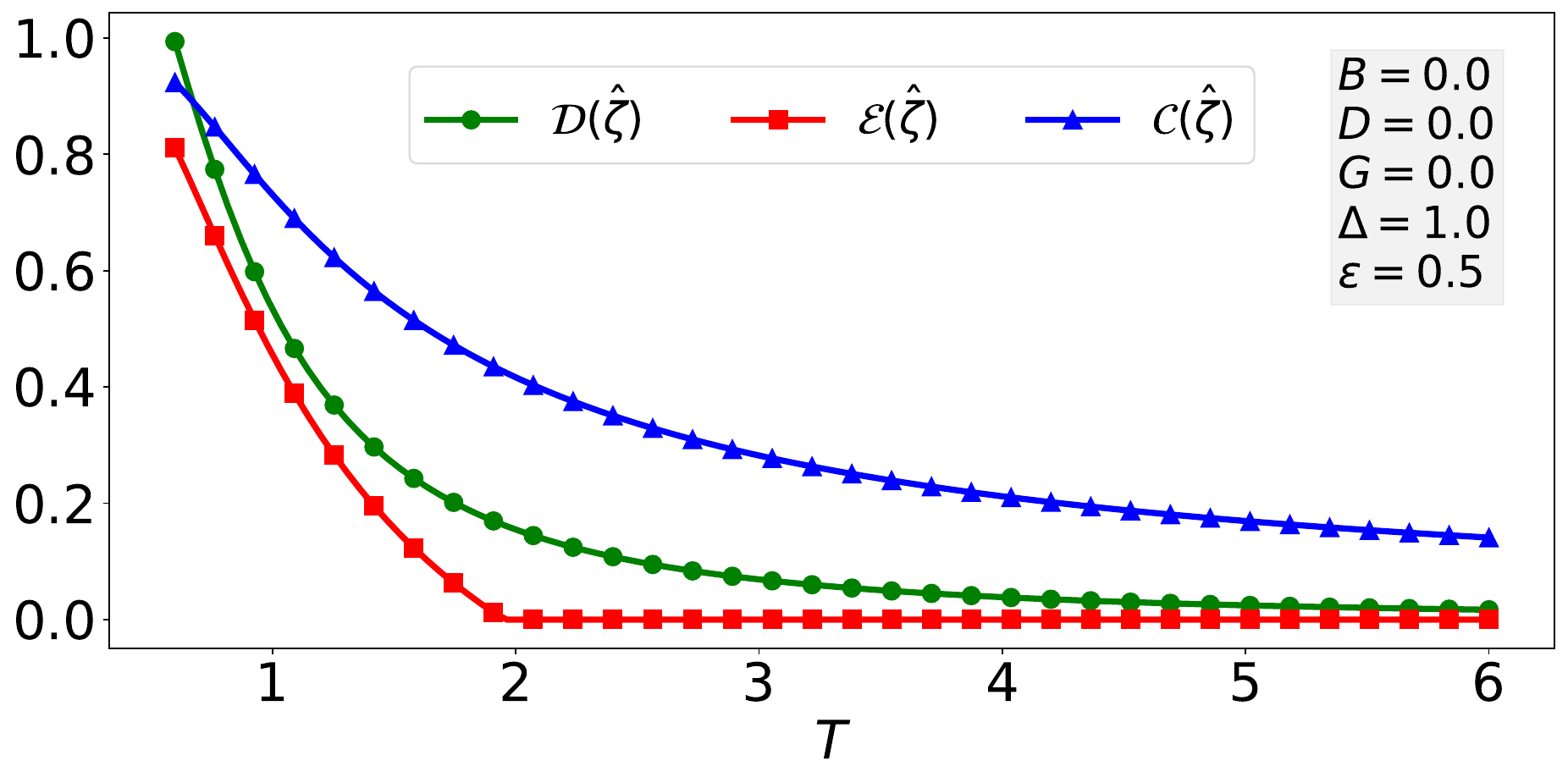}
        \put(-22,40){$(n)$}
    \end{minipage}%
    \begin{minipage}[b]{0.32\textwidth}
        \centering
        \includegraphics[width=\textwidth]{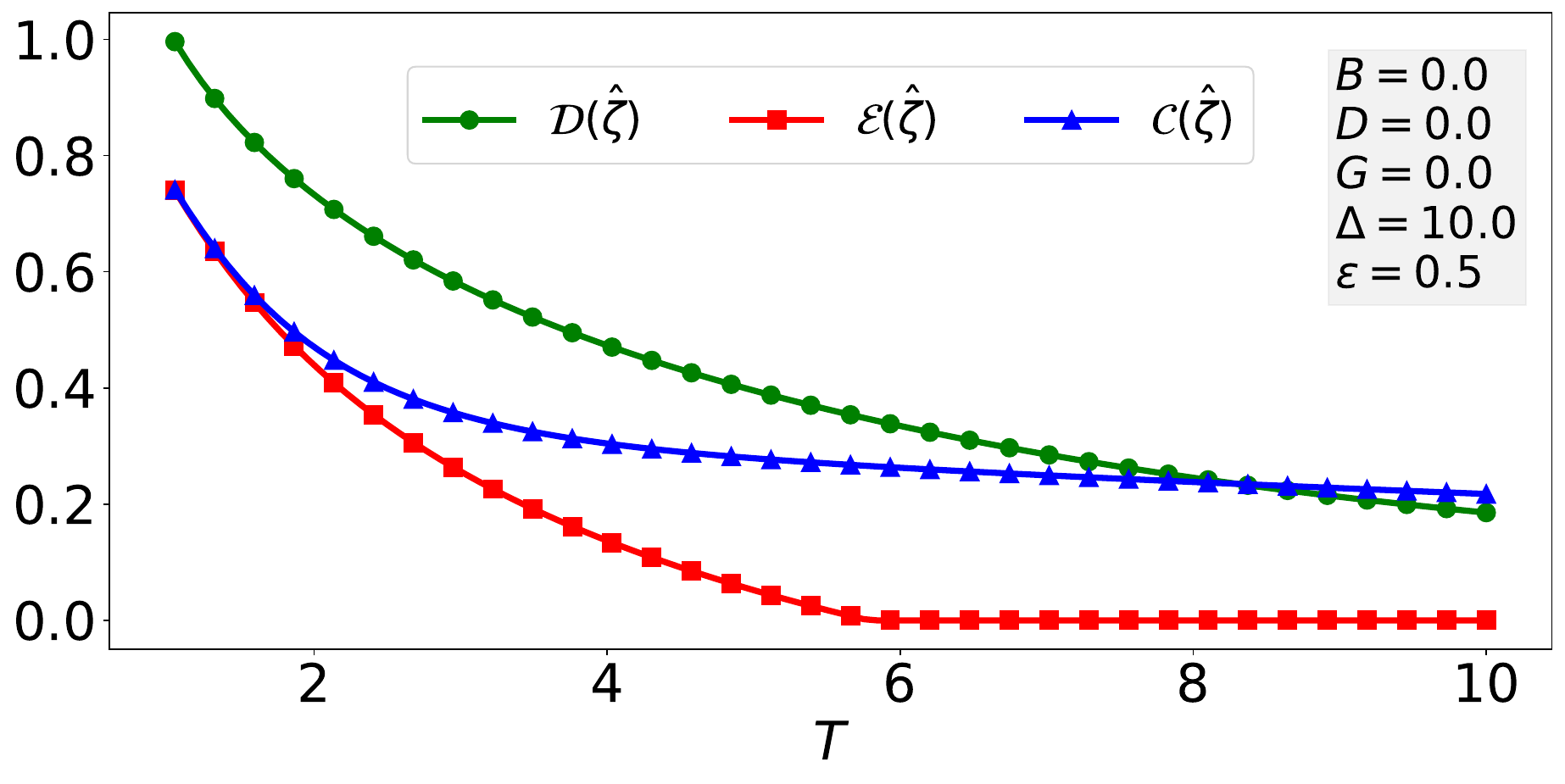}
        \put(-22,40){$(o)$}
    \end{minipage}
    \caption{Plots of $\mathcal{D}(\hat{\zeta})$, $\mathcal{E}(\hat{\zeta})$,  and $\mathcal{C}(\hat{\zeta})$ as a function of temperature $T$. Panels ($a$-$c$) show results for $\Delta = B = D = G = 0$, and rhombic parameter $\epsilon = 0.1$ ($a$), $\epsilon = 1.0$ ($b$), and $\epsilon = 10$ ($c$). Panels ($d$-$f$) present results for $\Delta = 1$, $\epsilon = 0.5$, $D = G = 0$, with magnetic field strengths $ B= 0.1$ ($d$), $ B= 1.0$ ($e$), and $ B= 10$ ($f$). Panels ($g$-$i$) depict results for $\Delta = 1$, $\epsilon = 0.5$, and $G = B = 0$ with $D = 0.1$ ($g$), $D = 1.0$ ($h$), and $D = 10$ ($i$). Panels ($j$-$l$) display results for $\Delta = 1$, $\epsilon = 0.5$, and $ B= D = 0$ with varying $G$ values: $G = 0.1$ ($j$), $G = 1.0$ ($k$), and $G = 10$ ($l$). Finally, panels ($m$-$o$) show results for $\epsilon = 0.5$ and $ B=  D = G = 0$ with $\Delta = 0.1$ ($m$), $\Delta = 1.0$ ($n$), and $\Delta = 10$ ($o$).}
    \label{figure4}
\end{figure*}    
    \item As illustrated in Fig.~\ref{figure3}($j$-$l$), increasing { the symmetric spin-orbit interaction $G$} leads to more rapid oscillations and higher peak values for all measures. Although the KSEA interaction generally enhances the peak values, it does not sustain nonzero correlation values over extended periods.
    
    \item As shown in Fig.~\ref{figure3}($m$-$o$), variations in $\Delta$ have no effect on correlations or . The measures $\mathcal{D}(\hat{\varrho})$, $\mathcal{E}(\hat{\varrho})$,  and $\mathcal{C}(\hat{\varrho})$ remain unchanged across different $\Delta$ values, indicating that $\Delta$ has no influence on  and correlations during dephasing.
\end{enumerate}

In summary, the $l_1$-norm of coherence demonstrates the greatest robustness, making it the primary focus for QB performance analysis. Increasing the rhombic parameter $\epsilon$ improves the robustness and longevity of quantum coherence. Higher $B$ reduces amplitude and increases oscillation frequency, negatively affecting all measures, especially $\mathcal{D}(\hat{\varrho})$ and $\mathcal{E}(\hat{\varrho})$. Asymmetric spin-orbit interaction $D$ and axial magnetic anisotropy $\Delta$ have minimal impact while increasing KSEA interaction $G$ accelerates oscillations and enhances initial magnitudes but does not sustain long-term correlations.

Now, we analyze the thermal behavior of quantum resources for the Gibbs state $\hat{\zeta}$ (see Eq. \eqref{EQ7}), as plotted in Fig. \ref{figure4}. 
\begin{enumerate}
\item Just as reported in Fig.~\ref{figure3}, Fig.~\ref{figure4} also shows that $\mathcal{C}(\hat{\zeta})$ is the most robust, followed by $\mathcal{D}(\hat{\zeta})$, with $\mathcal{E}(\hat{\zeta})$ being the least in thermodynamic equilibrium. 
\item Figure~\ref{figure4}($a$-$c$) shows that the rhombic parameter $\epsilon$ not only has a positive influence in generating all considered quantum resources but also makes them robust against temperature, a finding that was also advantageous in dephasing dynamics in Fig.~\ref{figure3}($a$-$c$). 
\item Stronger Zeeman fields reduce the peak values of all measures, as seen in Fig.~\ref{figure4}($d$-$f$). Concurrence $\mathcal{E}(\hat{\zeta})$ shows a notable decrease in peak value and increased cutoff temperature with stronger fields, indicating a trade-off between peak value and cutoff temperature.
\item Increasing DM interaction $D$ generally enhances both the peak values and robustness of quantum coherence, although, for moderate values, it may slightly reduce the peak correlations at $T \approx 0$, as seen in Fig.~\ref{figure4}($h$). This increase in $D$ raises both the peak and cutoff temperatures of $\mathcal{E}(\hat{\zeta})$, as shown in Fig.~\ref{figure4}($g$-$i$). In contrast, under dephasing dynamics, $D$ has no noticeable effect on quantum coherence and correlations over time, as indicated in Fig.~\ref{figure3}($g$-$i$).
\item Figure \ref{figure4}($j$-$l$) depicts that increasing KSEA interaction $G$ maintains the peak values of all measures while improving their robustness against temperature, which is also consistently true in Fig.~\ref{figure3}($j$-$l$) against time during dephasing dynamics.
\item { Unlike the dephasing scenario in Fig.~\ref{figure3}($m$-$o$), where the axial magnetic anisotropy $\Delta$ does not influence  and correlation metrics, increasing $\Delta$ in this context, as shown in Fig.~\ref{figure4}($m$-$o$), enhances the resilience of all quantum measures to temperature variations and raises the cutoff temperature at which entanglement death occurs, increasing the threshold temperature from which freezing of quantum discord starts.}
\end{enumerate}

{
\section{Magnetic Dipolar System as Working medium for the closed QB} \label{sec5}
} 
Reaching absolute zero is unfeasible for quantum applications, including QBs, as it requires infinite energy per the Third Law of Thermodynamics. At $T > 0$  close to zero, where quantum effects dominate, QBs can perform optimally without unattainable conditions. The Gibbs thermal state serves as a practical baseline for uncharged QBs \cite{binder2015quantacell,binder2015quantum,ghosh2020enhancement,ali2024ergotropy}, providing a low-energy starting point at realistic temperatures. This enables clear analysis of charging dynamics by minimizing interference from initial excitations and aligns with feasible thermodynamic conditions. Therefore, in our QB calculations as well, we consider the Gibbs thermal state as the uncharged state of our QB and charge it through a cyclic unitary evolution operator based on the NOT gate-based charging Hamiltonian \cite{ghosh2020enhancement,ali2024ergotropy}.

\subsection{Charging the QB}
\label{subsec3A}
The QB is charged with a constant magnetic field $\omega$ along the $x$-direction \cite{julia2020bounds,ghosh2022dimensional,ghosh2020enhancement}, described by  
\begin{equation}
\mathcal{H}_{ch} = \omega (\hat{\sigma}_{1}^{x} \otimes \hat{\mathbb{I}}_2 + \hat{\mathbb{I}}_1 \otimes \hat{\sigma}_{2}^{x}),
\end{equation}
where $\omega$ is the field strength. We assume a uniform field for simplicity. After reaching maximum energy, the $x$-axis field is disconnected to avoid cyclic reversion. The charging Hamiltonian acts as a NOT gate, using a uniform magnetic field perpendicular to the Zeeman field along the $z$-direction.

The battery's charging process can be described by a unitary operation:
\begin{equation}
\hat{U}_{ch}(t) = \exp(-i \mathcal{H}_{ch} t) = \left(
\begin{array}{cccc}
 a & c & c & b \\
 c & a & b & c \\
 c & b & a & c \\
 b & c & c & a \\
\end{array}
\right),
\end{equation}
where $a = \cos^2(\omega t)$, $ b=-\sin^2(\omega t)$, and $c = -\frac{i}{2} \sin(2 \omega t)$. %For convenience, we will set $\omega=1$  during the numerical analysis. 
In NMR settings using RF pulses, these unitaries can be easily implemented \cite{joseph2025decoupling}. 

{
\subsection{Performance indicators for the QB}\label{subsec:performance}
To evaluate the performance of our QB, we analyze a complete thermodynamic cycle consisting of three distinct phases: the initial uncharged state, the charging process, and the work extraction process. Let $\hat{\rho}$ represent the quantum state of a QB with Hamiltonian $\hat{\mathcal{H}}_{\text{QB}}$. The central question is: how much energy can be stored in and subsequently extracted from the QB through cyclic unitary processes? A cyclic process ensures that the Hamiltonian of the QB is identical at the start and end of the process, i.e., $\hat{\mathcal{H}}_{\text{QB}} = \hat{\mathcal{H}}_{\text{QB}}(0) = \hat{\mathcal{H}}_{\text{QB}}(\tau)$~\cite{binder2015quantum}. Since the process is unitary, any change in the internal energy $\langle \hat{\mathcal{H}}_{\text{QB}} \rangle$ reflects the work done on or by the QB.

We begin by implementing the eigen-spectral decomposition of the QB Hamiltonian \eqref{eq4} into its increasing spectral form:
\begin{equation}
\hat{\mathcal{H}}_{\text{QB}} = \sum_{\mu} \nu_\mu |\psi_\mu\rangle \langle \psi_\mu|, \quad \text{where } \nu_{\mu+1} \geq \nu_\mu \text{ for all } \mu,
\label{eq:hamiltonian_decomp}
\end{equation}
with $\nu_\mu$ being the energy eigenvalues and $|\psi_\mu\rangle$ the corresponding eigenstates as shown in Eq.~\eqref{eq2}. The generic state of the QB at any time $t$ can be similarly expressed in eigendecomposition form as:

\begin{equation}
\hat{\rho}(t) =\hat{U}_{ch}\hat{\zeta}\hat{U^{\dagger}_{ch}} =\sum_{\mu} \phi_{\mu} |\phi_\mu\rangle \langle \phi_\mu|.
\label{eq:rho_decomp}
\end{equation}

At $t = 0$, we assume the QB begins in the Gibbs thermal state $\hat{\zeta}$ expressed in Eq. \eqref{eq3}, which serves as our reference uncharged and passive QB state~\cite{binder2015quantum,bakhshinezhad2024trade}. The thermal (Gibbs) state is a special type of passive state: it is diagonal in the energy eigenbasis, with populations decreasing exponentially with increasing energy. As such, it represents a completely passive configuration from which no work can be extracted via unitary operations. Importantly, while all thermal states are passive, not all passive states are thermal \cite{koukoulekidis2021geometry,campaioli2019quantum}.

{During the charging phase, we apply a unitary operator 
\(\hat{U}_{ch} = \exp(-i \hat{\mathcal{H}}_{ch} t)\) that evolves the QB from its initial passive Gibbs state \(\hat{\zeta}\) to a time-dependent state 
\(\hat{\rho}(t) = \hat{U}_{ch} \hat{\zeta} \hat{U}_{ch}^\dagger\). 
At certain evolution times, \(\hat{\rho}(t)\) may coincide with the active state \(\hat{\eta}\), which is obtained by assigning the largest eigenvalues of \(\hat{\rho}(t)\) to the highest-energy eigenstates of the Hamiltonian \(\hat{\mathcal{H}}_{\text{QB}}\):}
\begin{equation}
\hat{\eta} = \sum_{\mu} \phi_\mu |\psi_\mu\rangle \langle \psi_\mu|, 
\quad \text{where } \phi_{\mu+1} \geq \phi_\mu \text{ for all } \mu.
\label{eq:active_state}
\end{equation}
In such cases, the stored energy corresponds to the maximum ergotropy with no contribution from anti-ergotropy. However, this alignment is not generic and depends on the dynamics dictated by the fixed form of the charging unitary.

% During the charging phase, we apply a charging unitary operator $\hat{U}_{\text{ch}}$ to transform the QB from its initial passive state $\hat{\zeta}$ to an active state $\hat{\eta}$. The active state is constructed by pairing the largest eigenvalues $\phi_\mu$ of the current state with the highest energy eigenstates $|\psi_\mu\rangle$ of the Hamiltonian:
% \begin{equation}
% \hat{\eta} = \sum_{\mu} \phi_\mu |\psi_\mu\rangle \langle \psi_\mu|, \quad \text{where } \phi_{\mu+1} \geq \phi_\mu \text{ for all } \mu.
% \label{eq:active_state}
% \end{equation}

The energy injection required to achieve this active state is quantified by the anti-ergotropy \cite{yadav2025thermo}, which represents the maximum amount of energy that can be injected via cyclic unitary processes \cite{yang2023battery,yadav2025thermo}
\begin{equation}
\mathcal{P}(t) = \text{Tr}\left[(\hat{\rho}(t) - \hat{\eta}) \hat{\mathcal{H}}_{\text{QB}}\right].
\label{eq:anti_ergotropy}
\end{equation}

Once the QB is fully charged to its active state, we can extract work from it until it reaches a passive state $\hat{\pi}$. The passive state is constructed by arranging the eigenvalues $\phi_\mu$ in decreasing order while pairing them with energy eigenstates $|\psi_\mu\rangle$ in increasing order of energy:
\begin{equation}
\hat{\pi} = \sum_{\mu} \phi_\mu |\psi_\mu\rangle \langle \psi_\mu|, \quad \text{where } \phi_{\mu+1} \leq \phi_\mu \text{ for all } \mu.
\label{eq:passive_state}
\end{equation}

A passive state, unique up to degeneracies in the Hamiltonian, is diagonal in the energy eigenbasis with populations that do not increase with energy~\cite{allahverdyan2004maximal,binder2015quantum}. This passive state is not generally thermal; rather, it is the optimal endpoint of a unitary work extraction process for a given state. Crucially, while the QB begins in a thermal state, the discharging process—being unitary and entropy-preserving—leads to a final passive state that is typically non-thermal. Therefore, the overall cycle is thermodynamically open unless a final non-unitary thermalization step is included to return the system to its initial Gibbs state. 

The maximum extractable work is quantified by the ergotropy \cite{yang2023battery,yadav2025thermo}
\begin{equation}
\xi(t) = \text{Tr}\left[(\hat{\rho}(t) - \hat{\zeta}) \hat{\mathcal{H}}_{\text{QB}}\right].
\label{eq:storable_energy}
\end{equation}
For our initial Gibbs state $\hat{\zeta}$, the ergotropy is initially zero, but after charging, it represents the maximum amount of work that can be extracted through unitary operations until the system reaches the passive (non-Gibbs) state $\hat{\pi}$.

The capacity of the QB is defined as the difference between the ergotropy and anti-ergotropy \cite{yang2023battery,yadav2025thermo}
\begin{equation}
\mathcal{Q} = \xi(t) - \mathcal{P}(t).
\label{eq:capacity}
\end{equation}
This capacity represents the total energy range over which the QB can operate during a complete thermodynamic cycle. The ergotropy and anti-ergotropy of a quantum system fluctuate during an isentropic thermodynamic cycle, but their difference remains constant during any unitary evolution, reflecting the fundamental energy span available for work transfer while preserving the unitary evolution of the battery.

To quantify the role of quantum coherence during the charging and discharging processes of the QB, we employ the $l_1$-norm of quantum coherence~\cite{baumgratz2014quantifying,HU20181}:
\begin{equation}
\mathcal{C}(\hat{\rho}(t)) = \sum_{i \neq j} |\langle i | \hat{\rho}(t) | j \rangle|,
\label{eq:QC_l1}
\end{equation}
where the sum is taken over all off-diagonal elements of the density matrix in the energy eigenbasis. This coherence measure captures the quantum superposition effects that enable enhanced performance beyond classical battery limitations.

% {Before moving ahead, we emphasize that the Pauli-X gate-based charging unitary is chosen for its experimental feasibility in NMR platforms (via RF pulses) and is not necessarily optimized to always achieve the active state. We note that an optimized unitary to achieve the maximum energy \(\max_{\hat{U}} \text{Tr}[\hat{\mathcal{H}}_{\text{QB}} \hat{U} \hat{\zeta} \hat{U}^\dagger]\) could be considered in future work but is beyond the scope of this study.}

The complete QB cycle can now be summarized as: (i) Initial State: $\hat{\rho}(0) = \hat{\zeta}$ (Gibbs thermal state, passive, no extractable work), (ii) Charging: $\hat{\zeta} \rightarrow \hat{\eta}$ (energy injection measured by anti-ergotropy $\mathcal{P}(t)$), (iii) Discharging: $\hat{\eta} \rightarrow \hat{\pi}$ (work extraction measured by ergotropy $\xi(t)$), and (iv) Capacity: $\mathcal{Q} = \xi(t) - \mathcal{P}(t)$ (total energy operating range). 

We emphasize that the final passive state $\hat{\pi}$ is generally not the same as the initial Gibbs state $\hat{\zeta}$, and thus, a fully closed thermodynamic cycle would require an additional thermal relaxation step to return the battery to its original uncharged configuration.
It is worth noting that, in principle, the initial passive state of the QB could have been constructed directly from the eigendecomposition of an arbitrary quantum state by reordering its populations to form a passive configuration. This would allow for a fully general definition of passive-to-active transitions (and vice versa) without reference to a thermal environment. However, in our study, we intentionally adopt the Gibbs thermal state as the initial uncharged and passive state. This choice is motivated by the desire to model the performance of the QB under physically realistic conditions—namely, finite temperature operation in contact with a thermal bath. By anchoring our protocol at a well-defined finite temperature, we provide a consistent and experimentally relevant framework for evaluating energy storage, extractable work, and coherence dynamics of the QB under realistic thermodynamic constraints.

\begin{figure}[!t]
\centering
\includegraphics[width=\columnwidth]{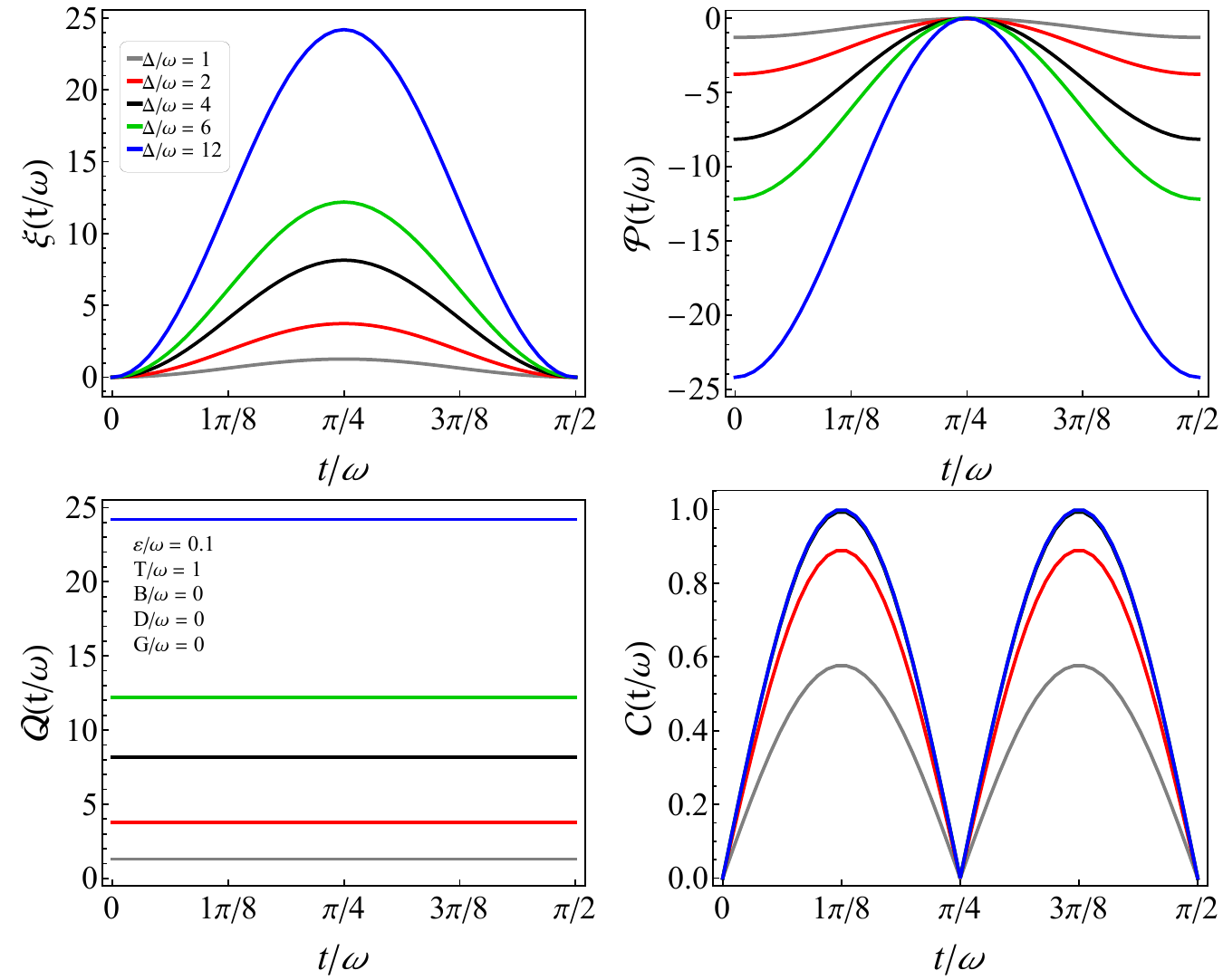}
\put(-247,188){(a)}
\put(-120,188){(b)}
\put(-247,90){(c)}
\put(-125,90){(d)}
\vspace{-3mm}
\caption{{Temporal evolution of QB performance metrics. (a) Ergotropy $\xi(t/\omega)$, (b) anti-ergotropy $\mathcal{P}(t/\omega)$, (c) capacity $\mathcal{Q}(t/\omega)$, and (d) $l_1$-norm of quantum coherence $\mathcal{C}(t/\omega)$ versus $t/\omega$ for different axial anisotropy values: $\Delta/\omega = 1$ (gray), 2 (red), 4 (black), 6 (green), and 12 (blue). Parameters: $G = D = B = 0$, $\epsilon/ \omega=0.1$, and $ T/\omega = 1$.}}
\label{F10}
\end{figure}
We now analyze the temporal evolution of key performance metrics for our closed QB system.}

{
\subsection{Axial anisotropy effects}
Figure~\ref{F10} presents the dynamics of ergotropy $\xi(t/\omega)$, anti-ergotropy $\mathcal{P}(t/\omega)$, capacity $\mathcal{Q}(t/\omega)$, and quantum coherence $\mathcal{C}(t/\omega)$ as functions of dimensionless time $t/\omega$ for various values of the axial anisotropy parameter $\Delta/\omega$.

The ergotropy $\xi(t/\omega)$ exhibits periodic oscillations with a characteristic period determined by the system's energy scale [Fig.~\ref{F10}(a)]. {For given values of the parameter chosen, the maximum extraction of work occurs at $t = \pi/(4\omega)$}. In the strongly anisotropic regime ($\Delta/\omega \gg \epsilon/\omega$), the maximum ergotropy scales linearly with the anisotropy parameter: $\xi_{\text{max}} \propto \Delta/\omega$. This scaling relationship directly reflects the $\Delta$-dependent eigenvalue spectrum of the QB Hamiltonian $\mathcal{\hat{H}}_{\text{QB}}$, where increasing $\Delta/\omega$ improves the energy gap $\nu_{max} - \nu_{min}$ between the highest and lowest energy eigenstates, thus expanding the operational range for work extraction.

The anti-ergotropy $\mathcal{P}(t/\omega)$ displays complementary temporal behavior to the ergotropy [Fig.~\ref{F10}(b)], establishing a fundamental trade-off between extractable work and required energy injection. This complementarity manifests as a reciprocal relationship: maximum ergotropy coincides with zero anti-ergotropy, indicating that a fully charged QB requires no additional energy input while providing maximum work output. In contrast, maximum energy injection occurs when the QB reaches complete depletion (passive state). Both quantities exhibit identical $\Delta$-dependence, confirming that enhanced storage capacity requires proportionally increased work injection to achieve full charging.

The QB capacity $\mathcal{Q}(t/\omega) = \xi(t/\omega) - \mathcal{P}(t/\omega)$ {being invariant in time}, satisfies the general inequality $\mathcal{Q}(t/\omega) \geq \xi(t/\omega)$ and demonstrates the same linear $\Delta/\omega$ scaling observed in its constituent quantities [Fig.~\ref{F10}(c)]. 

The quantum coherence dynamics reveal distinctly different characteristics [Fig.~\ref{F10}(d)]. The coherence measure $\mathcal{C}(t/\omega)$ reaches minimum values when the QB has charge injection (minimum anti-ergotropy or zero ergotropy) at $\omega t = 0$ and $t = \pi/(2\omega)$, while achieving maximum values $\mathcal{C}_{\text{max}} \approx 1$ when the QB is approximately 50\% charged. This phase relationship indicates that maximum coherence occurs at intermediate charging levels, specifically when $\xi(t/\omega) = \xi_{\text{max}}/2$ [equivalently, $\mathcal{P}(t/\omega) = \mathcal{P}_{\text{min}}/2$].
Notably, the quantum coherence oscillations exhibit negligible $\Delta/\omega$ dependence, contrasting sharply with the energy-related quantities. This independence indicates that coherence dynamics is mainly governed by the charging Hamiltonian $\mathcal{H}_{\text{ch}}$ rather than the intrinsic anisotropy parameters of the system.

The observed scaling behavior reveals a fundamental decoupling between quantum coherence dynamics and energy storage capacity. While $\mathcal{C}(t/\omega)$ saturates and becomes $\Delta/\omega$-independent for $\Delta/\omega \gtrsim 4$, both $\xi_{\text{max}}$ and $\mathcal{Q}(t/\omega)$ continue to scale linearly with $\Delta/\omega$. This decoupling demonstrates that enhanced ergotropy arises predominantly from spectral gap widening between extremal eigenstates rather than increased quantum coherence, indicating the emergence of classical (incoherent) ergotropy in strongly anisotropic systems.
The capacity scaling relationship $\mathcal{Q} \sim 2\Delta/\omega$ establishes that the QB's performance is fundamentally determined by the Hamiltonian's spectral properties. This result provides a clear design principle: maximizing the axial anisotropy parameter $\Delta/\omega$ directly enhances the QB's energy storage and extraction capabilities through purely spectral engineering, independent of quantum coherence resources.
\subsection{Magnetic field effects on QB performance}\label{subsec:magnetic_field}

We now examine the influence of an external magnetic field along the $z$-axis on the QB dynamics. Figure~\ref{F11} presents the temporal evolution of performance metrics for various magnetic field strengths $B/\omega$ ranging from $0$ to $2$.

\begin{figure}[!t]
\centering
\includegraphics[width=\columnwidth]{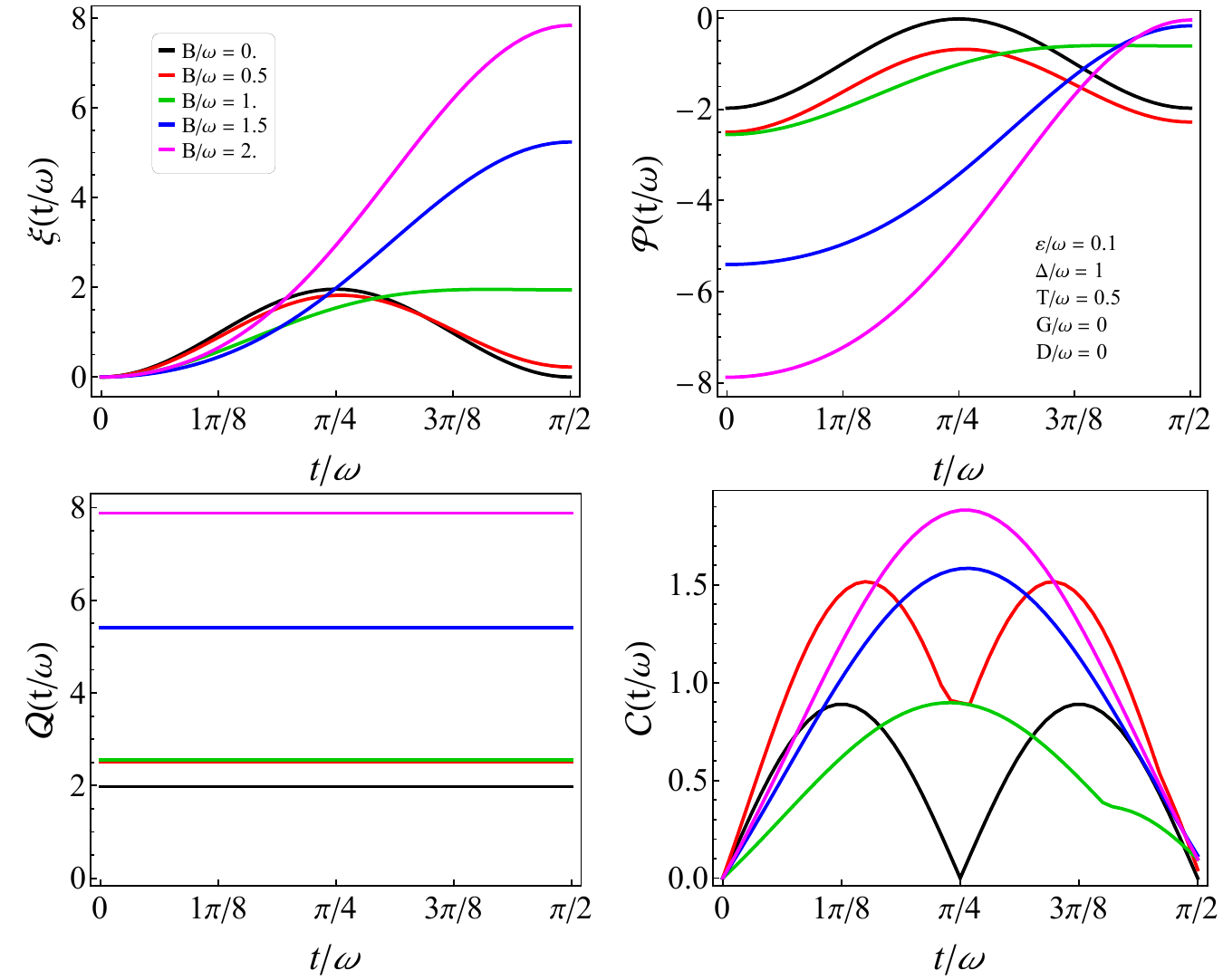}
\put(-247,188){(a)}
\put(-120,188){(b)}
\put(-247,88){(c)}
\put(-122,88){(d)}
\vspace{-3mm}
\caption{{Magnetic field dependence of QB performance metrics. (a) Ergotropy $\xi(t/\omega)$, (b) anti-ergotropy $\mathcal{P}(t/\omega)$, (c) capacity $\mathcal{Q}(t/\omega)$, and (d) $l_1$-norm of quantum coherence $\mathcal{C}(t/\omega)$ versus $t/\omega$ for different magnetic field strengths: $B/\omega = 0.0$ (black), $0.5$ (red), $1.0$ (green), $1.5$ (blue), and $2.0$ (magenta). Parameters: $G = D = 0$, $\Delta/\omega = 1$, $T/\omega = 0.5$, and $\epsilon/\omega = 0.1$.}}
\label{F11}
\end{figure}

The magnetic field introduces significant modifications to the QB's temporal characteristics [Figs.~\ref{F11}(a) and \ref{F11}(b)]. In the absence of a magnetic field ($B = 0$), the system exhibits rapid charging kinetics with maximum ergotropy $\xi_{\text{max}} \approx 2$ achieved at $t = \pi/(4\omega)$. As the magnetic field strength increases to $B = 2\omega$, the maximum ergotropy increases substantially to $\xi_{\text{max}} \approx 8$, but the optimal extraction time shifts to $t = \pi/(2\omega)$ {for a given charging unitary}, indicating a trade-off between {maximum ergotropy extraction (or anti-ergotropy injection)} and temporal efficiency.

The anti-ergotropy $\mathcal{P}(t/\omega)$ exhibits precisely complementary phase behavior [Fig.~\ref{F11}(b)], confirming that weak magnetic fields correspond to faster energy injection rates, while stronger fields result in prolonged injection timescales. This temporal retardation is compensated by significantly enhanced peak energy injection and extraction capabilities, demonstrating that magnetic field tuning provides control over both the magnitude and timing of QB operations.

The QB capacity analysis [Fig.~\ref{F11}(c)] confirms that increasing $B/\omega$ substantially enhances the energy storage range. Unlike the anisotropy-driven enhancement discussed previously, magnetic field effects exhibit distinct scaling behavior without apparent saturation within the examined parameter range.

Particularly noteworthy is the behavior of quantum coherence under magnetic field influence [Fig.~\ref{F11}(d)]. The peak coherence values increase proportionally with $B/\omega$ without exhibiting the saturation effects observed in the anisotropy case (Fig.~\ref{F10}). This contrasts markedly with the $\Delta/\omega$-driven enhancement, where quantum coherence plateaued beyond $\Delta/\omega \gtrsim 4$ while energy-related quantities continued to scale linearly.

The magnetic field presents a distinct optimization strategy that simultaneously enhances both energy storage capacity and quantum coherence generation. This dual enhancement mechanism differs fundamentally from anisotropy-based optimization, which primarily affects energy scales while leaving coherence dynamics largely unchanged. The absence of coherence saturation under magnetic field tuning suggests that $B$-field control accesses different degrees of freedom in the QB's Hilbert space, potentially enabling more efficient utilization of quantum resources.

These results establish magnetic field strength as a complementary control parameter to axial anisotropy, offering an alternative pathway for QB optimization that circumvents the coherence saturation limitations encountered in purely anisotropy-driven enhancement schemes.}

{
\subsection{Temperature effects}
The thermal environment plays a crucial role in determining QB performance. Figure~\ref{F12} examines the temperature dependence of all performance metrics across the range $T/\omega = 0.5$ to $T/\omega = 4.0$.
\begin{figure}[!t]
\centering
\includegraphics[width=\columnwidth]{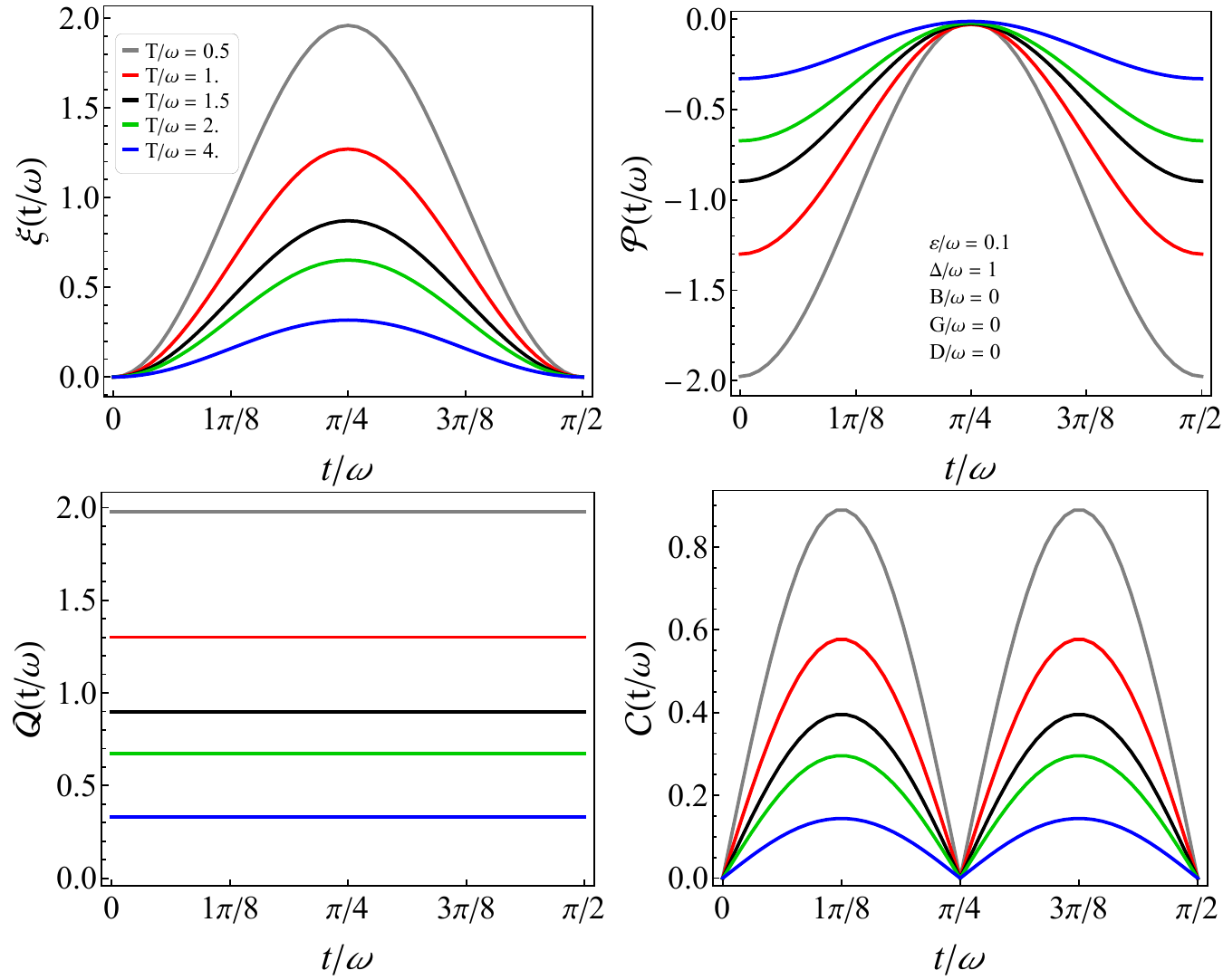}
\put(-247,188){(a)}
\put(-120,188){(b)}
\put(-247,88){(c)}
\put(-122,88){(d)}
\vspace{-3mm}
\caption{{Temperature dependence of QB performance metrics. (a) Ergotropy $\xi(t/\omega)$, (b) anti-ergotropy $\mathcal{P}(t/\omega)$, (c) capacity $\mathcal{Q}(t/\omega)$, and (d) $l_1$-norm of quantum coherence $\mathcal{C}(t/\omega)$ versus $t/\omega$ for different temperatures: $T/\omega = 0.5$ (gray), $1.0$ (red), $1.5$ (black), $2.0$ (green), and $4.0$ (blue). Parameters: $G = D = B = 0$, $\Delta/\omega = 1$, and $\epsilon/\omega = 0.1$.}}
\label{F12}
\end{figure}

The ergotropy evolution [Fig.~\ref{F12}(a)] exhibits strong temperature sensitivity with systematic amplitude suppression as thermal energy increases. At low temperature ($T/\omega = 0.5$), the system maintains robust oscillations with amplitude $\xi_{\text{max}} \approx 2.0$, while at elevated temperature ($T/\omega = 4.0$), thermal fluctuations substantially reduce the oscillation amplitude to $\xi_{\text{max}} \approx 0.25$. This temperature-induced degradation directly correlates with the thermal population of higher energy levels, which reduces the available energy difference between the ground state and excited states.

The anti-ergotropy dynamics [Fig.~\ref{F12}(b)] maintain the expected complementary relationship with ergotropy oscillations, confirming the fundamental thermodynamic consistency across all temperature regimes. The capacity analysis [Fig.~\ref{F12}(c)] demonstrates systematic reduction with increasing temperature, following the same scaling as the individual ergotropy and anti-ergotropy components.

The quantum coherence evolution [Fig.~\ref{F12}(d)] reveals particularly significant temperature dependence. The coherence amplitude decreases from $\mathcal{C}_{\text{max}} \approx 0.8$ at $T/\omega = 0.5$ to $\mathcal{C}_{\text{max}} \approx 0.1$ at $T/\omega = 4.0$, indicating that thermal fluctuations severely impact the quantum resources essential for QB operation. This coherence suppression occurs through thermal decoherence mechanisms that destroy the quantum superposition states responsible for enhanced energy storage capabilities.

Importantly, the phase relationship between coherence and energy-related quantities remains preserved across all temperatures, suggesting that while thermal effects uniformly suppress all quantum advantages, they do not fundamentally alter the underlying quantum correlations between different performance metrics.

These results establish clear operational constraints for practical QB implementations. The observed temperature dependence demonstrates that enhanced QB performance requires low temperature conditions to preserve both ergotropy and quantum coherence. The approximately linear relationship between temperature and performance degradation suggests that each order of magnitude increase in thermal energy results in proportional reduction of quantum advantages.

This temperature sensitivity contrasts with the enhancement mechanisms provided by anisotropy and magnetic field tuning, highlighting the fundamental trade-off between environmental control and parameter optimization in QB design. The results indicate that while structural parameters ($\Delta/\omega$, $B/\omega$) can enhance maximum performance, environmental temperature ultimately sets the practical limits for QB operation.
}
{
\subsection{KSEA interaction effect}
\begin{figure}[!t]
\centering
\includegraphics[width=\columnwidth]{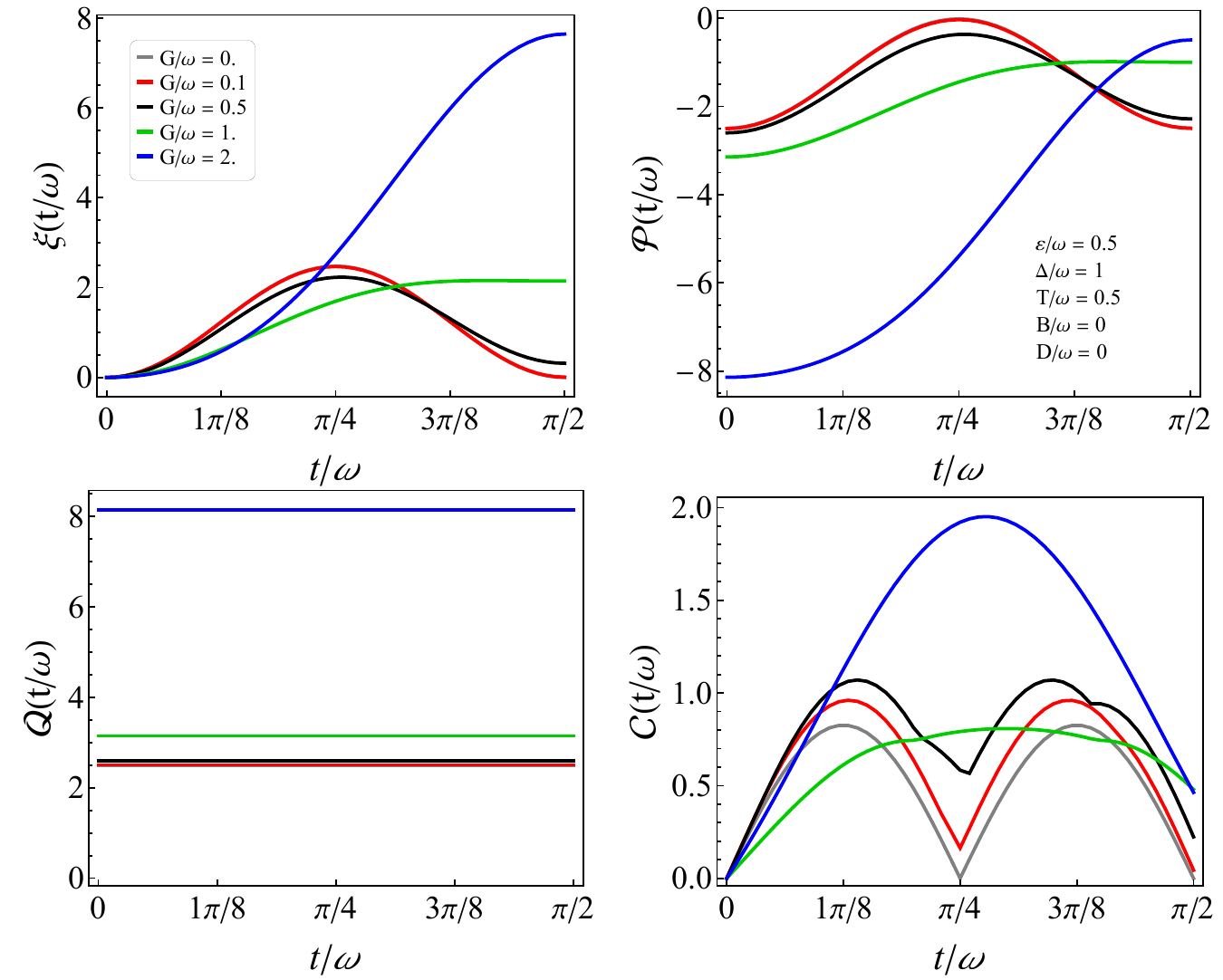}
\put(-247,188){(a)}
\put(-120,188){(b)}
\put(-247,88){(c)}
\put(-122,88){(d)}
\vspace{-3mm}
\caption{{ KSEA interaction effects on QB performance metrics. (a) Ergotropy $\xi(t/\omega)$, (b) anti-ergotropy $\mathcal{P}(t/\omega)$, (c) capacity $\mathcal{Q}(t/\omega)$, and (d) $l_1$-norm of quantum coherence $\mathcal{C}(t/\omega)$ versus $t/\omega$ for different KSEA interaction strengths: $G/\omega = 0.0$ (gray), $0.1$ (red), $0.5$ (black), $1.0$ (green), and $2.0$ (blue). Parameters: $D = B = 0$, $\Delta/\omega = 1$, and $T = \epsilon = \omega/2$.}}
\label{F13}
\end{figure}
The ergotropy evolution [Fig.~\ref{F13}(a)] demonstrates dramatic enhancement under KSEA interactions. In the absence of KSEA interactions ($G/\omega = 0$), the system exhibits moderate ergotropy oscillations with amplitude $\xi_{\text{max}} \approx 2$. As the KSEA interaction strength increases to $G/\omega = 2.0$, the oscillation amplitude grows substantially to $\xi_{\text{max}} \approx 8$, representing a four-fold enhancement in ergotropy extraction.

This enhancement mechanism operates through the collective nature of KSEA interactions, which modify the energy level structure by introducing symmetric exchange correlation-dependent energy shifts. The systematic scaling observed for intermediate interaction values ($G/\omega = 0.1, 0.5, 1.0$) indicates a continuous relationship between interaction strength and performance enhancement, suggesting that KSEA effects can be precisely tuned for optimal QB operation.

The anti-ergotropy dynamics [Fig.~\ref{F13}(b)] exhibit the expected complementary behavior, with oscillation amplitudes reaching $\mathcal{P}_{\text{min}} \approx -8$ at maximum collective interaction strength. This perfect phase opposition to ergotropy evolution confirms that KSEA interactions enhance both the system's ability to store energy and extract work, maintaining thermodynamic consistency throughout the interaction range.

The QB capacity analysis [Fig.~\ref{F13}(c)] reveals systematic enhancement following the same scaling as individual ergotropy and anti-ergotropy components. The capacity amplitude increases continuously from $\mathcal{Q} \approx 2$ at $G/\omega = 0$ to $\mathcal{Q} \approx 8$ at $G/\omega = 2.0$, demonstrating that KSEA interactions provide a robust mechanism for capacity enhancement without apparent saturation effects within the examined parameter range.

Particularly significant is the quantum coherence response to KSEA interactions [Fig.~\ref{F13}(d)]. Unlike the temperature effects that uniformly suppress coherence, KSEA interactions actively amplify quantum coherence, with values increasing from $\mathcal{C}_{\text{max}} \approx 0.5$ at weak interaction to $\mathcal{C}_{\text{max}} \approx 2.0$ at strong interaction. This coherence amplification exhibits non-linear dependence on $G/\omega$, suggesting that KSEA interactions access collective quantum states that are inherently more coherent than single-particle configurations.

The KSEA interaction results demonstrate that collective symmetric exchange interactions provide a distinct pathway for QB enhancement that simultaneously improves both QB performance metrics and quantum coherence. This dual enhancement distinguishes KSEA effects from the previously examined control parameters: while anisotropy primarily affects energy scales and magnetic fields provide capacity enhancement with coherence preservation, KSEA interactions actively generate quantum coherence through collective quantum correlations.

These findings establish KSEA interactions as a particularly promising mechanism for QB optimization, offering the potential to harness collective quantum effects for enhanced energy storage while simultaneously amplifying the quantum resources necessary for superior performance.}

{
\subsection{DM interaction effect}
We examine the influence of the $z$-component DM interaction on QB performance. Figure~\ref{F14} presents a systematic analysis of DM effects with interaction strength $D/\omega$ varying from $0$ to $2.0$. 
\begin{figure}[!t]
\centering
\includegraphics[width=\columnwidth]{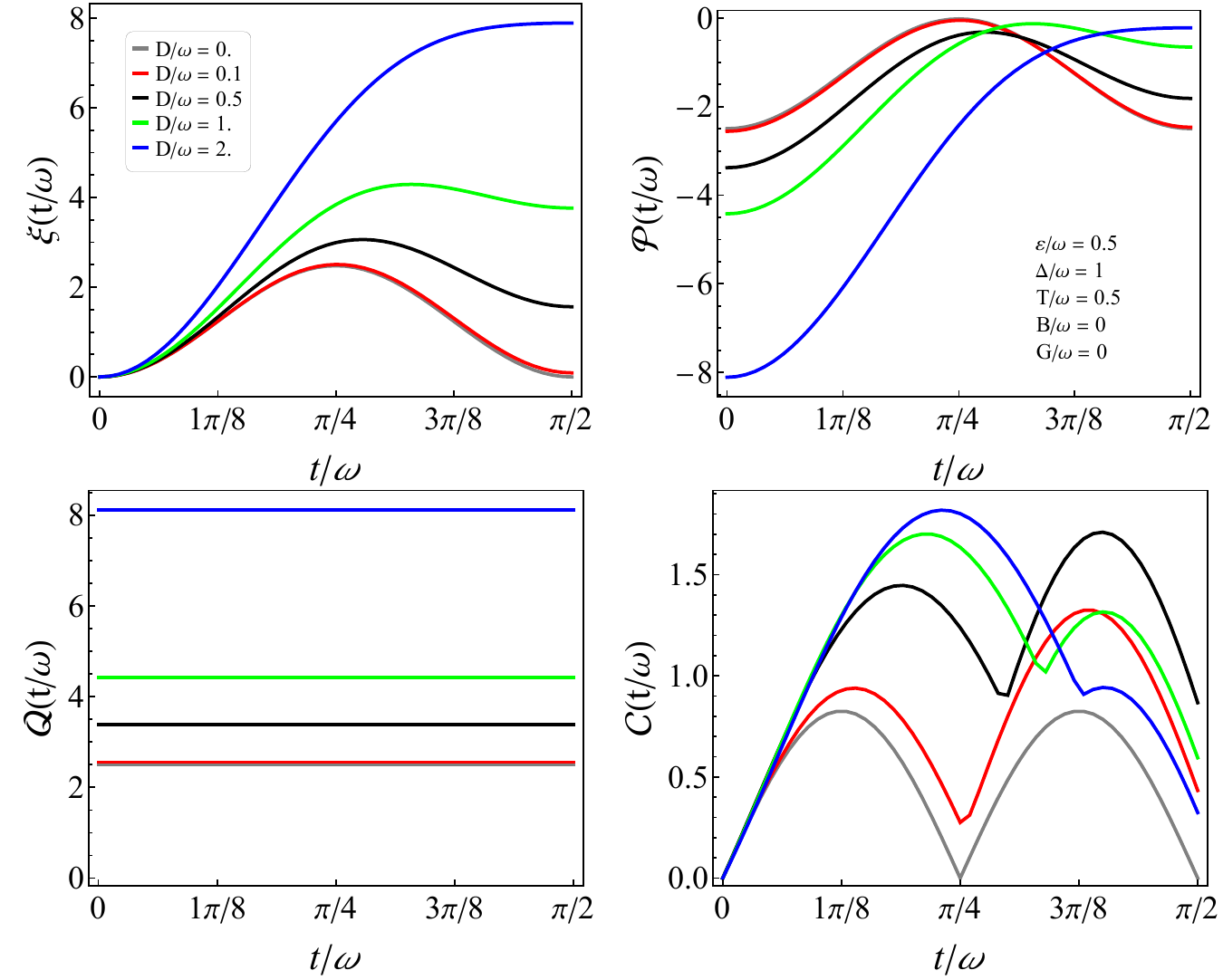}
\put(-247,188){(a)}
\put(-120,188){(b)}
\put(-247,88){(c)}
\put(-122,88){(d)}
\vspace{-3mm}
\caption{{ DM interaction effects on QB performance metrics. (a) Ergotropy $\xi(t/\omega)$, (b) anti-ergotropy $\mathcal{P}(t/\omega)$, (c) capacity $\mathcal{Q}(t/\omega)$, and (d) $l_1$-norm of quantum coherence $\mathcal{C}(t/\omega)$ versus $t/\omega$ for different DM interaction strengths: $D/\omega = 0.0$ (gray), $0.1$ (red), $0.5$ (black), $1.0$ (green), and $2.0$ (blue). Parameters: $G = B = 0$, $\Delta/\omega = 1$, and $T = \epsilon = \omega/2$.}}
\label{F14}
\end{figure}

The ergotropy dynamics [Fig.~\ref{F14}(a)] demonstrate substantial enhancement under DM interactions. In the absence of antisymmetric exchange interaction ($D/\omega = 0$), the system exhibits moderate ergotropy oscillations with amplitude $\xi_{\text{max}} \approx 2$. As the DM interaction strength increases to $D/\omega = 2.0$, the oscillation amplitude grows dramatically to $\xi_{\text{max}} \approx 8$, representing a four-fold enhancement in maximal extractable work.

This enhancement mechanism operates through the antisymmetric nature of DM interactions, which introduce spin-orbit coupling effects that modify the energy level structure and create new pathways for energy storage and extraction. The systematic scaling observed across intermediate interaction values demonstrates continuous tunability of QB performance through DM interaction strength.

The anti-ergotropy evolution [Fig.~\ref{F14}(b)] maintains perfect complementarity with ergotropy oscillations, reaching amplitudes of $\mathcal{P}_{\text{min}} \approx -8$ at maximum DM interaction. This behavior confirms that antisymmetric exchange interactions enhance both energy injection and work extraction capability while preserving fundamental thermodynamic relationships.

The QB capacity analysis [Fig.~\ref{F14}(c)] follows the same enhancement pattern as individual ergotropy components, with amplitude increasing systematically from $\mathcal{Q} \approx 2$ at $D/\omega = 0$ to $\mathcal{Q} \approx 8$ at $D/\omega = 2.0$. This scaling demonstrates that DM interactions provide relatively robust capacity enhancement comparable to KSEA effects.

However, the quantum coherence response [Fig.~\ref{F14}(d)] exhibits more constrained behavior under DM interactions. While coherence values increase from $\mathcal{C}_{\text{max}} \approx 0.5$ to $\mathcal{C}_{\text{max}} \approx 1.5$, this enhancement is more modest compared to the dramatic improvements in energy-related metrics. This suggests that DM interactions primarily affect the classical energy storage aspects while providing limited amplification of quantum coherence resources.

The DM interaction results establish antisymmetric exchange as an effective mechanism for QB enhancement, particularly for energy-related performance metrics. However, this enhancement involves temporal trade-offs: while increased DM interaction substantially amplifies peak energy extraction and storage capacity, it extends the characteristic charging and discharging timescales, similar to the behavior observed with magnetic field effects.

This temporal trade-off distinguishes DM interactions from KSEA effects, which provide both energy and coherence enhancement. The antisymmetric spin-orbit coupling introduced by DM interactions enables significant work extraction improvements while maintaining essential quantum coherence, though without the coherence amplification observed in collective KSEA interactions.

These findings position DM interactions as a valuable design parameter for QB optimization, particularly when maximum energy storage capacity is prioritized over charging speed or quantum coherence enhancement.}

{
\subsection{Impact of $\Delta$ versus $\epsilon$ on temporally optimized QB metrics }\label{subsec:optimization}

{After characterizing the temporal behavior of individual performance metrics, we now turn to the optimization of quantum battery performance through collective parameter variation. This analysis identifies parameter regimes that further enhance the time-optimized ergotropy, suppress the time-minimized anti-ergotropy, and strengthen the time-optimized quantum coherence under different control mechanisms. In contrast, the capacity is a time-independent quantity determined solely by the spectrum of the initial Gibbs state and the battery Hamiltonian. Since unitary charging preserves this spectrum, the capacity remains fixed in time, whereas ergotropy and anti-ergotropy evolve dynamically with the state.}

Figure~\ref{f15} presents a comprehensive analysis of temporally optimized {(maximized for ergotropy and quantum coherence, whereas minimized for anti-ergotropy)} performance metrics as functions of axial anisotropy $\Delta/\omega$ and transverse anisotropy $\epsilon/\omega$, with all exchange interactions disabled ($D = G = B = 0$) and fixed temperature $T/\omega = 1.0$. This parameter regime isolates the pure effects of magnetic anisotropy on QB functionality.

\begin{figure}[!t]
\centering
\includegraphics[width=0.95\columnwidth]{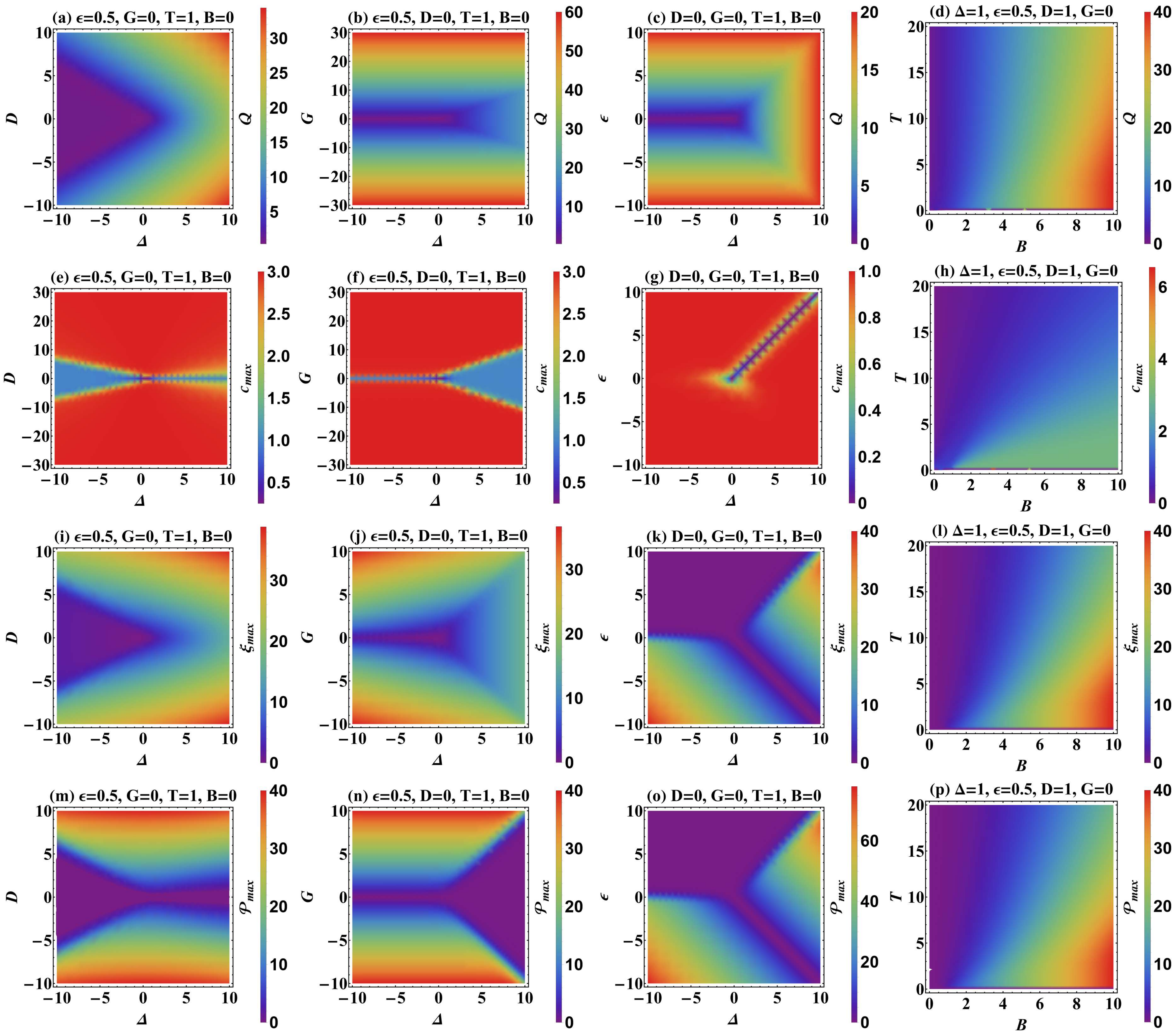}
\put(-165,188){(a)}
\put(-35,188){(b)}
\put(-165,85){(c)}
\put(-35,85){(d)}
\put(-130,140){\rotatebox{90}{${\xi_\text{max}}$}}
\put(0,140){\rotatebox{90}{${\mathcal{P}_\text{min}}$}}
\put(-130,35){\rotatebox{90}{${\mathcal{Q}}$}}
\put(0,35){\rotatebox{90}{${\mathcal{C}_\text{max}}$}}
\put(-240,140){\rotatebox{90}{${\epsilon/\omega}$}}
\put(-240,40){\rotatebox{90}{${\epsilon/\omega}$}}
\put(-105,142){\rotatebox{90}{${\epsilon/\omega}$}}
\put(-105,40){\rotatebox{90}{${\epsilon/\omega}$}}
\put(-195,95){${\Delta/\omega}$}
\put(-195,-5){${\Delta/\omega}$}
\put(-62,95){${\Delta/\omega}$}
\put(-62,-5){${\Delta/\omega}$}
\caption{{Density plots of (a) {temporally maximized ergotropy} $\xi_\text{max}$, (b) {temporally minimized} anti-ergotropy $\mathcal{P}_\text{min}$, (c) capacity $\mathcal{Q}$, and (d) {temporally maximized} quantum coherence $\mathcal{C}_\text{max}$ versus axial anisotropy $\Delta/\omega$ and transverse anisotropy $\epsilon/\omega$. Parameters: $D = G = B = 0$ and $T/\omega = 1.0$.}}
\label{f15}
\end{figure}

The {temporally maximized} ergotropy [Fig.~\ref{f15}(a)] reveals a highly structured parameter space with maximum values reaching $\xi_\text{max} \approx 30$ at specific combinations of anisotropy parameters. The enhancement exhibits pronounced asymmetry with respect to the sign of $\Delta/\omega$, indicating that spin alignment orientation—whether along the $z$-axis ($\Delta/\omega > 0$) or confined to the $xy$-plane ($\Delta/\omega < 0$)—fundamentally determines energy extraction capacity. Optimal ergotropy emerges at moderate values of both parameters rather than extremes, suggesting that balanced anisotropy configurations maximize extractable work.

The temporally minimized anti-ergotropy [Fig.~\ref{f15}(b)] displays complementary behavior with $\mathcal{P}_\text{min}$ values ranging from $-36$ to $0$, reflecting the energy injection requirements for transitioning from the passive Gibbs state to active configurations. The negative values confirm the fundamental thermodynamic constraint that energy investment is required to achieve maximum charging. The parameter dependence mirrors but inverts the ergotropy trends, validating the expected relationship between energy injection and extraction capabilities.

The capacity [Fig.~\ref{f15}(c)] synthesizes these complementary effects, showing $\mathcal{Q}$ values exceeding $36$ in optimal parameter regions. The strong correlation between optimal capacity and ergotropy zones demonstrates that maximizing work extraction inherently optimizes overall battery performance.

Particularly significant is the quantum coherence analysis [Fig.~\ref{f15}(d)], which reveals that maximum coherence ($\mathcal{C}_\text{max} \approx 1.8$) correlates strongly with enhanced QB performance regions. This correlation provides direct evidence for quantum advantage in energy storage, where coherence serves simultaneously as both a quantum resource and a performance indicator.

The parameter sensitivity analysis reveals that QB operates most efficiently when $\Delta/\omega$ and $\epsilon/\omega$ have comparable magnitudes with the same sign, creating balanced competition between axial and transverse anisotropies. This balance optimizes the energy level structure for maximum coherence generation while maintaining sufficient energy gaps for effective ergotropy extraction.
}
{
\subsection{Impact of DM and KSEA interactions on temporally optimized QB metrics }\label{subsec:optimization}

Figure~\ref{f16} presents density plots of temporally optimized performance metrics as functions of DM interaction $D/\omega$ and KSEA interaction $G/\omega$, with fixed parameters $\Delta/\omega = -1$, $\epsilon = \omega/2$, $T/\omega = 1.0$, and $B/\omega = 0$. This regime enables systematic investigation of the interplay between antisymmetric and collective exchange interactions.

\begin{figure}[!t]
\centering
\includegraphics[width=0.95\columnwidth]{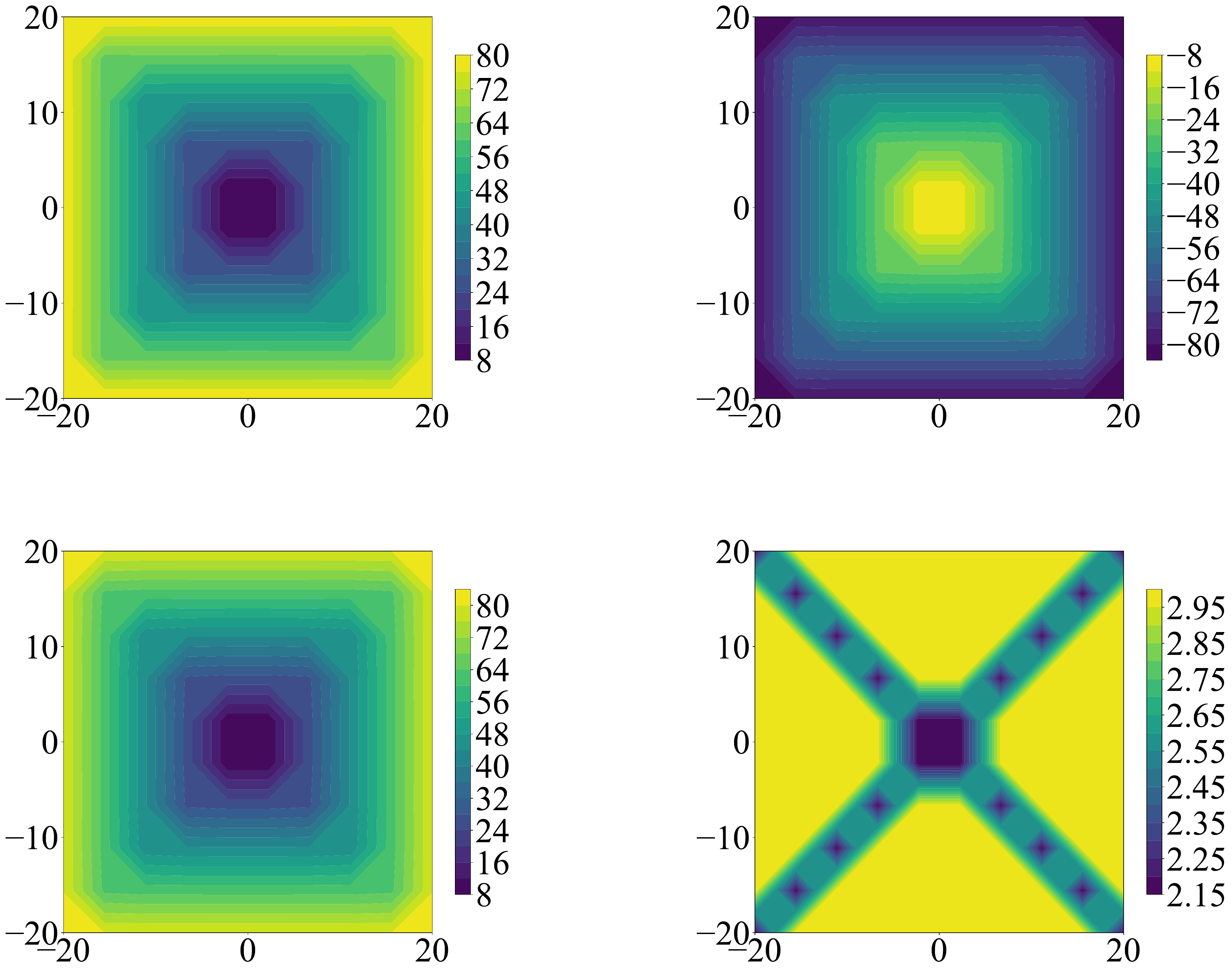}
\put(-165,188){(a)}
\put(-35,188){(b)}
\put(-165,85){(c)}
\put(-35,85){(d)}
\put(-130,140){\rotatebox{90}{${\xi_\text{max}}$}}
\put(0,140){\rotatebox{90}{${\mathcal{P}_\text{min}}$}}
\put(-130,35){\rotatebox{90}{${\mathcal{Q}}$}}
\put(0,35){\rotatebox{90}{${\mathcal{C}_\text{max}}$}}
\put(-240,140){\rotatebox{90}{${G/\omega}$}}
\put(-240,40){\rotatebox{90}{${G/\omega}$}}
\put(-105,142){\rotatebox{90}{${G/\omega}$}}
\put(-105,40){\rotatebox{90}{${G/\omega}$}}
\put(-195,95){${D/\omega}$}
\put(-195,-5){${D/\omega}$}
\put(-65,95){${D/\omega}$}
\put(-65,-5){${D/\omega}$}
\caption{{Density plots of (a) {temporally maximized} ergotropy $\xi_\text{max}$, (b) {temporally minimized} anti-ergotropy $\mathcal{P}_\text{min}$, (c) capacity $\mathcal{Q}$, and (d) {temporally maximized} quantum coherence $\mathcal{C}_\text{max}$ versus DM interaction $D/\omega$ and KSEA interaction $G/\omega$. Parameters: $\Delta/\omega = -1$, $\epsilon = \omega/2$, $T/\omega = 1.0$, and $B/\omega = 0$.}}
\label{f16}
\end{figure}

The {temporally} maximized ergotropy [Fig.~\ref{f16}(a)] exhibits remarkable enhancement with increasing magnitude of both interaction parameters, reaching values exceeding $80$ in optimal regions located at the corners of the parameter space. The enhancement pattern displays approximate reflection symmetry about both axes, indicating that interaction magnitudes $|D/\omega|$ and $|G/\omega|$ rather than their $+/-$ signs primarily govern energy extraction capability. This behavior arises from constructive contributions to the energy gap between extremal eigenstates, thereby enhancing maximum extractable work.

The {temporally} minimized anti-ergotropy [Fig.~\ref{f16}(b)] demonstrates complementary energy injection requirements with values ranging from $-80$ to $-8$, exhibiting inverse correlation with ergotropy enhancement. The capacity analysis [Fig.~\ref{f16}(c)] follows similar enhancement trends, with peak values exceeding $80$ in regions where both interaction parameters are maximized.

quantum coherence analysis [Fig.~\ref{f16}(d)] reveals $\mathcal{C}_\text{max}$ values ranging from $2.15$ to $2.95$, with the highest values strongly correlated with optimal battery performance regions. This correlation demonstrates that DM and KSEA interactions not only modify quantized energy level structure but also generate and sustain the quantum coherence essential for enhanced QB operation.

The synergistic enhancement across all metrics indicates that the DM and KSEA interactions operate constructively rather than additively. Optimal performance occurs where both parameters simultaneously achieve large magnitudes, suggesting that experimental implementations should focus on maximizing both antisymmetric and symmetric exchange interactions as well as other collective many-body effects.

\subsection{Impact of $B$ and $T$  on temporally optimized QB metrics }\label{subsec:optimization}

Figure~\ref{f17} demonstrates the temperature and magnetic field dependence of temporally optimized performance metrics for fixed parameters $\Delta/\omega = 0.1$, $\epsilon/\omega = 0.1$, and $D = G = 0$.

\begin{figure}[!t]
\centering
\includegraphics[width=0.95\columnwidth]{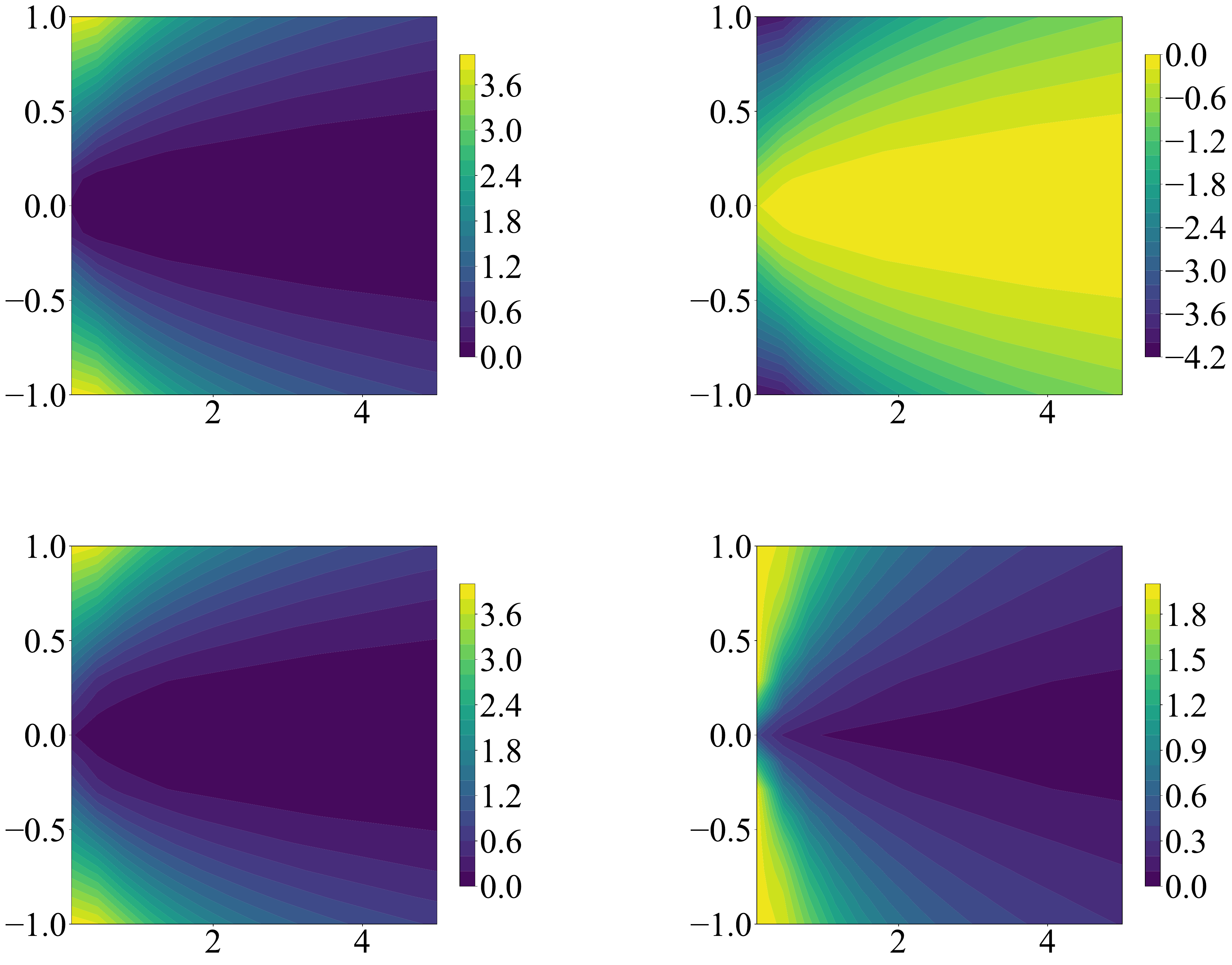}
\put(-165,188){(a)}
\put(-35,188){(b)}
\put(-165,85){(c)}
\put(-35,85){(d)}
\put(-130,140){\rotatebox{90}{${\xi_\text{max}}$}}
\put(0,140){\rotatebox{90}{${\mathcal{P}_\text{min}}$}}
\put(-130,35){\rotatebox{90}{${\mathcal{Q}}$}}
\put(0,35){\rotatebox{90}{${\mathcal{C}_\text{max}}$}}
\put(-240,140){\rotatebox{90}{${B/\omega}$}}
\put(-240,40){\rotatebox{90}{${B/\omega}$}}
\put(-105,142){\rotatebox{90}{${B/\omega}$}}
\put(-105,40){\rotatebox{90}{${B/\omega}$}}
\put(-195,95){${T/\omega}$}
\put(-195,-5){${T/\omega}$}
\put(-65,95){${T/\omega}$}
\put(-65,-5){${T/\omega}$}
\caption{{Density plots of (a) {temporally maximized} ergotropy $\xi_\text{max}$, (b) {temporally minimized} anti-ergotropy $\mathcal{P}_\text{min}$, (c) capacity $\mathcal{Q}$, and (d) { temporally maximized quantum coherence} $\mathcal{C}_\text{max}$ versus temperature $T/\omega$ and magnetic field $B/\omega$. Parameters: $\Delta/\omega = 0.1$, $\epsilon/\omega = 0.1$, and $D = G = 0$.}}
\label{f17}
\end{figure}

The analysis identifies optimal performance regions at low temperatures ($T/\omega \lesssim 2.0$) and moderate magnetic fields ($B/\omega \approx 0.5-1.0$), where maximum ergotropy reaches $\xi_\text{max} = 3.6$. The anti-ergotropy spans around $-4.2$ to $0$, displaying inverse correlation with ergotropy that confirms fundamental thermodynamic constraints. The capacity follows identical enhancement patterns, achieving peak values of $3.6$ in regions coincident with optimal ergotropy.

The maximum coherence reaches $\mathcal{C}_\text{max} = 1.8$, with peak coherence regions strongly correlating with optimal QB performance zones. The systematic degradation of all metrics with increasing temperature confirms that thermal fluctuations destroy essential conditions for enhanced operation, while magnetic fields provide counteracting enhancement through energy level structure modification. These optimization studies establish quantitative guidelines for QB design: anisotropy parameters should be balanced and moderate, exchange interactions should be maximized simultaneously, and operation requires low-temperature environments with moderate magnetic field enhancement.
}
{

{
\section{Towards Experimental Realization of Magnetic Dipolar QBs}
\label{sec:experimental}

The NMR platform provides a robust experimental framework for realizing the proposed magnetic dipolar QBs, leveraging precise control over nuclear spin dynamics to implement the theoretical model described in Sec.~\ref{sec2} \cite{micadei2019reversing,kuprov2023spin}. Our QB model utilizes a dipolar spin system with transverse Zeeman splitting, DM, and KSEA interactions, initialized in a Gibbs thermal state and charged via a cyclic unitary process driven by a Pauli-$X$ Hamiltonian. NMR's ability to manipulate nuclear spins with high-fidelity RF pulse sequences and measure quantum coherence through spectroscopy makes it ideal for validating the theoretical predictions of ergotropy, anti-ergotropy, capacity, and quantum coherence dynamics~\cite{micadei2019reversing,kuprov2023spin,levitt2008spin,cory2000nmr,jones2011quantum}.

To implement the magnetic dipolar QB, the experimental proposal begins by preparing the Gibbs thermal state, representing the uncharged, passive state. In NMR, the thermal equilibrium state of nuclear spins in a strong magnetic field naturally approximates the Gibbs state at ambient (liquid-state NMR at around 300 Kelvin) or cryogenic temperatures (solid-state NMR at below 300 Kelvin).

For a two-spin-1/2 system, such as $^{1}\text{H}$ or $^{13}\text{C}$ nuclei in molecules like chloroform or trans-crotonic acid~\cite{jones2011quantum}, the natural thermal state can be manipulated to closely match the desired Gibbs state using established techniques like the Broekaert-Jeener two-pulse sequence or spatial averaging methods~\cite{jeener1967nuclear,doronin2007dipolar}. These methods exploit RF pulses and gradient fields to adjust spin populations, ensuring that the initial state aligns with the theoretical Gibbs state at finite temperature. The choice of low temperature ensures that quantum effects dominate while remaining experimentally feasible.

The two-spin system can be realized using a heteronuclear spin pair in liquid-state NMR~\cite{jones2011quantum}, where the spins are coupled via scalar ($J$-coupling) and dipolar interactions. For instance, a molecule with two distinct nuclear spins (e.g., $^{1}\text{H}$ and $^{13}\text{C}$) in a high-field NMR spectrometer provides a controllable platform. The effective Hamiltonian can be engineered by combining the natural dipolar coupling, external Zeeman field, and tailored interactions to emulate DM and KSEA terms. The DM interaction can be simulated using effective Hamiltonians generated by time-dependent RF sequences, while the KSEA interaction can be approximated through coherent control of spin-spin couplings~\cite{micadei2019reversing,kuprov2023spin,levitt2008spin}. The axial and rhombic anisotropy parameters can be tuned by adjusting molecular geometry or applying field gradients to modulate the dipolar interaction tensor~\cite{joseph2025decoupling,rembold2020introduction}.

The QB charging process, driven by the Pauli-$X$ charging Hamiltonian, is implemented using RF pulses resonant with the Larmor frequencies of the two spins. A sequence of $\pi/2$ or $\pi$ pulses along the $x$-axis induces rotations on the Bloch sphere, effectively applying the charging unitary operator. For a given charging field strength, the pulse duration and amplitude are calibrated to achieve the desired rotation angle for optimal ergotropy. The charging process is cyclic, with the external field disconnected after reaching maximum ergotropy to prevent energy reversion.

Discharging involves applying a reverse unitary operation to extract work, transitioning the system from the active state to a passive state. This can be achieved by applying a conjugate sequence of RF pulses to align the state with the lowest-energy configuration of the QB Hamiltonian. The NMR platform allows precise control over these unitaries through pulse sequence design, with composite pulses (e.g., BB1 or CORPSE) ensuring robustness against pulse imperfections~\cite{joseph2025decoupling,ryan2010robust}.

To mitigate parasitic spin states that could degrade efficiency, dynamical decoupling sequences (e.g., CPMG or Uhrig) can be employed during idle periods to suppress unwanted coherences while preserving the desired quantum state~\cite{joseph2025decoupling,viola1999dynamical,uhrig2007keeping}. The buffer gate, critical for maintaining coherence, can be implemented using refocusing sequences like WAHUHA or MREV-8, which average out undesired internal Hamiltonian terms~\cite{haeberlen1968coherent}.

The Lindblad master equation with Pauli-$X$ dephasing models the dissipative dynamics observed in NMR experiments, particularly transverse relaxation. Pauli-$X$ dephasing can be experimentally introduced using magnetic field gradients to induce position-dependent phase accumulation, mimicking the collapse operators. The dephasing rate can be controlled by adjusting gradient strength or duration, allowing precise emulation of the theoretical noise model~\cite{pignol2024decoherence,bochkin2020many}. Longitudinal relaxation can be simulated by coupling spins to a controlled thermal bath, such as a lattice of auxiliary spins or an engineered noise source via acoustic modulation in solid-state NMR~\cite{krogmeier2024low}.

To study the impact of noise on QB performance, controlled decoherence can be introduced using randomized pulse sequences or gradient-induced dephasing, enabling the measurement of coherence decay rates and their effect on ergotropy and capacity. Average Liouvillian theory can be applied to model non-unitary dynamics arising from pulse imperfections or environmental noise, providing a framework to simulate realistic error models and optimize error correction strategies~\cite{ghose2000average}.

Real-time monitoring of the QB's charging and discharging dynamics is achieved through high-resolution NMR spectroscopy, which measures spin population dynamics and coherences via FID signals. Quantum state tomography provides complete reconstruction of the density matrix, enabling direct quantification of quantum coherence and performance metrics like ergotropy, anti-ergotropy, and capacity. QST in NMR typically involves a series of readout pulses to measure expectation values of Pauli operators, reconstructing the state with high fidelity~\cite{jones2011quantum}.

The theoretical findings of this work can be validated by systematically varying experimental parameters. For instance, the axial anisotropy $\Delta$ can be tuned by adjusting the dipolar coupling strength through molecular orientation or field gradients, while DM and KSEA interactions can be engineered via pulse sequences that mimic their effective Hamiltonians. The Zeeman field $B$ is controlled by the static magnetic field of the spectrometer. Temperature effects can be studied by varying the sample temperature in a cryostat, confirming the degradation of coherence and ergotropy at higher temperatures.

We anticipate that scaling this two-spin model to a larger number of spins will significantly enhance performance metrics, particularly ergotropy and charging power, due to the collective effects of spin-spin interactions and quantum correlations \cite{campaioli2024colloquium,quach2022superabsorption,gao2022scaling,rossini2020quantum}. We expect that increasing the number of spins will amplify quantum coherence, driven by the KSEA and DM interactions, potentially leading to superextensive scaling of power. We suggest that the considered collective interaction in this work will continue to boost ergotropy in larger systems, with the KSEA interaction enhancing quantum coherence and correlations, which are critical for QB efficiency. }
{
\section{Conclusion and Outlook}
\label{sec6}
This study presents a comprehensive investigation of a quantum battery (QB) based on a magnetic dipolar spin system, incorporating Zeeman splitting, Dzyaloshinskii-Moriya (DM), and Kaplan--Shekhtman--Entin-Wohlman--Aharony (KSEA) interactions. We systematically analyzed the dynamics of quantum resources, namely, the $l_1$-norm of quantum coherence, quantum discord, and concurrence under thermal equilibrium and Pauli-$X$ dephasing dynamics, using the Lindblad master equation. These resources were then correlated with the performance metrics of the QB, modeled with the Gibbs thermal state as the uncharged state and charged via a cyclic unitary process driven by a Pauli-$X$ Hamiltonian. Our findings reveal intricate relationships between system parameters, quantum resources, and QB performance, offering new insights into quantum energy storage.

Key results demonstrate that the $z$-component Zeeman splitting field suppresses quantum resources in both thermal and dephasing scenarios, yet significantly enhances QB performance metrics, ergotropy, anti-ergotropy, capacity, and quantum coherence during cyclic charging (discharging), achieving up to a four-fold increase in extractable work. The axial anisotropy parameter ($\Delta$) has minimal impact on quantum resources under dephasing but bolsters their thermal robustness and markedly improves QB metrics, with ergotropy scaling linearly with $\Delta/\omega$. Notably,  saturates at high $\Delta$, indicating that increased ergotropy stems from incoherent contributions, a phenomenon consistent with incoherent energy storage mechanisms \cite{francica2020quantum}. The $z$-component KSEA interaction consistently amplifies quantum resources and QB performance, driving nonlinear quantum coherence enhancement through collective effects. In contrast, the $z$-component DM interaction enhances QB metrics and thermal resource resilience but shows negligible influence on dephasing dynamics. The rhombic parameter $\epsilon$ universally improves quantum resources and QB performance across all scenarios.

Temporal optimization studies of various QB metrics reveal that balanced $\Delta$ and $\epsilon$, maximized DM and KSEA, and low temperatures with moderate $B$ yield peak QB performance. Intriguingly,  quantum coherence does not always guarantee ergotropy, and high ergotropy can occur with minimal quantum coherence, underscoring the complex interplay between quantum and classical contributions to energy storage. These findings establish magnetic dipolar systems as a versatile platform for QBs, with tunable interactions enabling tailored performance.

Looking ahead, we propose the NMR platform as a promising testbed for experimental realization, leveraging RF pulse sequences and quantum state tomography to monitor charging dynamics and validate theoretical predictions. Future research could explore non-unitary charging protocols to reduce decoherence, investigate environmental effects on QB stability, and optimize pulse designs for greater efficiency. Additionally, extending the model to larger spin systems could examine superextensive scaling and super-absorption, potentially unlocking quantum advantages for practical applications. This work lays a strong foundation for advancing QB technology, bridging theoretical insights with experimental feasibility and opening avenues for innovative energy storage solutions in quantum systems.
}
\appendix
\section{Elements of thermal state, eigenvalues, and eigenstates}
\label{App0}
{The elements of the thermal state Eq. \eqref{EQ7} read

\begin{equation}
\zeta_{11} = \frac{\cosh{\mathcal{J}} - \frac{2B\,\sinh{\mathcal{J}}}{\mathcal{J} \,T}}{2(e^{\frac{4\Delta}{3T}}\cosh{\mathcal{S}} + \cosh{\mathcal{J}})},
\end{equation}
\begin{equation}
\zeta_{14} = \frac{i(G + i\,\epsilon)\sinh{\mathcal{J}}}{\mathcal{J}\,T(e^{\frac{4\Delta}{3T}}\cosh{\mathcal{S}} + \cosh{\mathcal{J}})},
\end{equation}
\begin{equation}
\zeta_{22} = \frac{1}{2} + \frac{1}{-2 - 2\,e^{\frac{4\Delta}{3T}}\cosh{\mathcal{S}}\,\mathrm{sech}{\mathcal{J}}},
\end{equation}
\begin{equation}
\zeta_{23} = \frac{e^{\frac{4\Delta}{3T}}(-3\,i\,D + \Delta)\sinh{\mathcal{S}}}{3\,T\,\mathcal{S}(e^{\frac{4\Delta}{3T}}\cosh{\mathcal{S}} + \cosh{\mathcal{J}})},
\end{equation}
\begin{equation}
\zeta_{44} = \frac{\cosh{\mathcal{J}} + \frac{2B\,\sinh{\mathcal{J}}}{\mathcal{J} \,T}}{2(e^{\frac{4\Delta}{3T}}\cosh{\mathcal{S}} + \cosh{\mathcal{J}})},
\end{equation}
The eigenvalues of thermal state Eq. \eqref{EQ7}  are obtained as
\begin{eqnarray}
\phi_{1,2} &=& \frac{e^{\frac{4\Delta}{3T}}\cosh{\mathcal{S}} \mp \frac{\mathcal{T}}{\kappa_{1} \kappa_{2}}}{2(e^{\frac{4\Delta}{3T}}\cosh{\mathcal{S}} + \cosh{\mathcal{J}})} \label{eigen01_1}
\end{eqnarray}
and
\begin{eqnarray}
\phi_{3,4} &=& \frac{\cosh{\mathcal{J}} \mp \frac{\sqrt{\kappa_{1}^{2}\kappa_{2}^{2}\sinh{\mathcal{J}}^{2}}}{\kappa_{1} \kappa_{2}}}{2(e^{\frac{4\Delta}{3T}}\cosh{\mathcal{S}} + \cosh{\mathcal{J}})}.\label{eigen01_2}
\end{eqnarray}
The corresponding orthonormal eigenstates are respectively given by
\begin{eqnarray}
\ket{\phi_1} &=& \Xi\,\left[ \alpha\left| 01\right\rangle +\left| 10\right\rangle \right], \label{phi1}
\end{eqnarray}
\begin{eqnarray}
\ket{\phi_2} &=& \Xi\,\left[-\alpha\left| 01\right\rangle +\left| 10\right\rangle \right], \label{phi2}
\end{eqnarray}
\begin{eqnarray}
\ket{\phi_3} &=& \Lambda_{+} \left[-\frac{i\left( B+\mathcal{L}\right) }{G-i\,\epsilon}\left| 00\right\rangle +\left| 11\right\rangle\right], \label{phi3}
\end{eqnarray}
\begin{eqnarray}
\ket{\phi_4} &=& \Lambda_{-} \left[ -\frac{i\left( B-\mathcal{L}\right) }{G-i\,\epsilon}\left| 00\right\rangle +\left| 11\right\rangle\right],\label{phi4}
\end{eqnarray}
In the above equations we set $\mathcal{J}=\frac{2\kappa_{2}}{T}$,  $\mathcal{S}=\frac{2\kappa_{1}}{3T}$,
%================================================================================
$\alpha=\frac{e^{\frac{4\Delta}{3T}}\eta \kappa_{1}\kappa_{2}\sinh{\mathcal{S}}}{\mathcal{T}}$,
%================================================================================
 $\mathcal{T}=\sqrt{e^{\frac{8\Delta}{3T}}\kappa_{1}^2\kappa_{2}^2\sinh{\mathcal{S}}^{2}}$ and
 %================================================================================
  $\mathcal{L}=\frac{\mathcal{T}\csch{\mathcal{J}}}{\kappa_{1}}$ with the normalization factors as
%===================================================================================
 $\Xi=\frac{1}{\sqrt{1+|\alpha|^{2}}}$ and
%===================================================================================
$\Lambda_{\pm}=\frac{1}{\sqrt{1+\left|\frac{ B\pm\mathcal{L}}{G-i\,\epsilon} \right|^{2}}}$.

Note that these expressions are used throughout the paper to derive quantum resources and performance metrics of magnetic dipolar QB.

\section{$l_1$-norm of quantum coherence}
\label{App1}
In quantum information theory, quantum coherence is a valuable resource with broad applications \cite{schumacher1998quantum, cwiklinski2015limitations, huelga2013vibrations, garttner2018relating}. Although it gained prominence with the work of Plenio \textit{et al.}, who introduced its quantification via the $l_1$-norm and relative entropy of quantum coherence \cite{baumgratz2014quantifying},  quantum coherence derives from the superposition principle and is crucial for phenomena not possible in classical mechanics. It underpins multipartite interference and quantum correlation. The resource theory of quantum coherence \cite{streltsov2017colloquium,baumgratz2014quantifying} identifies incoherent states $\delta$ as those diagonal in a reference basis $\{|i\rangle\}$, defined by
\begin{equation}
\nabla \in \delta \iff \nabla = \sum_i \nabla_i |i\rangle\langle i|.
\label{eq}
\end{equation}
Incoherent operations map incoherent states to incoherent states. Baumgratz \textit{et al.} \cite{baumgratz2014quantifying} proposed the $l_1$ norm of quantum coherence as a quantifier:
\begin{equation}\label{B2}
\mathcal{C}(\rho) = \sum_{i\neq j} |\langle i|\rho|j\rangle| = \sum_{i,j} |\rho_{ij}| - \sum_i |\rho_{ii}|,  
\end{equation}
where $\rho$ is the density operator of the system.
\section{Quantum discord}
\label{App2}
Quantum discord, a novel quantum correlation, was introduced by Olivier and Zurek \cite{ollivier2001quantum,zurek2003quantum} and detailed by Vedral \textit{et al.} \cite{henderson2001classical}. For a bipartite state $\rho \in \mathcal{H}_A \otimes \mathcal{H}_B$, the quantum mutual information is
\begin{equation}
\mathcal{I}(\rho) = S(\rho_A) + S(\rho_B) - S(\rho),
\end{equation}
where $S(\rho) = -\text{Tr}(\rho \log_2 \rho)$ is the von Neumann entropy, and $\rho_A$ ($\rho_B$) denotes the reduced density matrix in $\mathcal{H}_A$ ($\mathcal{H}_B$). For a von Neumann measurement on subsystem $A$ with projective operators ${E_i^A}$, the conditional state $\rho_i$ is
\begin{equation}
\rho_i = \frac{1}{p_i}(E_i^A \otimes \hat{\mathbb{I}})\rho(E_i^A \otimes \hat{\mathbb{I}}),
\end{equation}
where $p_i = \text{Tr}[(E_i^A \otimes \hat{\mathbb{I}})\rho(E_i^A \otimes \hat{\mathbb{I}})]$. The post-measurement quantum mutual information is
\begin{equation}
\mathcal{I}(\rho|{E_i^A}) = S(\rho_B) - S(\rho|{E_i^A}),
\end{equation}
with $S(\rho|{E_i^A}) = \sum_i p_i S(\rho_i)$ being the quantum conditional entropy. The classical correlation is \cite{zurek2003quantum,henderson2001classical}
\begin{equation}
C_A(\rho) = \sup_{{E_i^A}} \mathcal{I}(\rho|{E_i^A}),
\end{equation}
and the quantum discord is
\begin{equation}
\mathcal{D}(\rho) = \mathcal{I}(\rho) - C_A(\rho).
\end{equation}
\section{Concurrence}
\label{App3}
Concurrence is an entanglement monotone for quantifying the entanglement in bipartite states \cite{wootters1998entanglement}. For a density matrix $\rho$, the concurrence $\mathcal{E}(\rho)$ is defined as
\begin{equation}
\mathcal{E}(\rho) = \max\{0, \lambda_1 - \lambda_2 - \lambda_3 - \lambda_4\},
\label{ce1}
\end{equation}
where $\lambda_i$ are the square roots of the eigenvalues of the matrix $\rho (\hat{\sigma}_y \otimes \hat{\sigma}_y) \rho^* (\hat{\sigma}_y \otimes \hat{\sigma}_y)$ in decreasing order. Here, $\hat{\sigma}_y$ is the $y$-component of the Pauli matrices, and $\rho^*$ is the complex conjugate of $\rho$. The measure \eqref{ce1} ranges from 0 to 1, with $\mathcal{E}(\rho) = 0$ indicating no entanglement and $\mathcal{E}(\rho) = 1$ indicating maximal entanglement.

\vspace{0.5cm}

% \section{Closed-form expressions for performance metrics of magnetic dipolar QB}
% \label{App4}
% The expressions for capacity  $\mathcal{Q}(\omega)$ and  ergotropy $\xi(\omega)$ (also the anti-ergotropy $\mathcal{P}=d\xi/dt$) turn out to be 
% \begin{equation}
% \small
% \mathcal{Q}=\frac{2 U_1 (3 B+2 \Delta ) e^{\frac{4 \Delta }{3 T}}+6 B V_2+2 Q_1 U_2 e^{\frac{4 \Delta }{3 T}}+6 Q_2 V_1}{3 \left(U_1 e^{\frac{4 \Delta }{3 T}}+V_2\right)}
% \end{equation}
% and
% \begin{widetext}
% \begin{align}
% \xi = \frac{2 e^{-\frac{2 \Delta}{3 T}}}{Z} & \bigg[ 
% (-\Delta + i D + \epsilon) \sin^2(2 \omega t) V_2  \nonumber + (\Delta - i D - \epsilon) e^{\frac{2 (\Delta - i D)}{T}} \sin^2(2 \omega t)\\
% &  + \frac{2 \sin^2(\omega t) V_1 \left(2 B^2 + \epsilon (-\Delta + i D + \epsilon) + \epsilon (-\Delta + i D + \epsilon) \cos(2 \omega t) + 2 G^2\right)}{Q_2}    \bigg],
% \end{align}
% where 
% \begin{align}
% U_1 &= \cosh \left(\frac{2 \sqrt{\Delta^2 + 9 D^2}}{3T}\right), \quad
% U_2 = \sinh \left(\frac{2 \sqrt{\Delta^2 + 9 D^2}}{3T}\right), \quad
% V_1 = \sinh \left(\frac{2 \sqrt{B^2 + G^2 + \epsilon^2}}{T}\right), \nonumber \\
% V_2 &= \cosh \left(\frac{2 \sqrt{B^2 + G^2 + \epsilon^2}}{T}\right), \quad
% Q_1 = \sqrt{\Delta^2 + 9 D^2},  \quad
% Q_2 = \sqrt{B^2 + G^2 + \epsilon^2}.
% \end{align}
% \end{widetext}

%\vspace{8cm}

\section*{Disclosures}
The authors declare that they have no known competing financial interests.

\section*{Data availability}
No datasets were generated or analyzed during the current study.

\bibliography{Amref} 
\end{document}